\documentclass[11pt,a4paper]{article}
\pdfoutput=1
\usepackage{jheppub}
\usepackage{amsmath}
\usepackage{epsfig}
\usepackage{amssymb}
\usepackage{graphics}
\usepackage[active]{srcltx}
\usepackage{epstopdf}

\newcommand{\bea}{\begin{eqnarray}}
\newcommand{\eea}{\end{eqnarray}}
\newcommand{\bean}{\begin{eqnarray*}}
\newcommand{\eean}{\end{eqnarray*}}

\def\O #1{\overline{#1}}

\def\W #1{\widetilde{#1}}

\def\eref#1{(\ref{#1})}

\newcommand{\IC}{\mathbb{C}}

\def\Label#1{\label{#1}%
  \smash{\hbox to0pt{\raise1ex\hbox{\tiny[#1]}\hss}}}

\def\ruleI#1{\mathcal{R}_{{\scriptsize\mbox{ule}}}^{{\tiny\mbox{I}}}[#1]}
\def\ruleII#1{\mathcal{R}_{{\scriptsize\mbox{ule}}}^{{\tiny\mbox{II}}}[#1]}
\def\ruleIII#1{\mathcal{R}_{{\scriptsize\mbox{ule}}}^{{\tiny\mbox{III}}}[#1]}
\def\ruleIX#1{\mathcal{R}_{{\scriptsize\mbox{ule}}}^{{\tiny\mbox{IX}}}[#1]}

\def\rulei{\mathcal{R}_{{\scriptsize\mbox{ule}}}^{{\tiny\mbox{I}}}}
\def\ruleii{\mathcal{R}_{{\scriptsize\mbox{ule}}}^{{\tiny\mbox{II}}}}
\def\ruleiii{\mathcal{R}_{{\scriptsize\mbox{ule}}}^{{\tiny\mbox{III}}}}
\def\ruleix{\mathcal{R}_{{\scriptsize\mbox{ule}}}^{{\tiny\mbox{IX}}}}

\def\dunderline#1{\underline{\underline{#1}}}

\title{Feynman Rules of Higher-order Poles in CHY Construction}
\author[a]{Rijun Huang\footnote{The
unconventional ordering is to let authors get proper
recognition of contributions under the outdated practice in China.},}
\author[a,b]{Bo Feng,}
\author[a]{Ming-xing Luo}
\author[c]{and  Chuan-Jie Zhu}

\affiliation[a]{Zhejiang Institute of Modern Physics, Department of Physics,
 Zhejiang University, Hangzhou, 310027, P.R. China}
\affiliation[b]{Center of Mathematical Science,
  Zhejiang University, Hangzhou, 310027, P.R. China}
\affiliation[c]{Department of Physics, Renmin University of China, Beijing, 100872, P.R. China}

\emailAdd{huang@nbi.dk}
\emailAdd{fengbo@zju.edu.cn}
\emailAdd{mingxingluo@zju.edu.cn}
\emailAdd{zhucj@ruc.edu.cn}

\date{\today}
\abstract{
In this paper, we generalize the integration rules for scattering equations to situations
where higher-order poles are present. We describe the strategy to deduce the Feynman rules
of higher-order poles from known analytic results of simple CHY-integrands, and propose the
Feynman rules for single double pole and triple pole as well as duplex-double pole and
triplex-double pole structures. We demonstrate the validation and strength of these rules by ample non-trivial
examples.
}
\keywords{Scattering Equation, Amplitude, Integration Rule}

\begin{document}
\maketitle \flushbottom

\section{Introduction}

In the last few years, a new formulation for tree-level amplitudes
of massless theories has been presented by Cachazo, He and Yuan
[CHY] in a series of papers \cite{Cachazo:2013gna,Cachazo:2013hca,
Cachazo:2013iea, Cachazo:2014nsa,Cachazo:2014xea}. The formula is
given by
\bea
    {\cal A}_n& = & \int {\left(\prod_{i=1}^n dz_i\right)\over {\rm
vol}(SL(2,\IC))} \Omega({\cal E}) {\cal I}(z)
    =  \int {\left(\prod_{i=1}^n dz_i\right)\over d\omega}  \Omega({\cal E}) {\cal I} \ ,
    ~~\label{gen-A}
\eea
where $z_i$ are puncture locations of $n$ external particles in
${\rm CP}^1$ and $d\omega={d z_r d z_s d z_t\over z_{rs} z_{st}
z_{tr}}$ comes after we use the M\"obius $SL(2,\IC)$ symmetry to fix
the location of three of the variables $z_r, z_s, z_t$ by the
Faddeev-Popov method. The $\Omega$ is defined by
\bea \Omega({\cal E})\equiv z_{ij}z_{jk}z_{ki} \prod_{a\neq
i,j,k}\delta\left( {\cal E}_a\right) \ , ~~\label{measure-Omega}
\eea
where ${\cal E}_a$'s are the {\sl scattering equations} defined as
\bea
    {\cal E}_a\equiv \sum_{b\neq a} {s_{ab}\over z_a-z_b}=0~~,~~a=1,2,...,n \
    .
    ~~\label{SE-def}
\eea
As in the formula \eref{gen-A}, the CHY construction involves two
parts: {\sl solving the scattering equations} ${\cal E}_a$ which are
universal for all theories,  and {\sl formulating the CHY-integrand}
${\cal I}$ for a given theory.

Although conceptually, the CHY approach is remarkable and very
useful for many theoretical studies on the properties of scattering
amplitudes, when applying to practical evaluation, one confronts the
problem of solving scattering equations. The scattering equation
problem can be related to solving a polynomial equation system of
degree $(n-3)!$ \cite{Dolan:2014ega}. It is well-known that when
$n\geq 6$, in general there is no analytic way to solve
it\footnote{There is some discussion on the special solutions in
four-dimension \cite{Weinzierl:2014vwa,Du:2016blz}}. To deal with
this problem, many methods have been proposed. In
\cite{Kalousios:2015fya}, using classical formulas of Vieta, which
relates the sums of roots of polynomials to the coefficients of
these polynomials, analytic expression can be obtained without
solving roots explicitly. The view of algebraic geometry has been
further developed in several works. In \cite{Huang:2015yka}  the
so-called companion matrix method from computational algebraic
geometry, tailored for zero-dimensional ideals, to study the
scattering equations has been proposed. In \cite{Sogaard:2015dba},
the Bezoutian matrix method has been used to facilitate the
calculation of amplitudes. In \cite{Cardona:2015ouc,
Dolan:2015iln,Cardona:2015eba}, the elimination theory has been
exploited to great details for the group of scattering equations
with $(n-3)$ variables. In \cite{Lam:2015sqb,Lam:2016tlk} an
efficient method has been developed to evaluate the Yang-Mills
theory.

Different from the algebraic geometry methods, two more generic
algorithms have been proposed. In one approach
\cite{Cachazo:2015nwa}, using known results for scalar $\phi^3$
theory, one can iteratively decompose the {\sl 4-regular} graph
determined by the corresponding CHY-integrand to building blocks
related to $\phi^3$ theory, Which ends the evaluation. In another
approach \cite{Baadsgaard:2015voa, Baadsgaard:2015ifa}, by careful
analysis of pole structures, the authors wrote down an integration
rule, so that from the related CHY-integrand, one can read out
contributions from the corresponding Feynman diagrams directly if
there are only simple poles appear. The later is really an efficient
tool, however when higher-order poles are presented in a
CHY-integrand, it can not proceed furthermore because of lacking
Feynman rules for them.

In this paper, we would like to improve the efficiency of solving
scattering equations by generalizing {\sl integration rule} from
only simple poles to higher-order poles, in the sprint of
\cite{Baadsgaard:2015voa, Baadsgaard:2015ifa}. The key-point of our
generalization is to realize that, even for higher-order poles, one
can find properly-defined Feynman rules. Although in current stage,
we are not able to derive these generalized Feynman rules for
higher-order poles, we have provided substantial evidences to
support our claim.

This paper is organized as follows. In \S\ref{secReview}, we provide
a review of integration rule for simple poles. In \S\ref{secRule},
we state the Feynman rules for some higher-order poles, with
discussion on how to deduce these rules. Followed in
\S\ref{secSupporting}, we provide ample examples to support the
proposed rules. In the last section, discussion and conclusion are
provided.

{\bf Note added:} During the preparation of this paper, a new
formalism named $\Lambda$-algorithm has been presented by Gomez in
\cite{Gomez:2016bmv}, which provides an alternative algorithm to
deal with higher-order poles.

\section{Review of integration algorithm}
\label{secReview}

Before proposing the Feynman rules for higher-order poles, let us
for reader's convenience review the {\sl integration algorithm}
presented in
\cite{Baadsgaard:2015voa,Baadsgaard:2015ifa,Baadsgaard:2015hia}\footnote{In
these literatures, it is called {\sl integration rule}. To not get
confused with Feynman rules, we will call it {\sl integration
algorithm}.}. The whole algorithm can be roughly described as two
parts: (1) given a CHY-integrand, find out the corresponding Feynman
diagrams; (2) work out corresponding Feynman rules (or {\sl
generalized} Feynman rule while considering higher-order poles),
with which write down the result directly for given Feynman
diagrams.

\begin{figure}
  \centering
  \includegraphics[width=7in]{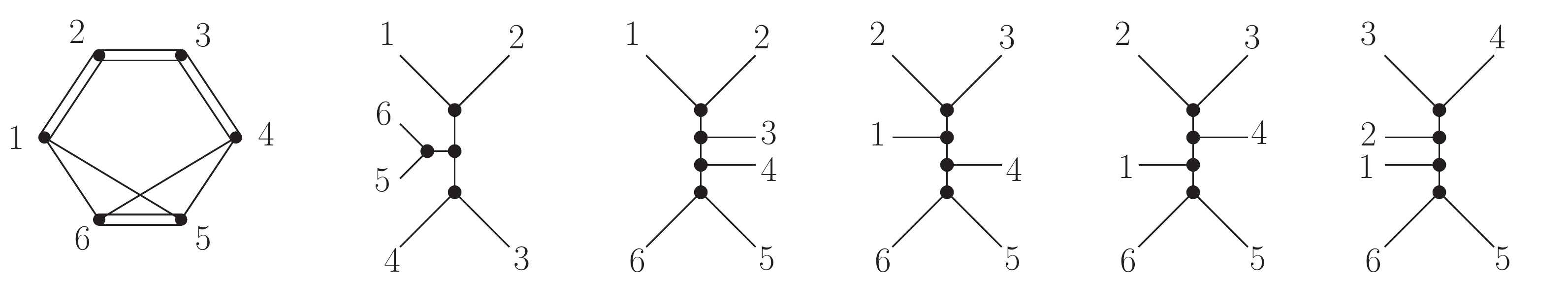}\\
  \caption{The left-most graph is an example of {\sl 4-regular} graph for certain CHY-integrand.
   The remaining five
  diagrams are contributing Feynman diagrams for amplitude of that CHY-integrand.}\label{Figsimplepole}
\end{figure}

For a $n$-point amplitude, the CHY-system is described by $n$
complex variables $\{z_1,z_2,\ldots, z_n\}$ before gauge fixing. The
CHY-integrand, as a rational function of $z_{ij}\equiv z_i-z_j$, can
be represented by a graph of $n$ nodes with many lines connecting
them. A factor $z_{ij}$ in denominator(or numerator) can be
represented by a solid line(or dashed line) connecting nodes $i$ and
$j$ with an arrow pointing from $i$ to $j$. A CHY-integrand which is
invariant under M\"obius transformation is then represented by a
{\sl 4-regular} graph, i.e., a graph such that for each node, the
number of attached solid lines minus the number of attached dashed
lines is four. This is illustrated by an example with only simple
poles as shown in Figure \ref{Figsimplepole}.
The corresponding CHY-integrand for this {\sl 4-regular}
graph\footnote{Throughout this paper we have omitted the arrows of
lines in the graph for simplicity. Also when we read out result
using {\sl integration algorithm}, there is an overall sign one need
to pay attention to. } is
\bea {1\over (1,2,3,4,5,6)(1,2,3,4,6,5)}={1\over
z_{12}^2z_{23}^2z_{34}^2z_{56}^2z_{45}z_{61}z_{46}z_{15}}~.~~~\label{simplepoleEx1}\eea

The purpose of {\sl integration algorithm} is to {\sl read out} the
result of a given CHY-integrand directly from the {\sl 4-regular}
graph without solving any scattering equations. As described in
\cite{Baadsgaard:2015voa, Baadsgaard:2015ifa,Baadsgaard:2015hia},
this is readily programmed for situations where only simple poles
are present. Now we present a review on {\sl integration algorithm}.
For a $n$-point system, let
$A=\{a_1,a_2,\ldots,a_m\}\subset\{1,2,\ldots, n\}$, which is a
subset of nodes. For each subset, let us define following {\bf pole
index}
\bea \chi(A):=\mathbb{L}[A]-2(|A|-1)~,~~~\label{pole-degree}\eea
where $\mathbb{L}[A]$ is the number\footnote{More accurately, it is
the difference of number between solid and dashed lines.} of lines
connecting these nodes inside $A$, and $|A|$ is the number of nodes.
The condition
\bea \chi(A)\geq 0~~~~\label{Pole-cond}\eea
will be called the {\bf pole condition} for a given subset.
Explicitly, each subset gives a non-zero pole contribution if and
only if $\chi(A)\geq 0$. The pole will be in the form ${1\over
s_A^{\chi(A)+1}}$, where $s_{A}=(p_{a_1}+p_{a_2}+\cdots+p_{a_m})^2$
for massless momentum $p_i$. Also, because of the momentum
conservation, we will treat the subset $A$ to be equivalent to its
complement $\O A=\{1,2,\ldots, n\}-A$, thus we can impose the length
condition $2\leq |A|\leq \lfloor {n\over 2}\rfloor$.

Once the pole structures of subsets are clear, we define the {\bf
compatible condition} for two subsets $A_1,A_2$: two subsets are
compatible to each other if one subset is completely contained
inside another subset or the intersection of two subsets is empty.

Now we state the {\sl integration algorithm}:
\begin{itemize}

\item (1) Given a CHY-integrand, draw the corresponding {\sl
  4-regular} graph. For example, for the CHY-integrand
  \eref{simplepoleEx1} we draw the graph in Figure
\ref{Figsimplepole}.

\item (2) Find all subsets $A$ with $\chi(A)\geq 0$. For simple
pole, we have $\chi(A)= 0$. For higher-order poles as considered
in this paper, we have $\chi(A)>0$.

For example \eref{simplepoleEx1}  (see Figure
\ref{Figsimplepole}), since $n=6$, we have $2\leq |A|\leq 3$.
For subsets with two nodes, it is easy to find that subsets
$\{1,2\},\{2,3\},\{3,4\}$ and  $\{5,6\}$ having $\chi=0$, thus
we should find simple poles $s_{12},s_{23},s_{34},s_{56}$. For
subsets with three nodes, $\{1,2,3\}\equiv \{4,5,6\}$ or
$\{2,3,4\}=\{1,5,6\}$ having $\chi=0$, thus we should also find
poles $s_{123},s_{234}$.

\item (3) Find all maximum {\sl compatible combinations}, i.e., the
combination of subsets with largest number such that each pair
in the combination is compatible.

Going back to our example, the combination
$\{\{1,2\},\{2,3\},\{5,6\}\}$ is not a compatible combination
since $\{1,2\}$ and $\{2,3\}$ do not satisfy compatible
condition. For this example, we can find five maximum compatible
combinations
\bea
&&\{\{1,2\},\{3,4\},\{5,6\}\}~~,~~\{\{1,2,3\},\{1,2\},\{5,6\}\}~~,
~~\{\{1,2,3\},\{2,3\},\{5,6\}\}~,~~~\nonumber\\
&&\{\{2,3,4\},\{2,3\},\{5,6\}\}~~,~~\{\{2,3,4\},\{3,4\},\{5,6\}\}~.~~~\label{Ex1-5comb}\eea

\item (4) {\bf Length condition for combination}: For each maximum combination with $m$ subsets, it gives
non-zero contribution if and only if $m=n-3$. In the current
example, all five combinations \eref{Ex1-5comb} give non-zero
contributions.

\item (5) For each compatible combination which has non-zero contribution, we can
draw the corresponding Feynman diagram with only cubic vertices,
and the propagators respecting the pole structures of compatible
combination. Then we can use the (generalized) Feynman rule to
write down results. For the above five compatible combinations
given in \eref{Ex1-5comb}, the corresponding Feynman diagrams
are presented in Figure \ref{Figsimplepole}. Using the rule for
simple pole, we can write down result immediately as
\bea {1\over s_{12}s_{34}s_{56}}+{1\over
s_{123}s_{12}s_{56}}+{1\over s_{123}s_{23}s_{56}}+{1\over
s_{234}s_{23}s_{56}}+{1\over s_{234}s_{34}s_{56}}~.~~~\eea
\end{itemize}

From above {\sl integration algorithm}, one can see that all the
five steps are in fact computer task except one, i.e., the
determination of Feynman rules. For simple pole, the Feynman rule is
nothing but the usual Feynman propagator ${1\over s_A}$. For
higher-order poles, the story becomes complicated, and they are the
main context we would discuss in the following sections.

Before going on, there are some general remarks. The first is, why
we believe there are Feynman rules for higher-order poles? The first
hint comes from the {\sl pinching} picture given in
\cite{Feng:2016nrf}. When discussing contributions with a given pole
structure, it is found that we can group some nodes together to
reduce a CHY-graph to two CHY-graphes with fewer nodes. Thus the
contribution can be roughly written as $T_{s_A}\sim T_L \times
{1\over s_A} \times T_R$, where $T_L, T_R$ have natural tree
amplitude structure. Iterating the pinching procedure we can reduce
a big CHY-graph to some building blocks. This picture is nothing but
the familiar one when cutting a propagator in Feynman diagram to
reduce a big diagram to two sub-diagrams, while the building block
is nothing but the Feynman rules.

Keeping above picture in mind, now we present a more detailed
discussion on the {\sl pinching} operation. Let us divide $n$ nodes
to the subset $A\equiv \{z_{a_1},z_{a_2},\ldots,z_{a_m}\}$ and its
complement $B\equiv\{z_{a_{m+1}},z_{a_2},\ldots,z_{a_n}\}$. The
pinching corresponds to representing the whole subset by a new node
and then adding this new node to the graph, which means we will get
two new sets $\W A\equiv \{z_{a_1},z_{a_2},\ldots,z_{a_m}, z_B\}$
and $\W B\equiv\{z_A,z_{a_{m+1}},z_{a_2},\ldots,z_{a_n}\}$.
Furthermore, all lines $z_{ab}$ with $z_a\in A$ and $z_b\in B$ will
be modified as follows: $z_{ab}\to z_{aB}= z_a -z_B$ in the new set
$\W A$ and $z_{ab}\to z_{Ab}=z_A- z_b$ in the new set $\W B$. Now we
want each new set $\W A, \W B$ to give legitimate CHY-graphs, i.e.,
{\sl 4-regular} graph. To see the condition, let us use
$\mathbb{L}[A], \mathbb{L}[B],\mathbb{L}[A,B]$ to denote the number
of lines inside the subset $A$, $B$ and $\{A,B\}$ respectively. Then
we have
\bea 2n= \mathbb{L}[A]+ \mathbb{L}[B]+\mathbb{L}[A,B]~~,~~
2(m+1)=\mathbb{L}[A]+ \mathbb{L}[A,B]~~,~~2(n-m+1)=
\mathbb{L}[B]+\mathbb{L}[A,B]~.~~~\eea
From these relations, we find
\bea \mathbb{L}[A,B]=4~.~~~\eea
This condition leads to $\chi(A)=0$, i.e., the subset contributes to
simple pole ${1\over s_{A}}$. It means we can only pinch nodes with
simple poles.

Above pinching condition has an important implication. Using
pinching, we can reduce a bigger CHY-graph to a smaller one as shown
in following figure:
\begin{center}
\includegraphics[width=3in]{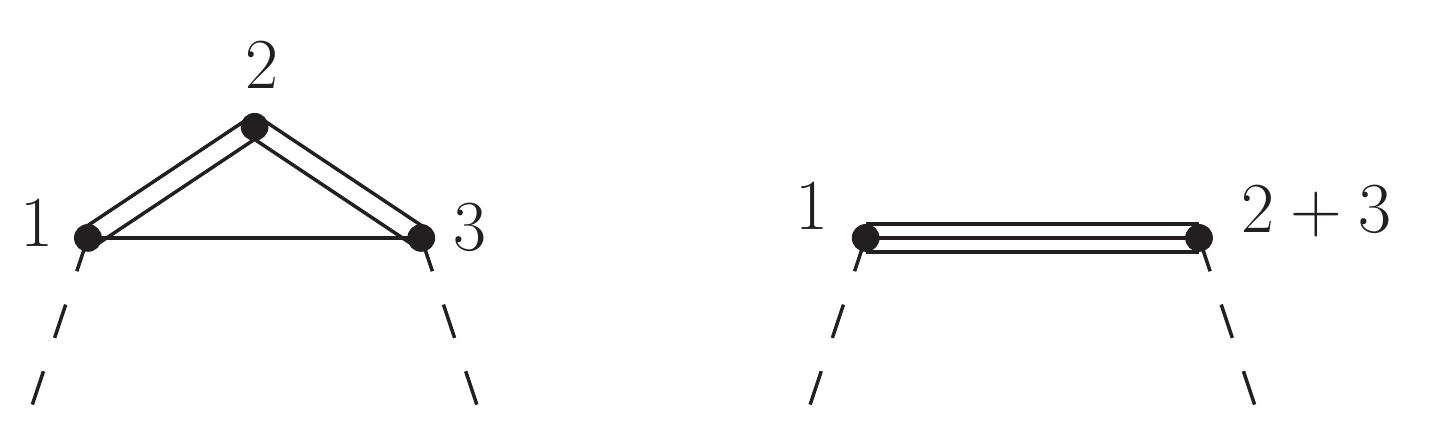}
\end{center}
Iterating the procedure, we can reduce any graphs to a much simpler
primary graph. No matter what the reduced graph looks like, its
higher pole structure is invariant, i.e., the number and the degree
of higher-order poles are not changing. The pinching picture has
also inspired us to find the Feynman rule for a given higher-order
pole with the strategy: {\sl solving the simplest CHY-integrand
containing such pole structure and then properly generalizing it}!
This will be the approach we use in the paper.

Our result of Feynman rules for higher-order poles has also revealed
that, while the propagator of simple pole structure should be
considered to be complete local (i.e., its Feynman rule ${1\over
s_A}$ depends only on the momentum flowing in between), the
propagator of higher-order poles should be properly considered as
{\sl quasi-local}, i.e., its rule will depend also on the momentum
as well as the types of poles that connecting to the four corners of
propagators.

\section{The Feynman rules for higher-order poles}
\label{secRule}

As reviewed in \S\ref{secReview}, the {\sl integration algorithm}
can be readily generalized to CHY-integrand with higher-order poles
if we have corresponding Feynman rules for them. Once the Feynman
rules of higher-order poles are in order, we can produce the results
of any CHY-integrands instantly,  following the standard five steps
of {\sl integration algorithm}.

In this section, we will provide Feynman rules for some higher-order
poles. The general strategy of deducing these Feynman rules are
described as follows. Firstly, we find out the simplest
CHY-integrand which would contain the required higher-order pole,
and obtain the analytic result by any reliable method. In many
cases, one would find more than one Feynman diagram for a given
CHY-integrand, and from them we need to isolate the contribution
that only generated by Feynman rule of that higher-order pole. Then
we generalize the result to generic situations by considering the
symmetry structures of poles, the kinematics as well as when the
external legs are massive. This involves {\sl trial-and-error} by
some other CHY-integrands to fix ambiguities. Once the result is
formulated as a {\sl rule}, then it can be applied in the
computation of any CHY-integrands that containing corresponding
higher-order poles.

\subsection{The Feynman rule $\rulei$ for a single double pole}

%
\begin{figure}[h]
  \centering
  \includegraphics[width=5.5in]{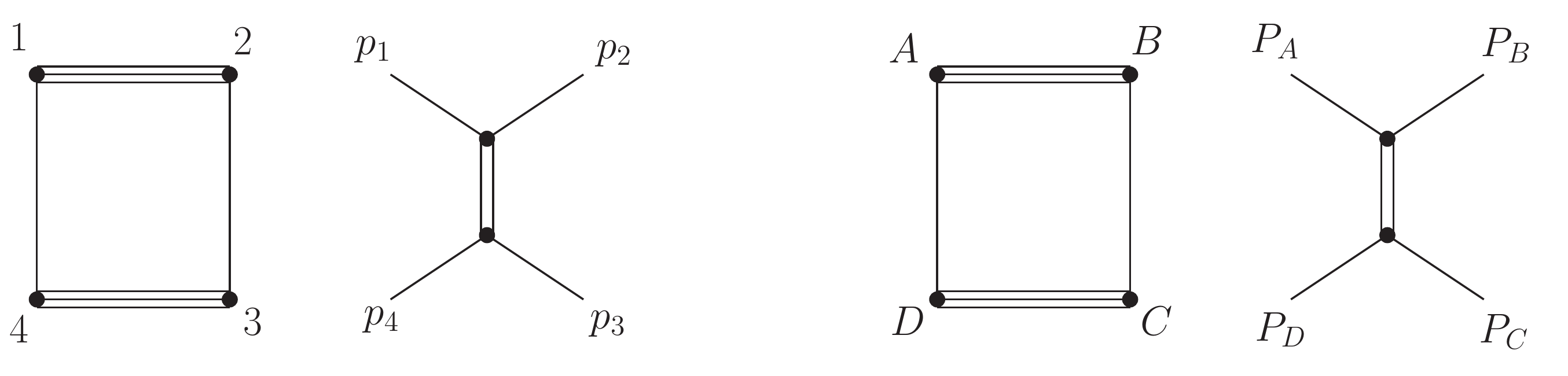}\\
  \caption{Left: The {\sl 4-regular} graph of a 4-point CHY-integrand and its
  corresponding Feynman diagram. The {\sl double-line} propagator denotes a double pole of this CHY-integrand.
  Right: Labels of external momenta for the Feynman rule $\rulei$ when legs are massive.}\label{Figrule1}
\end{figure}
%

\subsection*{The Feynman rule}

For external legs as labeled in the right-most diagram of Figure
\ref{Figrule1}, the Feynman rule for a single double pole is
formulated as
\bea \ruleI{P_A,P_B,P_C,P_D}={2P_AP_C+2P_BP_D\over
2s_{AB}^2}~,~~~\label{rule1} \eea
where $P$ with capital letters as subscripts denotes a sum of
several massless legs, e.g., $P_A=\sum_{i=1}^mp_{a_i}$. Terms such
as $2P_AP_B$ is understood to be the Minkowski products $2P_A\cdot
P_B=s_{AB}-P_A^2-P_B^2$. We could notice the difference between
$2P_AP_B$ and $s_{AB}$ when external legs are massive.

Two remarks before proceed. Firstly, the Feynman diagram in Figure
\ref{Figrule1} is blind with ordering. But for a given {\sl
4-regular} graph of CHY-integrand as shown in Figure \ref{Figrule1},
especially paying attention to the lines connecting nodes $A,D$ and
nodes $B,C$, the resulting CHY-integrand do depends on the ordering
of nodes, e.g., the ordering $\{A,B,C,D\}$ gives different result
against ordering $\{A,B,D,C\}$. This point is very important when
applying the rules of higher-order poles. One should be very careful
when drawing the Feynman diagrams for CHY-integrands, especially for
those legs connecting to the propagators of double poles. Secondly,
different from the simple pole which is completely local, rule for
higher-order pole depends not only on the total momentum $P_A+P_B$
flowing through the propagator, but also momenta $P_A,P_B,P_C,P_D$
at four corners. Hence we call this propagator to be {\sl
quasi-local}. As we shall see very soon, when two propagators of
double poles are connected by a vertex, the Feynman rule for this
kind of pole structure will be different, i.e., we can not apply
rule (\ref{rule1}) to those two propagators, but consider them as a
single object. This quasi-local property is the reason that deducing
Feynman rules for higher-order poles is quite difficult.

\subsection*{Deducing the Feynman rule}

As mentioned, the strategy is to find out the simplest CHY-integrand
that containing the required higher-order pole structure and obtain
the analytic result by any means. Thus we start from the simple
four-point CHY-integrand,
\bea {\cal I}= -{1\over z_{12}^3 z_{23} z_{34}^3
z_{41}}~.~~~\label{single-double-4p-1}\eea
It can be trivially computed by directly solving scattering
equations, with the result
\bea {s_{13}\over s_{12}^2}~.~~\label{single-double-4p-1-1}\eea

Now we try to reproduce above result by {\sl integration algorithm}
reviewed in previous section. We draw the {\sl 4-regular} graph for
this CHY-integrand, as shown in Figure \ref{Figrule1}. It is easy to
see that there are only two independent subsets of nodes, i.e.,
$\{1,2\}$ and $\{2,3\}$. While $\chi[\{2,3\}]=1-2(2-1)=-1$, and
$\chi[\{1,2\}]=3-2(2-1)=1$, we conclude that the only compatible
combination is $\{\underline{1,2}\}$, where an underline is to
emphasize that it corresponds to a double pole. There is only one
Feynman diagram corresponding to this four-point CHY-integrand, as
shown in Figure \ref{Figrule1} besides the {\sl 4-regular} graph,
where we use a double line to denote the double pole.

In order to deduce the Feynman rule for this double pole, let us try
to get some hints from the result (\ref{single-double-4p-1-1}). Note
that the {\sl 4-regular} graph (originates from the CHY-integrand)
apparently possesses symmetries under exchanging of nodes
\bea  \{1,2,3,4\}\leftrightarrow
\{4,3,2,1\}~~~\mbox{and}~~~\{1,2,3,4\}\leftrightarrow
\{2,1,4,3\}~,~~~\eea
while under such exchanging, the kinematic variables are changed as
\bea s_{13}\leftrightarrow s_{24}~.~~~\eea
It means that the numerator of (\ref{single-double-4p-1-1}) is not
invariant under above exchanging. We would like to symmetrize it
as\footnote{In fact, if we want to get a totally symmetric
expression under above exchanging, the denominator $s_{12}^2$ should
be rewritten as $s_{12}s_{34}$. However, since we know poles are
always in the form $((\sum_i p_i)^2)^a$, and by momentum
conservation we have $s_{12}=s_{34}$. So we can keep denominator as
$s_{12}^2$. However, for numerator the story would be quite
different.}
\bea  {s_{13}+s_{24}\over 2 s_{12}^2}~.~~~\label{DoublePole-4p-2}
\eea

However, expression \eref{DoublePole-4p-2} is not the final answer,
since $s_{12}$ is generally not equal to $2p_1\cdot p_2$ if $p_1$,
$p_2$ are massive. Since the numerator is not constrained by any
considerations, both $s_{ij}$ and $2p_i\cdot p_j$ could be possible
candidates. This motives us to consider the case when external
momenta are massive as shown in the right-most diagram of Figure
\ref{Figrule1}, which will be a sum over several massless momenta
for generic CHY-integrands. It forces us to determine a proper
formulation for (\ref{DoublePole-4p-2}) that can be applied to the
computation of more generic CHY-integrands. The denominator, since
contains only physical poles, should be $s_{AB}$ but not $2P_AP_B$.
In this case, $s_{AB}=s_{CD}$ because of momentum conservation and
we can take $s_{AB}^2$ as denominator. While for the numerator, a
five-point example of CHY-integrand would suffice to tell us which
one to choose. With {\sl trial-and-error} method, one find that for
massive case, we should use $2P_AP_B$ instead of $s_{AB}$ in the
numerator. Although it has no difference in the massless limit(i.e.,
the four-point CHY-integrand), it indeed introduces important
corrections to guarantee the correctness for generic CHY-integrands.
This leads us to formulate the Feynman rule of a single double pole
as in (\ref{rule1}).

\subsection*{An illustrative example}
%
\begin{figure}
  \centering
  \includegraphics[width=6in]{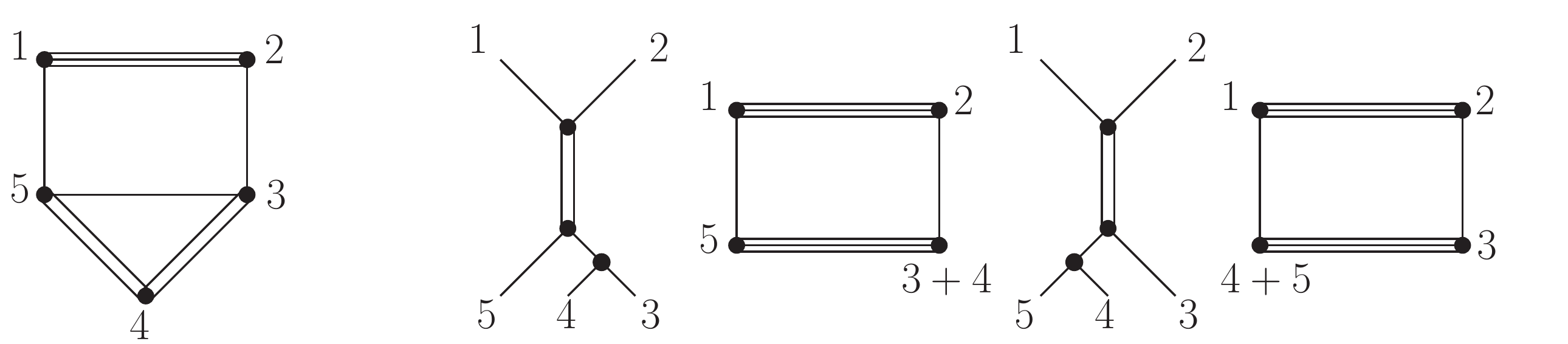}\\
  \caption{The {\sl 4-regular} graph of a given five-point CHY-integrand with
  a double pole, and two Feynman diagrams corresponding to this CHY-integrand. To indicate
  the ordering of two Feynman diagrams, we have drawn two pinched CHY-graphs. }\label{FigA52}
\end{figure}
Let us illustrate the Feynman rule $\rulei$ by a five-point
CHY-integrand
\bea -{1\over z_{12}^3z_{23}
z_{34}^2z_{45}^2z_{53}z_{51}}~,~~~\nonumber\eea
which is also studied in \cite{Cachazo:2015nwa} by standard KLT
construction. We can draw the {\sl 4-regular} graph for this
CHY-integrand, as shown in Figure \ref{FigA52}. By counting the
number of lines that connecting subsets of nodes, it is trivial to
find that $\{\underline{1,2}\}$ contributes to a double pole, while
$\{3,4\}$, $\{4,5\}$ contribute to a simple pole. The compatible
combinations consist $5-3=2$ subsets of nodes that satisfying
compatible condition, and here we have two compatible combinations
as
\bea
\{\{\underline{1,2}\},\{3,4\}\}~~~,~~~\{\{\underline{1,2}\},\{4,5\}\}\eea
From them we can draw two Feynman diagrams which have the required
propagators as shown in Figure \ref{FigA52}. Note that for simple
poles, the Feynman rule is not affected by the ordering of external
legs attached to the propagator. For example, switching leg 3 and
leg 4 of the first Feynman diagram in Figure \ref{FigA52}, we still
get ${1\over s_{34}}$ for the propagator. However, the Feynman rules
of higher-order poles do depend on the ordering of external legs, as
can be seen in the definition (\ref{rule1}). The ordering of legs
can be traced back to the four-point {\sl 4-regular} graph. For
example, legs 3 and 4 in the first Feynman diagram are combined
together and become a massive leg of the double pole, and the
ordering of legs are determined by the ordering of nodes in the
four-point {\sl 4-regular} graph besides the Feynman diagram. When
applying the Feynman rule $\rulei$ for the double pole, we should
refer to the ordering marked by the {\sl 4-regular} graph. In order
to use the Feynman rule of double pole directly in the Feynman
diagram, we should draw the external legs in a definite ordering by
using the four-point {\sl 4-regular} graph as assist. Hereafter, we
will always draw Feynman diagrams with definite ordering of legs
attached to the higher-order poles so that we can read out the rules
directly from them, but keep in mind that this ordering of legs is
determined by the {\sl 4-regular} graph.

Now we can write down the final result by summing contributions of
two Feynman diagrams. It is
\bea &&{1\over s_{34}}\ruleI{\{1\},\{2\},\{3,4\},\{5\}}+{1\over
s_{45}}\ruleI{\{1\},\{2\},\{3\},\{4,5\}}\nonumber\\
&=&{1\over s_{34}}{2p_1p_{34}+2p_2p_5\over 2s_{12}^2}+{1\over
s_{45}}{2p_1p_{3}+2p_2p_{45}\over 2s_{12}^2}={s_{25}\over
s_{34}s_{12}^2}+{s_{13}\over s_{45}s_{12}^2}-{1\over
s_{12}^2}~.~~~\eea
Note that the last term $-{1\over s_{12}^2}$ in the result is in
fact a correction term introduced by using $2P_AP_B$ for massive
legs instead of $s_{AB}$ in the Feynman rule. This result agrees
with that in \cite{Cachazo:2015nwa}.

\subsection{The Feynman rule $\ruleii$ for a single triple pole}

%
\begin{figure}[h]
  \centering
  \includegraphics[width=5.5in]{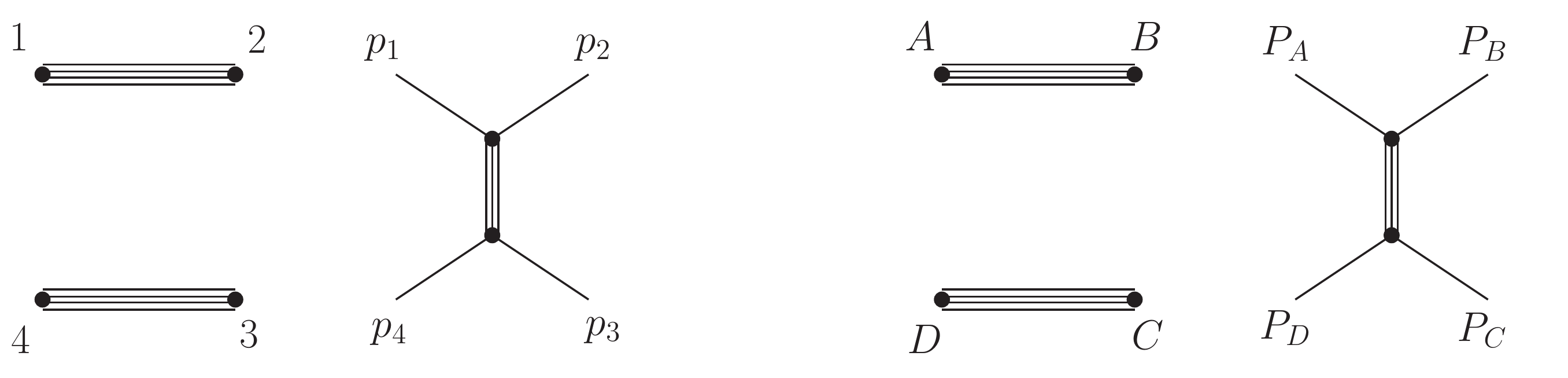}\\
  \caption{Left: The {\sl 4-regular} graph of a given CHY-integrand and its
  corresponding Feynman diagram. The {\sl triple-line} propagator denotes a triple pole of this CHY-integrand.
  Right: Labels of external momenta for the Feynman rule $\ruleii$ when legs are massive.}\label{Figrule2}
\end{figure}
%
\subsection*{The Feynman rule}

For external legs as labeled in the right-most diagram of Figure
\ref{Figrule2}, the Feynman rule for a single triple pole is
formulated as
\bea&&\ruleII{P_A,P_B,P_C,P_D}~~~~\label{rule2}\\
&=&{(2P_AP_C)(2P_AP_D)+(2P_BP_C)(2P_BP_D)+(2P_CP_A)(2P_CP_B)+(2P_DP_A)(2P_DP_B)\over
4s_{AB}^3}\nonumber\\
&&-{(P_A^2-P_B^2)^2+(P_C^2-P_D^2)^2\over 4s_{AB}^3}+{2\over
9}{(P_A^2+P_B^2)(P_C^2+P_D^2)\over 4s_{AB}^3}~.~~~\nonumber\eea
Note that terms in the second line are zero when all external legs
are massless, so they can not be deduced from the result of
four-point CHY-integrand. Nevertheless, it is very important in
order to produce correct answer for general situations.

\subsection*{Deducing the Feynman rule}

Again let us start from a simple four-point CHY-integrand, which
contains this pole structure and is given by
\bea {\cal I}= {1\over z_{12}^4
z_{34}^4}~.~~~\label{triple-4p-1}\eea
The result can be trivially obtained by solving scattering equations
as
\bea {s_{14}s_{13}\over s_{12}^3}~.~~\label{triple-4p-1-1}\eea

Now we follow the {\sl integration algorithm} for this example. In
the {\sl 4-regular} graph as shown in Figure \ref{Figrule2}, it is
easy to see that $\chi[\{2,3\}]=0-2(2-1)=-2$ while
$\chi[\{1,2\}]=4-2(2-1)=2$, which means there is only one subset
$\{\dunderline{1,2}\}$ corresponding to a triple pole (a two-fold
underline is introduced to emphasize it). So we can draw the Feynman
diagram for this CHY-integrand, as shown in Figure \ref{Figrule2}
besides the {\sl 4-regular} graph, where a triple line is introduced
to denote the triple pole.

To deduce (but not derive!) the Feynman rule for a single triple
pole from result (\ref{triple-4p-1-1}), we follow the similar
considerations as in previous subsection. Apparently, the {\sl
4-regular} graph is invariant under exchanging of nodes $1,2$ as
well as exchanging of nodes $3,4$. It is also invariant under
exchanging $i\to \mbox{Mod}[i+2,4]$. From these considerations,
intuitively we can propose a symmetrization for
(\ref{triple-4p-1-1}) as
\bea &&{s_{14}s_{13}+s_{24}s_{23}+s_{41}s_{42}+s_{31}s_{32}\over
4s_{12}^3}\nonumber\\
&&~~\to
\mathcal{R}'={(2P_AP_D)(2P_AP_C)+(2P_BP_D)(2P_BP_C)+(2P_DP_A)(2P_DP_B)+(2P_CP_A)(2P_CP_B)\over
4s_{AB}^3}~,~~\label{triple-guess1}\eea
using identities $s_{13}=s_{24},s_{14}=s_{23}$, and also the rule
shown above. However, when computing the five-point CHY-integrand as
shown in Figure \ref{FigA53}, the above rule fails to produce
correct answer. But another rule
\bea \mathcal{R}''={(2P_AP_D)(2P_AP_C)+(2P_BP_D)(2P_BP_C)\over
2s_{AB}^3}~,~~~\label{triple-guess2}\eea
although not preserving the required symmetries, can produce correct
answer. Anyway, we need a rule that preserving the symmetries of
{\sl 4-regular} graph, and this motives us to inspect the difference
between above two expressions (\ref{triple-guess1}),
(\ref{triple-guess2}). Notice that for the five-point CHY-integrand
in Figure \ref{FigA53}, we have  applied the rule under the
condition that $P_A=p_1, P_B=p_2$ are massless, while $P_C$, $P_D$
could be massive. After taking the difference of
(\ref{triple-guess2}) and (\ref{triple-guess1}), we get
\bea R''-R'&=&-{(P_C^2-P_D^2)^2\over 4s_{AB}^2}~.~~~\eea
This indicates that $\mathcal{R}'$ plus the extra term would produce
correct result for CHY-integrand in Figure \ref{FigA53}. Considering
symmetries of {\sl 4-regular} graph, it is reasonable to refine the
rule as\footnote{We comment that, if only relying on the symmetry
consideration, there would be many possible terms to be added in the
rule, for example
$${(P_A^2+P_B^2)^2+(P_C^2+P_D^2)^2\over 4s_{AB}^3}~~,~~{(P_A^2+P_B^2+P_C^2+P_D^2)^2\over 4s_{AB}^3}~~,~~\mbox{etc.}$$
So if there is no {\sl trial-and-error} process with other
CHY-integrands, it is in fact difficult to formulate a valid rule
from these possible terms due to too many ambiguities.}
\bea \mathcal{R}'-{(P_A^2-P_B^2)^2+(P_C^2-P_D^2)^2\over
4s_{AB}^3}~.~~~\label{triple-guess3}\eea
However, this is still not the complete rule. For CHY-integrand in
Figure \ref{FigA53}, we do not confront situations that two or more
external legs are massive. We found that, when both $P_A,P_B$ or
both $P_C,P_D$ are on-shell, the rule (\ref{triple-guess3}) still
works, otherwise it fails. This motive us to add another correction
term, with the property that it becomes zero when $P_A^2=P_B^2=0$ or
$P_C^2=P_D^2=0$, but non-zero when $P_A^2=P_C^2=0$ or
$P_B^2=P_D^2=0$. An immediate candidate is
$(P_A^2+P_B^2)(P_C^2+P_D^2)$, and after {\sl trial-and-error}, we
find that a correction term
\bea {2\over 9}{(P_A^2+P_B^2)(P_C^2+P_D^2)\over 4s_{AB}^3}\eea
would suffice to produce correct results for applying the rule of
triple pole to all possible situations of massive legs. Then we
formulated the Feynman rule $\ruleii$ for triple pole as shown in
(\ref{rule2}).

\subsection*{An illustrative example}
%
\begin{figure}
  \centering
  \includegraphics[width=5in]{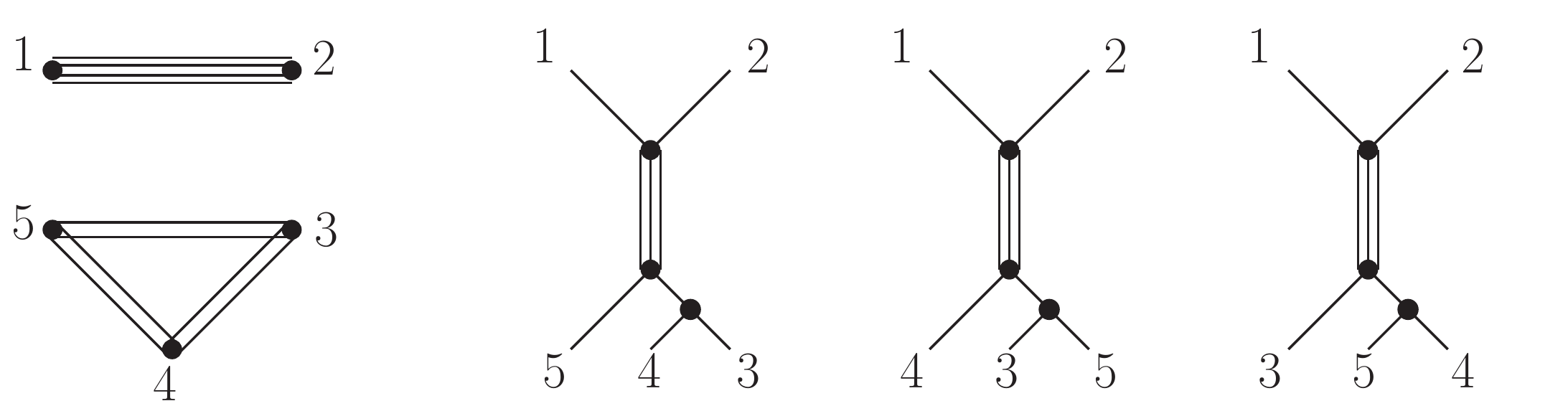}\\
  \caption{The {\sl 4-regular} graph of a given five-point CHY-integrand with
  a triple pole, and two Feynman diagrams corresponding to this CHY-integrand.}\label{FigA53}
\end{figure}
Let us illustrate the Feynman rule $\ruleii$ by a five-point
CHY-integrand
\bea -{1\over z_{12}^4 z_{34}^2z_{45}^2z_{53}^2}~.~~~\nonumber\eea
The {\sl 4-regular} graph for this CHY-integrand is drawn in Figure
\ref{FigA53}. By counting the number of lines connecting subsets of
nodes, we find that $\{\dunderline{1,2}\}$ is associated with a
triple pole, while $\{3,4\}, \{4,5\},\{5,3\}$ are associated with
simple poles. We need to select $5-3=2$ subsets to construct the
compatible combinations, and there are three,
\bea \{\{\dunderline{1,2}\}, \{3,4\}\}~~~,~~~\{\{\dunderline{1,2}\},
\{4,5\}\}~~~,~~~\{\{\dunderline{1,2}\}, \{3,5\}\}~.~~~\eea
From them we can draw three Feynman diagrams, as shown in Figure
\ref{FigA53} besides {\sl 4-regular} graph. Note that in the
definition of $\ruleii$, the exchanging $P_A\leftrightarrow P_B$ or
$P_C\leftrightarrow P_D$ will not affect the rule( which is not true
for $\rulei$). So there is no definite ordering for external legs
attached to the same end of triple pole propagator. Then it is
straightforward to write down the result by applying Feynman rules
of triple pole as
\bea &&{1\over s_{34}}\ruleII{\{1\},\{2\},\{3,4\},\{5\}}+{1\over
s_{45}}\ruleII{\{1\},\{2\},\{4,5\},\{3\}}+{1\over
s_{35}}\ruleII{\{1\},\{2\},\{3,5\},\{4\}}\nonumber\\
&=&{(2p_1p_{34})(2p_1p_5)+(2p_2p_{34})(2p_2p_5)+(2p_{34}p_{1})(2p_{34}p_2)+(2p_5p_{1})(2p_5p_2)-((p_3+p_4)^2)^2\over
4s_{12}^3s_{34}}\nonumber\\
&&+{(2p_1p_{45})(2p_1p_3)+(2p_2p_{45})(2p_2p_3)+(2p_{45}p_{1})(2p_{45}p_2)+(2p_3p_{1})(2p_3p_2)-((p_4+p_5)^2)^2\over
4s_{12}^3s_{45}}\nonumber\\
&&+{(2p_1p_{35})(2p_1p_4)+(2p_2p_{35})(2p_2p_4)+(2p_{35}p_{1})(2p_{35}p_2)+(2p_4p_{1})(2p_4p_2)-((p_3+p_5)^2)^2\over
4s_{12}^3s_{35}}\nonumber\\
&=&\frac{s_{15} s_{25}}{s_{12}^3 s_{34}}+\frac{s_{13}
s_{23}}{s_{12}^3 s_{45}}+\frac{s_{14} s_{24}}{s_{12}^3
s_{35}}+\frac{1}{s_{12}^2}~.~~~\nonumber\eea

\subsection{The Feynman rule $\ruleiii$ for duplex-double pole}

%
\begin{figure}[h]
  \centering
  \includegraphics[width=5.5in]{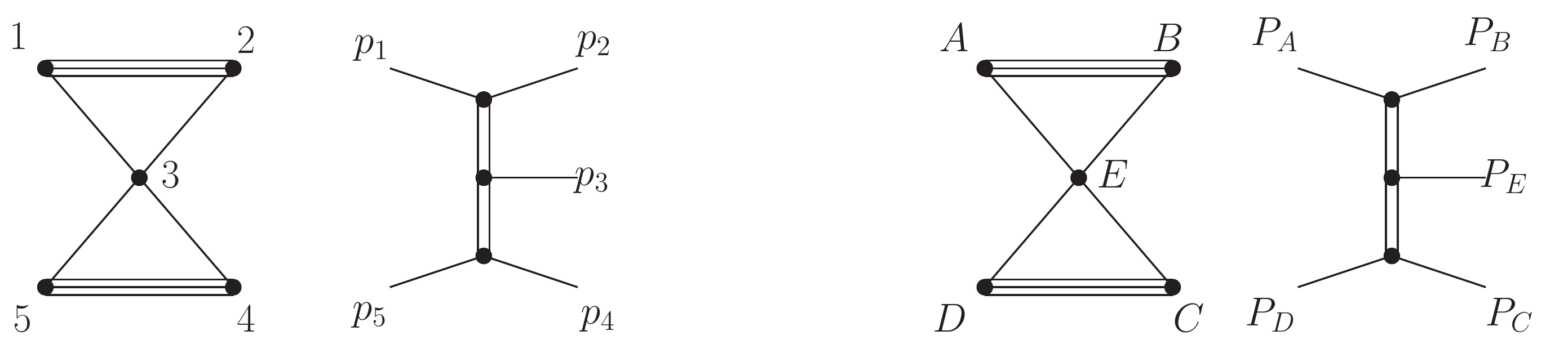}\\
  \caption{Left: The {\sl 4-regular} graph of a given CHY-integrand and its
  corresponding Feynman diagram. The {\sl double-line} propagator denotes a double pole of this CHY-integrand. Since two
  propagators of double poles are connected at one vertex, we define it as duplex-double pole.
  Right: Labels of external momenta for the Feynman rule $\ruleiii$ when external legs are massive.}\label{Figrule3}
\end{figure}
%
\subsection*{The Feynman rule}

For external legs as labeled in the right-most diagram of Figure
\ref{Figrule3}, the Feynman rule for duplex-double pole is
formulated as
\bea
\ruleIII{P_A,P_B,P_E,P_C,P_D}&=&{(2P_AP_D)(2P_BP_C)-(2P_AP_C)(2P_BP_D)\over
s_{AB}^2s_{CD}^2}\nonumber\\
&&-{(P_E^2)(2P_AP_D+2P_BP_C-2P_AP_C-2P_BP_D)\over
4s_{AB}^2s_{CD}^2}~.~~~\label{rule3}\eea
Notice that in the five-point CHY-integrand, we have
$P_E^2=p_3^2=0$, which means that the second term in fact can not be
deduced from five-point result. But it is non-zero for generic
CHY-integrands, and is crucial in order to produce the correct
answer.

\subsection*{Deducing the Feynman rule}

Although the Feynman diagram in Figure \ref{Figrule3} contains only
double pole, these two propagators of double poles are connected at
a vertex. Practically, we find it impossible to use the Feynman rule
$\rulei$ separately for each double pole, and should treat them as a
single object which we call duplex-double pole. In order to deduce
the Feynman rule for this duplex-double pole, let us start from a
simple five-point CHY-integrand,
\bea {\cal I}= -{1\over z_{12}^3 z_{23}z_{34} z_{45}^3
z_{53}z_{31}}~.~~~\label{duplex-double-5p-1}\eea
Although it is not as trivial as the four-point CHY-integrands, we
can still compute it by solving scattering equations directly. After
some simplification, we can write the result as
\bea {s_{15}s_{24}-s_{14}s_{25}\over
s_{12}^2s_{45}^2}~.~~\label{duplex-double-5p-1-1}\eea
Of course we can rewrite it in other equivalent forms by using
momentum conservation, but at this point let us just take this
expression.

The {\sl 4-regular} graph of this CHY-integrand is shown in Figure
\ref{Figrule3}. It is easy to figure out that the possible subsets
of nodes that contributing to poles are $\{\underline{1,2}\}$ and
$\{\underline{4,5}\}$. Since $\chi[\{1,2\}]=\chi[\{4,5\}]=1$, both
of them are double poles. They can only form one compatible
combination $\{\{\underline{1,2}\},\{\underline{4,5}\}\}$, and from
which we draw the Feynman diagrams as shown besides the {\sl
4-regular} graph in Figure \ref{Figrule3}. This means that we should
deduce the Feynman rule for duplex-double pole from result
(\ref{duplex-double-5p-1-1}).

Again considering the symmetries among nodes of the {\sl 4-regular}
graph(or the CHY-integrand), clearly it is
\bea &&\mbox{anti-symmetric~under}~~~~~~\{1,2\}\leftrightarrow
\{2,1\}~~\mbox{or}~~\{4,5\}\leftrightarrow
\{5,4\}~,~~~\nonumber\\
&&\mbox{symmetric~under}~~~~~\{1,2,3,4,5\}\leftrightarrow
\{4,5,3,1,2\}~~\mbox{or}~~\{1,2,3,4,5\}\leftrightarrow
\{2,1,3,5,4\}~.~~~\nonumber\eea
By inspecting the kinematic variables, it is easy to see that the
result (\ref{duplex-double-5p-1-1}) already possesses above
anti-symmetries as well as symmetries, so it is not necessary to
further symmetrize it. With experiences of previous Feynman rules of
higher-order poles, it is straightforwardly to propose a Feynman
rule
\bea {(2P_AP_D)(2P_BP_C)-(2P_AP_C)(2P_BP_D)\over
s_{AB}^2s_{CD}^2}~~~~\eea
for massive legs as shown in the right-most diagram of Figure
\ref{Figrule3}. However, it is not yet the complete rule. After {\sl
trial-and-error}, we find that the second term in $\ruleiii$
(\ref{rule3}) should be included. In the five-point CHY-integrand,
since $P_E^2=p_3^2=0$, it vanishes and we can never deduce it from
(\ref{duplex-double-5p-1-1}). But it is crucial in order to produce
the correct result for other generic CHY-integrands, and we will
show this in \S\ref{secSupportingmixed} with examples containing
mixed types of higher order poles.

\subsection*{An illustrative example}

\begin{figure}
  \centering
  \includegraphics[width=5in]{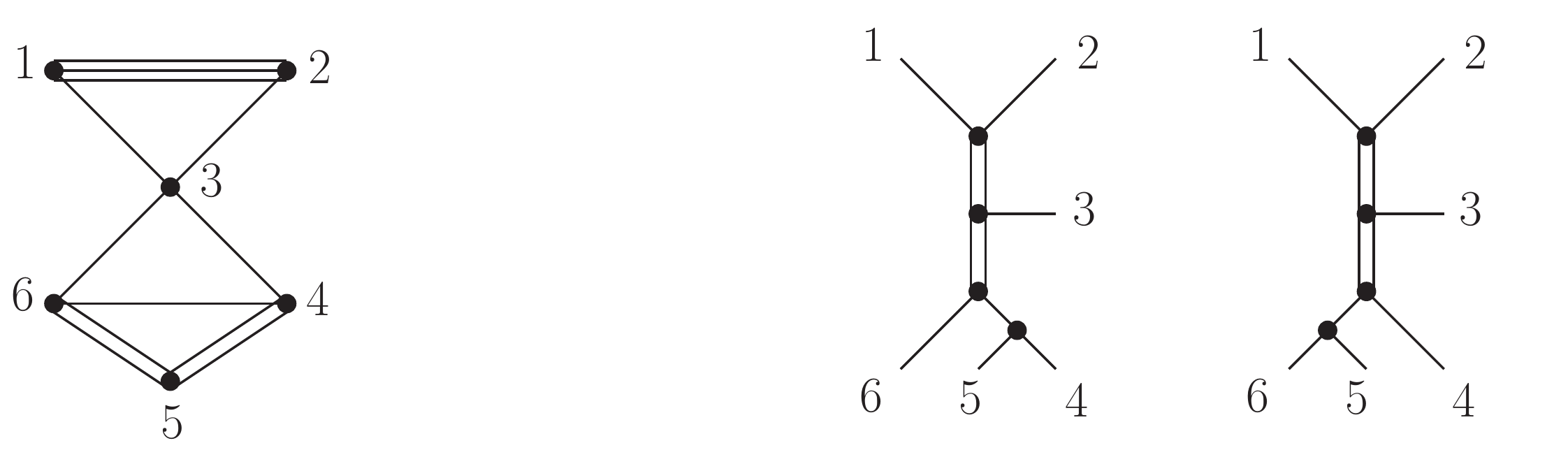}\\
  \caption{The {\sl 4-regular} graph of a given CHY-integrand with duplex-double poles, and two
  Feynman diagrams corresponding to this CHY-integrand.}\label{FigA68}
\end{figure}

We will illustrate the Feynman rule $\ruleiii$ by a six-point
CHY-integrand
\bea -{1\over z_{12}^3
z_{23}z_{34}z_{45}^2z_{56}^2z_{64}z_{63}z_{31}}~.~~~\eea
The {\sl 4-regular} graph of this CHY-integrand is shown in Figure
\ref{FigA68}.
Following the {\sl integration algorithm}, firstly we list all the
subsets of nodes that contribute to either simple pole or
higher-order poles, as
\bea
\{\underline{1,2}\}~~~,~~~\{\underline{4,5,6}\}~~~,~~~\{4,5\}~~~,~~~\{5,6\}~.~~~\nonumber\eea
Secondly, we need to select $6-3=3$ subsets out of above four
subsets to construct compatible combinations. They are given by
\bea &&\{\{\underline{1,2}\}, \{\underline{4,5,6}\},
\{4,5\}\}~~~,~~~\{\{\underline{1,2}\}, \{\underline{4,5,6}\},
\{5,6\}\}~.~~~\nonumber\eea
Immediately we draw two Feynman diagrams as shown in Figure
\ref{FigA68}, which have the required propagators according to
compatible combinations respectively. We have intentionally labeled
the external legs in certain ordering such that we can directly read
out the Feynman rules from the diagram. Then according to the rules,
we get the result
\bea &&{1\over s_{45}}
\ruleIII{\{1\},\{2\},\{3\},\{4,5\},\{6\}}+{1\over s_{56}}
\ruleIII{\{1\},\{2\},\{3\},\{4\},\{5,6\}}\nonumber\\
&=&{(2p_1p_{45})(2p_2p_6)-(2p_1p_6)(2p_2p_{45})\over
s_{45}s_{12}^2s_{456}^2}+{(2p_1p_{4})(2p_2p_{56})-(2p_1p_{56})(2p_2p_{4})\over
s_{56}s_{12}^2s_{456}^2}\nonumber\\
&=&{s_{145}s_{26}-s_{16}s_{245}\over
s_{45}s_{12}^2s_{456}^2}+{s_{14}s_{256}-s_{156}s_{24}\over
s_{56}s_{12}^2s_{456}^2}+{s_{16}-s_{26}+s_{24}-s_{14}\over
s_{12}^2s_{456}^2}~.~~~\eea
The last term is in fact the correction term introduced by using
$2P_AP_B$ instead of $s_{AB}$ for massive legs in the Feynman rules.
This result is confirmed numerically.

\subsection{The Feynman rule $\ruleix$ for triplex-double pole}
We have presented the Feynman rules $\rulei,\ruleii,\ruleiii$ for
double pole, triple pole and duplex-double pole. They are reasonably
simple. But when going to the higher-point CHY-integrands, new
structures of poles would appear. Deducing rules for them could be
very difficult, and the rules themselves could be complicated. To
illustrate, let us present another Feynman rule $\ruleix$ for a pole
structure that first appear in six-point CHY-integrand.

\begin{figure}[h]
  \centering
  \includegraphics[width=6.5in]{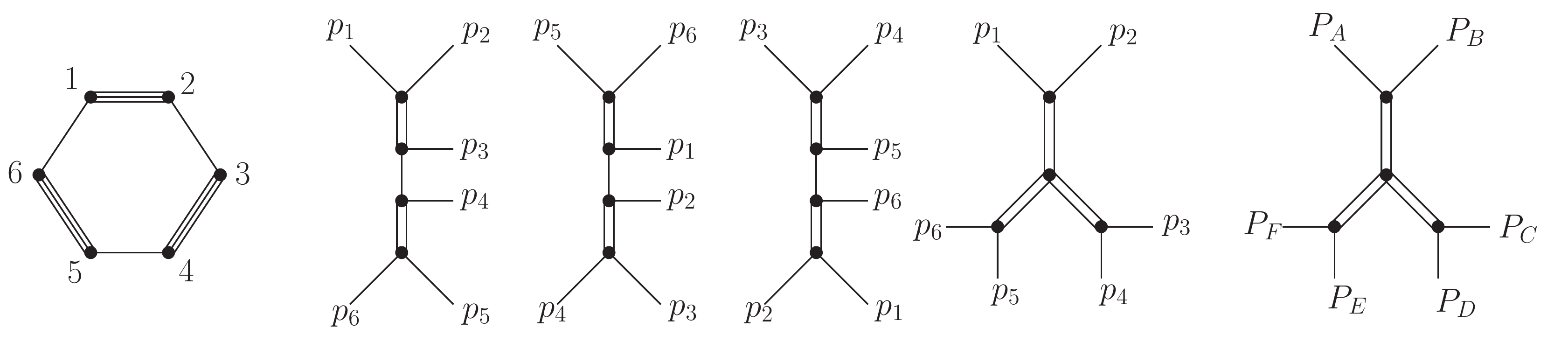}\\
  \caption{Left-most: The {\sl 4-regular} graph of a given CHY-integrand. Middle-four: Its
  corresponding four Feynman diagrams. The {\sl double-line} propagator denotes a double pole of this CHY-integrand.
  In the fourth Feynman diagram, three double poles are connected at one vertex, and we define it as triplex-double pole.
  Right-most: Labels of external momenta for the Feynman rule $\ruleix$ when legs are massive.}\label{Figrule4}
\end{figure}
%
\subsection*{The Feynman rule}
For external legs as labeled in the right-most diagram of Figure
\ref{Figrule4}, the Feynman rule for triplex-double pole is
formulated as (for a better presentation, we define the {\sl
stripped} Mandelstam variables $\widetilde{s}_{AB}=2P_AP_B$ as well
as $\widetilde{s}_{ABC}=2P_AP_B+2P_AP_C+2P_BP_C$\footnote{When
$A,B$, etc., contains only one massless leg, $\widetilde{s}=s$,
otherwise they differs. For example, if $A=\{1,2\},B=\{3\}$,
$s_{AB}=(p_1+p_2+p_3)^2=2p_1p_2+2p_1p_3+2p_2p_3$ but
$\widetilde{s}_{AB}=2p_{12}p_3=2p_1p_3+2p_2p_3$.})
\bea &&\ruleIX{P_A,P_B,P_C,P_D,P_E,P_F}~~~~~\label{rule4}\\
&=&\left({\mathcal{R}_{11}\over
8s_{AB}^2s_{CD}^2}+{\mathcal{R}_{12}\over
8s_{CD}^2s_{EF}^2}+{\mathcal{R}_{13}\over
8s_{EF}^2s_{AB}^2}\right)+\left({\mathcal{R}_{21}\over
2s_{AB}^2s_{CD}s_{EF}}+{\mathcal{R}_{22}\over
2s_{CD}^2s_{EF}s_{AB}}+{\mathcal{R}_{23}\over
2s_{EF}^2s_{AB}s_{CD}}\right)\nonumber\\
&&+\left({\mathcal{R}_{31}\over
2s_{AB}^2s_{CD}^2s_{EF}}+{\mathcal{R}_{32}\over
2s_{CD}^2s_{EF}^2s_{AB}}+{\mathcal{R}_{33}\over
2s_{EF}^2s_{AB}^2s_{CD}}\right)+{\mathcal{R}_{4}\over
2s_{AB}^2s_{CD}^2s_{EF}^2}+{1\over
s_{AB}s_{CD}s_{EF}}\nonumber\\
&&~~~~-(P_A^2+P_B^2+P_C^2+P_D^2+P_E^2+P_F^2)\left({1\over
4s_{AB}^2s_{CD}^2}+{1\over 4s_{AB}^2s_{EF}^2}+{1\over
4s_{CD}^2s_{EF}^2}\right)~,~~~\nonumber\eea
where
\bea
\mathcal{R}_{11}&=&2(\widetilde{s}_{EC}+\widetilde{s}_{FB}-\widetilde{s}_{EB}-\widetilde{s}_{FC})-(\widetilde{s}_{ABC}+\widetilde{s}_{BCD}+\widetilde{s}_{CDE}+\widetilde{s}_{DEF}+\widetilde{s}_{EFA}+\widetilde{s}_{FAB})~,~~~\\
\mathcal{R}_{12}&=&2(\widetilde{s}_{AE}+\widetilde{s}_{BD}-\widetilde{s}_{AD}-\widetilde{s}_{BE})-(\widetilde{s}_{ABC}+\widetilde{s}_{BCD}+\widetilde{s}_{CDE}+\widetilde{s}_{DEF}+\widetilde{s}_{EFA}+\widetilde{s}_{FAB})~,~~~\\
\mathcal{R}_{13}&=&2(\widetilde{s}_{CA}+\widetilde{s}_{DF}-\widetilde{s}_{CF}-\widetilde{s}_{DA})-(\widetilde{s}_{ABC}+\widetilde{s}_{BCD}+\widetilde{s}_{CDE}+\widetilde{s}_{DEF}+\widetilde{s}_{EFA}+\widetilde{s}_{FAB})~,~~~\\
\mathcal{R}_{21}&=&\widetilde{s}_{AF}+\widetilde{s}_{BC}+\widetilde{s}_{AC}+\widetilde{s}_{BF}-(\widetilde{s}_{ACE}+\widetilde{s}_{BDF})~,~~~\\
\mathcal{R}_{22}&=&\widetilde{s}_{CB}+\widetilde{s}_{DE}+\widetilde{s}_{CE}+\widetilde{s}_{DB}-(\widetilde{s}_{ACE}+\widetilde{s}_{BDF})~,~~~\\
\mathcal{R}_{23}&=&\widetilde{s}_{ED}+\widetilde{s}_{FA}+\widetilde{s}_{EA}+\widetilde{s}_{FD}-(\widetilde{s}_{ACE}+\widetilde{s}_{BDF})~,~~~\eea
and
\bea
\mathcal{R}_{31}&=&\widetilde{s}_{BC}(\widetilde{s}_{ED}+\widetilde{s}_{FA}-\widetilde{s}_{EC}-\widetilde{s}_{FB})\nonumber\\
&&~~~~~~+(\widetilde{s}_{CA}\widetilde{s}_{DE}+\widetilde{s}_{BD}\widetilde{s}_{AF}-\widetilde{s}_{BF}\widetilde{s}_{CA}-\widetilde{s}_{CE}\widetilde{s}_{BD})+(\widetilde{s}_{CA}-\widetilde{s}_{BD})(\widetilde{s}_{CE}-\widetilde{s}_{BF})~,~~~\\
\mathcal{R}_{32}&=&\widetilde{s}_{DE}(\widetilde{s}_{AF}+\widetilde{s}_{BC}-\widetilde{s}_{AE}-\widetilde{s}_{BD})\nonumber\\
&&~~~~~~+(\widetilde{s}_{EC}\widetilde{s}_{FA}+\widetilde{s}_{DF}\widetilde{s}_{CB}-\widetilde{s}_{DB}\widetilde{s}_{EC}-\widetilde{s}_{EA}\widetilde{s}_{DF})+(\widetilde{s}_{EC}-\widetilde{s}_{DF})(\widetilde{s}_{EA}-\widetilde{s}_{DB})~,~~~\\
\mathcal{R}_{33}&=&\widetilde{s}_{FA}(\widetilde{s}_{CB}+\widetilde{s}_{DE}-\widetilde{s}_{CA}-\widetilde{s}_{DF})\nonumber\\
&&~~~~~~+(\widetilde{s}_{AE}\widetilde{s}_{BC}+\widetilde{s}_{FB}\widetilde{s}_{ED}-\widetilde{s}_{FD}\widetilde{s}_{AE}-\widetilde{s}_{AC}\widetilde{s}_{FB})+(\widetilde{s}_{AE}-\widetilde{s}_{FB})(\widetilde{s}_{AC}-\widetilde{s}_{FD})~,~~~\\
\mathcal{R}_4&=&\widetilde{s}_{BC}(\widetilde{s}_{CE}\widetilde{s}_{EA}+\widetilde{s}_{BF}\widetilde{s}_{FD})+\widetilde{s}_{DE}(\widetilde{s}_{EA}\widetilde{s}_{AC}+\widetilde{s}_{DB}\widetilde{s}_{BF})+\widetilde{s}_{FA}(\widetilde{s}_{AC}\widetilde{s}_{CE}+\widetilde{s}_{FD}\widetilde{s}_{DB})\nonumber\\
&&+\widetilde{s}_{BC}\widetilde{s}_{DE}(\widetilde{s}_{EA}+\widetilde{s}_{BF})+\widetilde{s}_{DE}\widetilde{s}_{FA}(\widetilde{s}_{AC}+\widetilde{s}_{DB})+\widetilde{s}_{FA}\widetilde{s}_{BC}(\widetilde{s}_{CE}+\widetilde{s}_{FD})+2\widetilde{s}_{BC}\widetilde{s}_{DE}\widetilde{s}_{FA}~.~~~\eea
Note again that the last term in (\ref{rule4}) vanishes for
six-point CHY-integrand, and it can not be deduced from there.
However it indeed contributes when any of external legs is sum of
more than one massless momenta. We need this correction term to
produce correct answer.

\subsection*{Deducing the Feynman rule}

When applying {\sl integration algorithm} to a six-point
CHY-integrand
\bea {1\over z_{12}^3z_{34}^3z_{56}^3
z_{23}z_{45}z_{61}}~,~~~\label{triplexpoleCHY}\eea
with its {\sl 4-regular} graph drawn in Figure \ref{Figrule4}, we
find six subsets of nodes
\bea
\{\underline{1,2}\}~~~,~~~\{\underline{3,4}\}~~~,~~~\{\underline{5,6}\}~~~,~~~\{1,2,3\}~~~,~~~\{2,3,4\}~~~,~~~\{3,4,5\}~~~\eea
that will contribute to poles, of which three are associated to
double poles. From them we can construct four compatible
combinations, and the corresponding four Feynman diagrams are shown
in Figure \ref{Figrule4}. The first three Feynman diagrams can be
computed by rule for simple poles and $\rulei$. However, for the
last Feynman diagram, three double poles are connected to one
vertex, and we can neither use rule for double pole nor rule for
duplex-double pole to compute it. Therefore, we need to create a
rule $\ruleix$ for this pole structure, which we call triplex-double
pole. It has $1/(s_{12}^2s_{34}^2s_{56}^2)$ pole structure.

Recall that when deducing Feynman rules $\rulei,\ruleii,\ruleiii$
for double, triple and duplex-double poles, we always start from
known results of the simplest CHY-integrands which exactly contain
one Feynman diagram of that pole structure. However, the six-point
CHY-integrand (\ref{triplexpoleCHY}) is the first one that appears
the triplex-double pole, and it has four Feynman diagrams. It is
impossible to find another CHY-integrand that contains only one
Feynman diagram, of which is exactly the triplex-double pole. So in
order to deduce the Feynman rule $\ruleix$, we have no choice but
play with the result of CHY-integrand (\ref{triplexpoleCHY}).

In order to get the result of (\ref{triplexpoleCHY}), one can use
the Pfaffian identity, as is already shown in
\cite{Baadsgaard:2015voa, Baadsgaard:2015ifa}. Starting from a
template of {\sl 3-regular} graph,
\begin{center}
\includegraphics[width=1.5in]{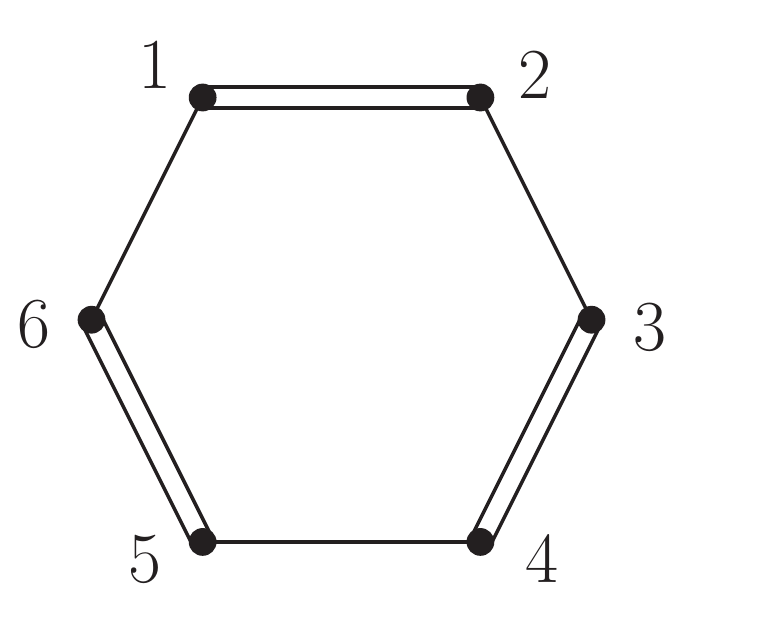}
\end{center}
we have the Pfaffian identity, for example,
\begin{center}
\includegraphics[width=6.5in]{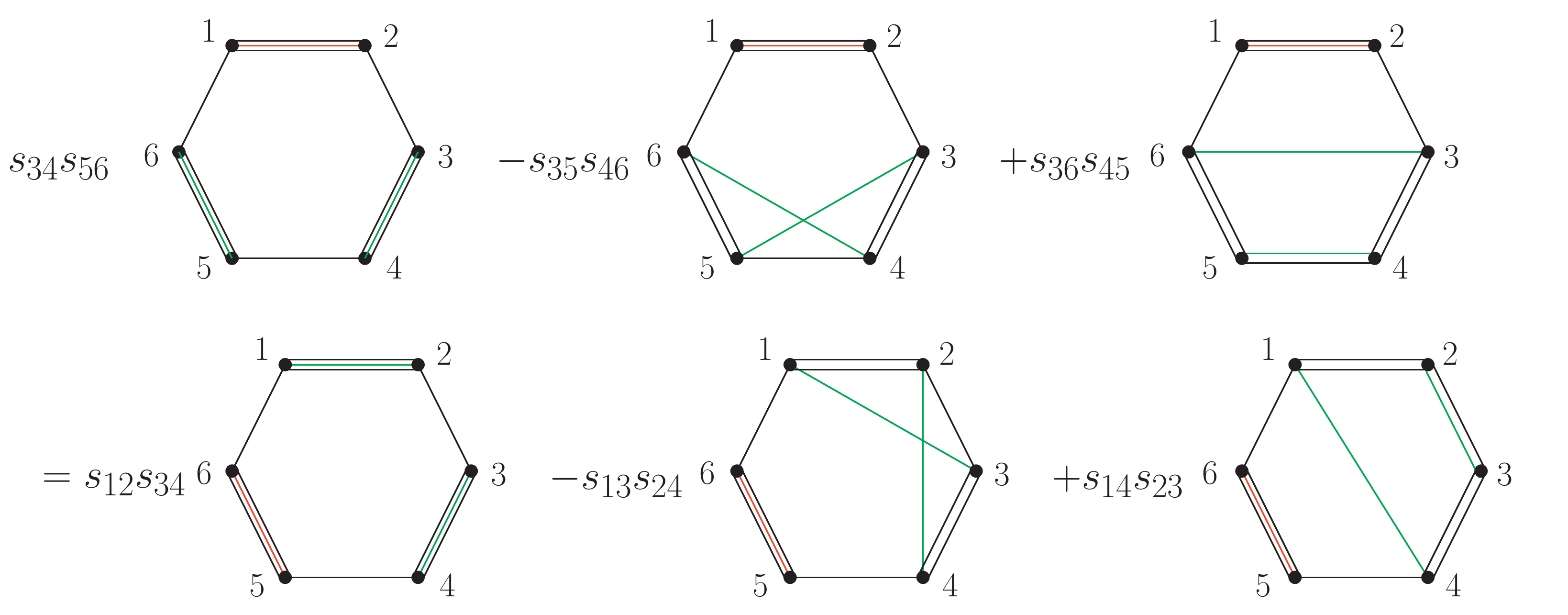}
\end{center}
The first diagram in the first line and second line is the one we
want to compute, and now we can instead compute the remaining four
in the identity, which could be CHY-integrands with only simple
poles or higher-order poles. In paper \cite{Baadsgaard:2015voa,
Baadsgaard:2015ifa}, one need more Pfaffian identities if some of
the remaining CHY-integrands still possess higher-order poles.
However, since now we have the Feynman rule for double pole, the
remaining four CHY-integrands can be instantly computed, without
reducing to those only with simple poles. Anyway, we can phrase the
result as
\bea \mathcal{A}_{{\tiny \mbox{triplex}}}=T_1+T_2~,~~~\eea
where
\bea T_1={s_{13}s_{46}\over s_{12}^2s_{56}^2s_{123}}+{
s_{15}s_{24}\over s_{34}^2s_{56}^2s_{234}}+{ s_{26}s_{35}\over
s_{12}^2s_{34}^2s_{345}}~,~~~\eea
which contains simple poles $s_{123},s_{234},s_{345}$, and
\bea T_2&=&-\frac{s_{13}}{s_{12} s_{34} s_{56}^2}+\frac{s_{36}
s_{13}}{s_{12}^2 s_{34} s_{56}^2}+\frac{s_{46} s_{13}}{s_{12}^2
s_{34} s_{56}^2}+\frac{s_{123}}{s_{12} s_{34} s_{56}^2}+\frac{s_{36}
s_{45} s_{134}}{s_{12}^2 s_{34}^2 s_{56}^2}-\frac{s_{26}}{s_{12}
s_{34}^2 s_{56}}\nonumber\\
&&+\frac{s_{45}}{s_{12} s_{34}^2 s_{56}}+\frac{s_{26}
s_{35}}{s_{12}^2 s_{34}^2 s_{56}}+\frac{s_{26} s_{36}}{s_{12}^2
s_{34}^2 s_{56}}-\frac{s_{36} s_{45}}{s_{12}^2 s_{34}^2
s_{56}}+\frac{s_{35} s_{46}}{s_{12}^2 s_{34}^2
s_{56}}-\frac{s_{46}}{s_{12} s_{34} s_{56}^2}\nonumber\\
&&+\frac{s_{35} s_{46}}{s_{12}^2 s_{34} s_{56}^2}+\frac{s_{36}
s_{46}}{s_{12}^2 s_{34} s_{56}^2}-\frac{s_{36} s_{123}}{s_{12}^2
s_{34} s_{56}^2}+\frac{s_{15} s_{24}}{s_{12} s_{34}^2
s_{56}^2}-\frac{s_{14} s_{25}}{s_{12} s_{34}^2
s_{56}^2}-\frac{s_{45} s_{134}}{s_{12} s_{34}^2
s_{56}^2}-\frac{s_{35} s_{46} s_{134}}{s_{12}^2 s_{34}^2
s_{56}^2}~.~~~\eea
On the other hand, summing over all four Feynman diagrams in Figure
\ref{Figrule4}, we should also get the result
\bea \mathcal{A}_{{\tiny \mbox{triplex}}}=F_1+F_2+F_3+F_4~,~~~\eea
where
\bea F_1&=&{1\over
s_{123}}\ruleI{\{1\},\{2\},\{3\},\{4,5,6\}}\ruleI{\{1,2,3\},\{4\},\{5\},\{6\}}\nonumber\\
&=&{s_{13}s_{46}\over
s_{12}^2s_{56}^2s_{123}}-{2s_{13}+2s_{46}-s_{123}\over
4s_{12}^2s_{56}^2}~,~~~\eea
\bea F_2&=&{1\over
s_{234}}\ruleI{\{5\},\{6\},\{1\},\{2,3,4\}}\ruleI{\{5,6,1\},\{2\},\{3\},\{4\}}\nonumber\\
&=&{s_{15}s_{24}\over
s_{34}^2s_{56}^2s_{234}}-{2s_{15}+2s_{24}-s_{234}\over
4s_{34}^2s_{56}^2}~,~~~\eea
and
\bea F_3&=&{1\over
s_{345}}\ruleI{\{3\},\{4\},\{5\},\{6,1,2\}}\ruleI{\{3,4,5\},\{6\},\{1\},\{2\}}\nonumber\\
&=&{s_{35}s_{26}\over
s_{12}^2s_{34}^2s_{345}}-{2s_{35}+2s_{26}-s_{345}\over
4s_{12}^2s_{34}^2}~.~~~\eea
$F_4$ should be generated by the rule of triplex-double pole we want
to deduce.

From results $F_1,F_2,F_3$ it is immediately to see that, the terms
$T_1$, i.e., with simple poles $s_{123},s_{234},s_{345}$, is really
computed by $F_1,F_2,F_3$. This is consistent with the fact that the
triplex-double pole can not produce simple poles
$s_{123},s_{234},s_{345}$. The equality
\bea T_1+T_2=F_1+F_2+F_3+F_4\eea
then provides us an expression for $F_4$, or the Feynman rule
$\ruleix$, with only double poles of $s_{12},s_{34},s_{56}$. We can
rephrase the expression as
\bea F_4&=&{s_{12}^2(s_{15}+s_{24}-s_{14}-s_{25})\over
4s_{12}^2s_{34}^2s_{56}^2}+{s_{34}^2(s_{13}+s_{46}-s_{14}-s_{36})\over
4s_{12}^2s_{34}^2s_{56}^2}+{s_{56}^2(s_{26}+s_{35}-s_{25}-s_{36})\over
4s_{12}^2s_{34}^2s_{56}^2}\nonumber\\
&&-{(s_{12}^2+s_{34}^2+s_{56}^2)(s_{123}+s_{234}+s_{345})\over
4s_{12}^2s_{34}^2s_{56}^2}\nonumber\\
&&+{s_{12}s_{56}(s_{15}+s_{26})\over
s_{12}^2s_{34}^2s_{56}^2}+{s_{34}s_{56}(s_{35}+s_{46})\over
s_{12}^2s_{34}^2s_{56}^2}+{s_{12}s_{34}(s_{13}+s_{24})\over
s_{12}^2s_{34}^2s_{56}^2}\nonumber\\
&&-{(s_{12}s_{15}+s_{26}s_{56})s_{123}\over
s_{12}^2s_{34}^2s_{56}^2}-{(s_{34}s_{46}+s_{35}s_{56})s_{234}\over
s_{12}^2s_{34}^2s_{56}^2}-{(s_{12}s_{24}+s_{13}s_{34})s_{345}\over
s_{12}^2s_{34}^2s_{56}^2}\nonumber\\
&&-{(s_{12}s_{34}+s_{12}s_{56}+s_{34}s_{56})s_{135}\over
s_{12}^2s_{34}^2s_{56}^2}+{s_{123}s_{234}s_{345}\over
s_{12}^2s_{34}^2s_{56}^2}~.~~~\eea
To deduce the Feynman rule from such a complicated result is quite
difficult. While considering the symmetries, apparently, the {\sl
4-regular} graph in Figure \ref{Figrule4} is invariant under
exchanging
\bea i\to
\mbox{Mod}[i+2,6]~~~\mbox{as~well~as}~~~\{1,2,3,4,5,6\}\leftrightarrow\{2,1,6,5,4,3\}~,~~~\eea
and similarly other exchanging of legs. We should reformulate $F_4$
in such a way that these symmetries are still kept. However, there
are too many possibilities to reformulate $F_4$ due to the momentum
conservation and massless conditions, and we have no strategy to
choose the one that can be generalized as rule. There is another
very important difference here compared to the previous rules. For
$\rulei,\ruleii,\ruleiii$, it is possible to reformulate the
numerator of the known results such that the $s$ variable of
higher-order pole do not present in the numerator. So we do not need
to consider the possibility that whether $s$ in the numerator would
cancel a factor of higher-order pole or not. Although not explicitly
mentioned, this is a guide line to deduce the Feynman rules
$\rulei,\ruleii,\ruleiii$. However, here it seems impossible to
completely eliminate $s_{12},s_{34},s_{56}$ dependence in the
numerator of $F_4$ by momentum conservation, thus it is also a
possibility that they would cancel a factor of double poles. All
these ambiguities makes things difficult. By extensive {\sl
trial-and-error} method with hints from a seven-point CHY-integrand,
we finally formulate a valid Feynman rule $\ruleix$ for
triplex-double pole as presented in (\ref{rule4}), and at least it
works for the examples we checked up to nine points.

\section{Supporting examples}
\label{secSupporting}

%
\begin{figure}[h]
  \centering
  \includegraphics[width=5.5in]{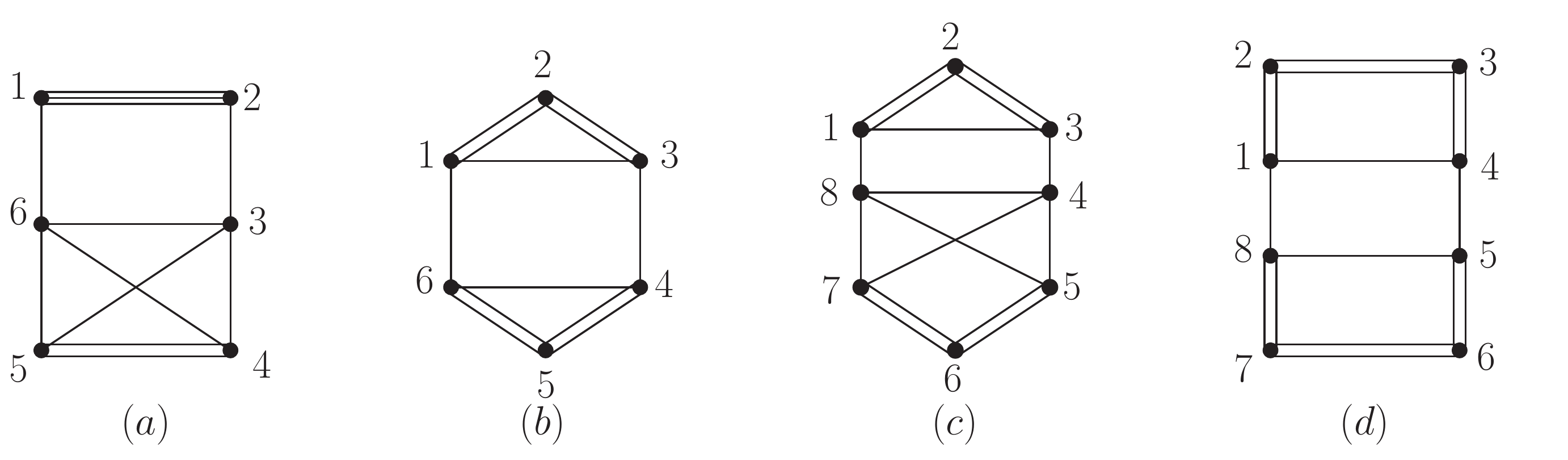}\\
  \caption{The {\sl 4-regular} graphs of supporting examples with only double poles and simple poles.}\label{Figsupport1}
  \includegraphics[width=5.5in]{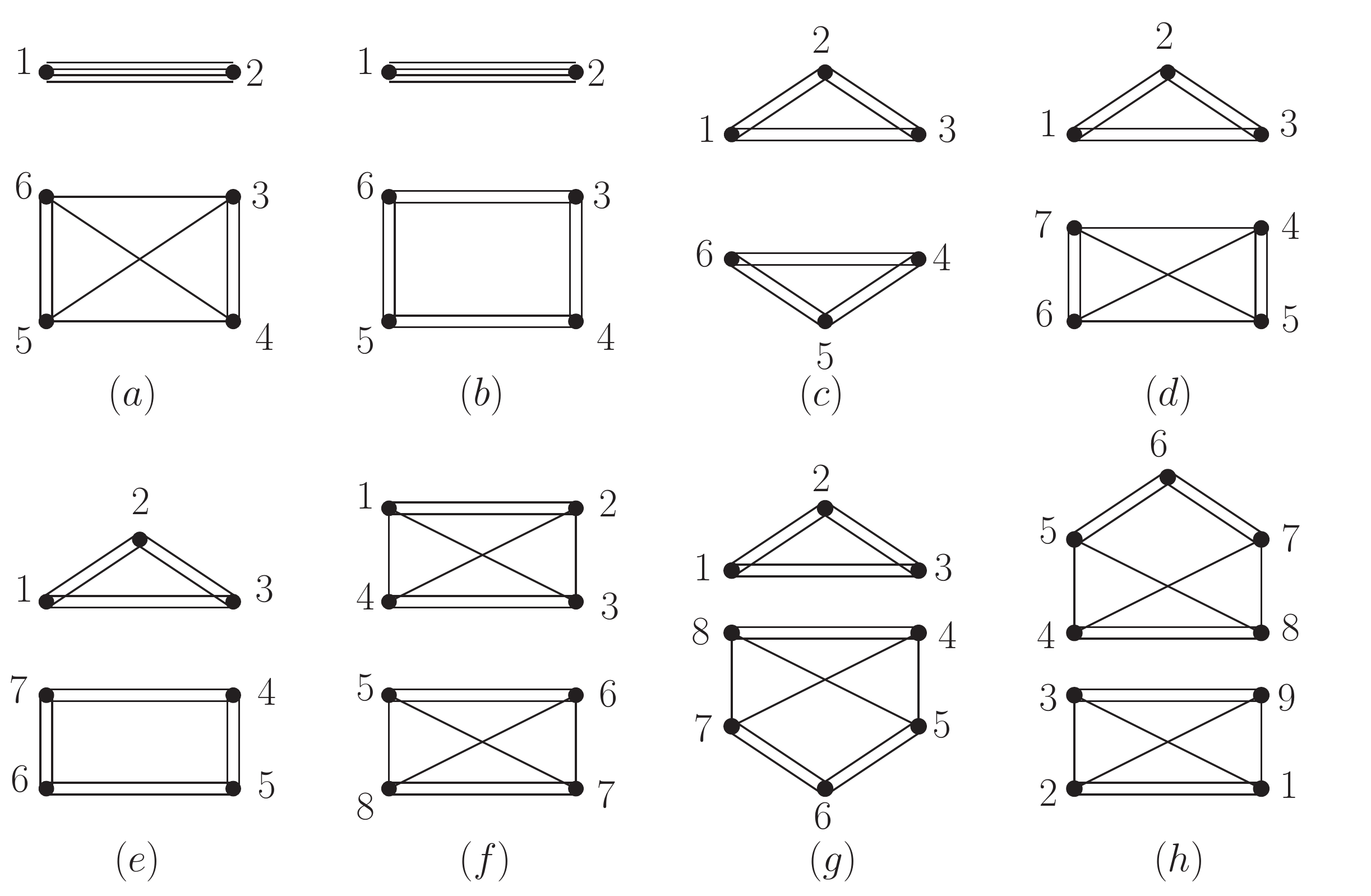}\\
  \caption{The {\sl 4-regular} graphs of supporting examples with only triple poles and simple poles.}\label{Figsupport2}
\end{figure}
\begin{figure}[h]
  \centering
 \includegraphics[width=5.5in]{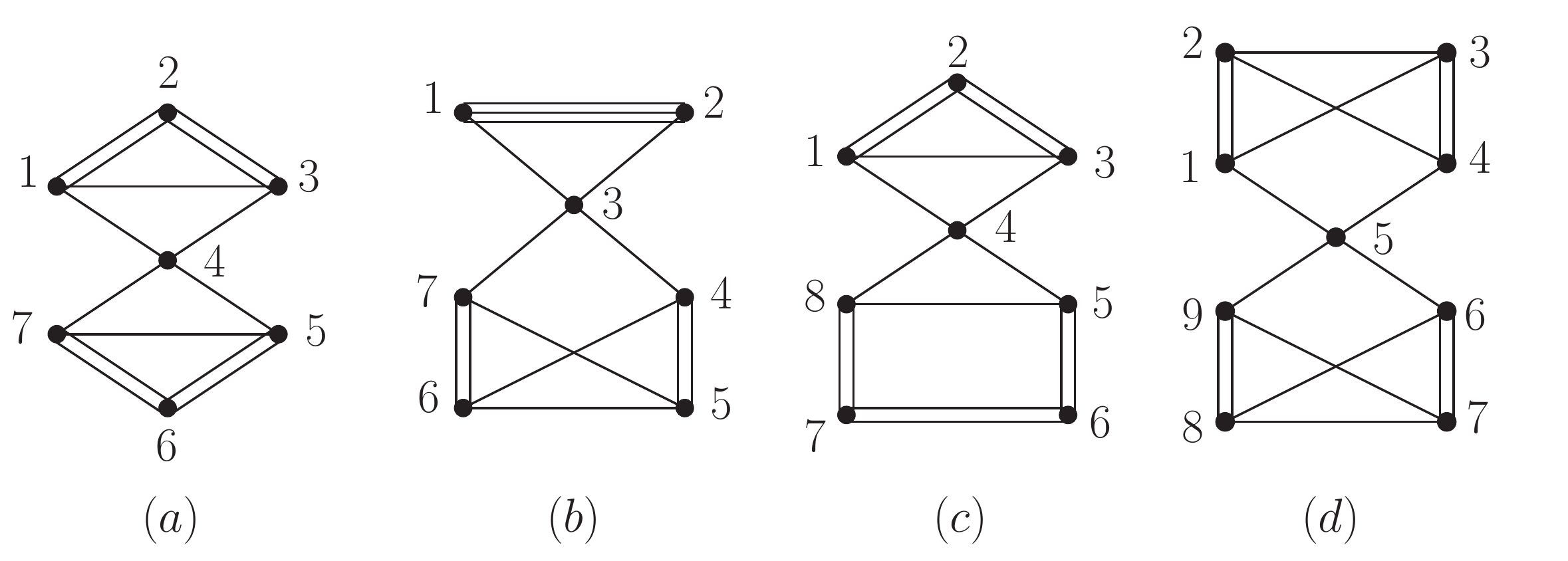}\\
  \caption{The {\sl 4-regular} graphs of supporting examples with only duplex-double poles and simple poles.}\label{Figsupport3}
  \includegraphics[width=5.5in]{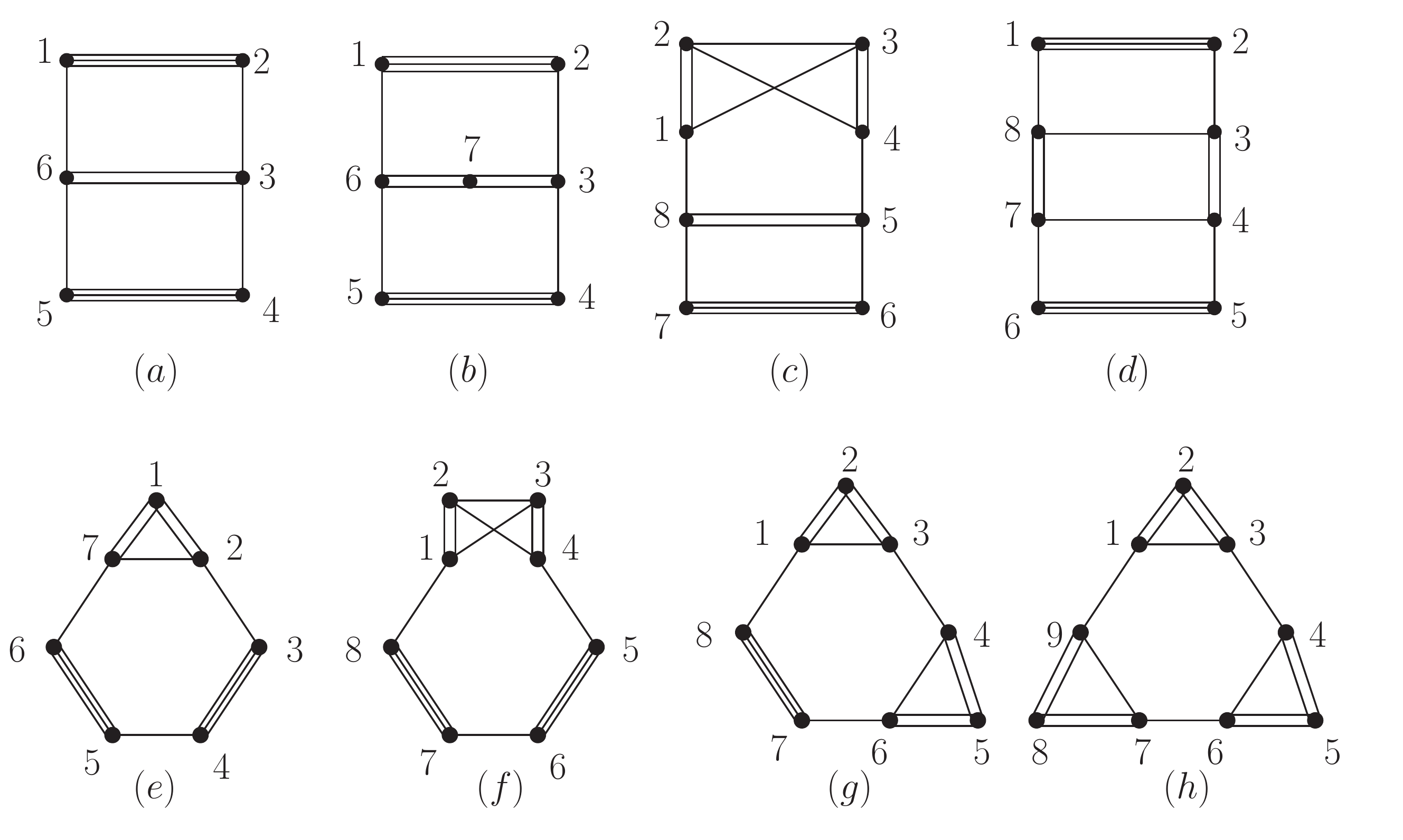}\\
  \caption{The {\sl 4-regular} graphs of supporting examples with mixed types of higher-order poles and simple poles. CHY-integrands in the first line contain
  double poles and duplex-double poles, while those in the second line contain double poles and triplex-double poles.}\label{Figsupportmix}
\end{figure}

The Feynman rules $\rulei$, $\ruleii$, $\ruleiii$, $\ruleix$ are
deduced but not derived, and their validation is supported by ample
examples. The {\sl integration algorithm}, as reviewed in
\S\ref{secReview}, involves nothing but selecting subsets of nodes
according to pole condition, constructing compatible combinations
and performing the pattern matching to apply the rules. All these
steps can be implemented by a few lines of codes in {\bf
Mathematica}. One is able to produce the analytic result within
seconds, while if trying to obtain a result from solving scattering
equations, even numerically it would take for example two or more
hours for nine-point CHY-integrands in a laptop(this is the reason
why we do not check the rules by ten or even more point examples).¡¡

For those who are not familiar with {\sl integration algorithm}, we
present here the results of various CHY-integrands by rules of
higher-order poles, and all have been checked numerically. For
reader's convenience, we collect all the {\sl 4-regular} graphs of
examples in Figure \ref{Figsupport1}, Figure \ref{Figsupport2},
Figure \ref{Figsupport3} and Figure \ref{Figsupportmix}. In case
that some readers are specially interested in certain CHY-integrand,
they can find the {\sl 4-regular} graph in these figures and jump to
corresponding subsection for details.

\subsection{Examples with only double poles and simple poles}
\label{secSupportingdouble}

In this subsection, we will examine the Feynman rule $\rulei$ of
double pole by four examples in Figure \ref{Figsupport1}.

\subsection*{Example Figure \ref{Figsupport1}.a:}

This is a six-point example with CHY-integrand
\bea {1\over z_{12}^3 z_{45}^2 z_{23}
z_{16}z_{36}z_{35}z_{46}z_{34}z_{56}}~.~~~\eea
The possible subsets for poles are $\underline{\{1,2\}}$, $\{4,5\}$,
$\{3,4,5\}$, $\{4,5,6\}$, so from them we can only construct two
compatible combinations of subsets as
\bea \{\underline{\{1,2\}}, \{4,5\},
\{3,4,5\}\}~~~,~~~\{\underline{\{1,2\}},  \{4,5\},
\{4,5,6\}\}~.~~~\nonumber\eea
\begin{figure}
  \centering
  \includegraphics[width=4.5in]{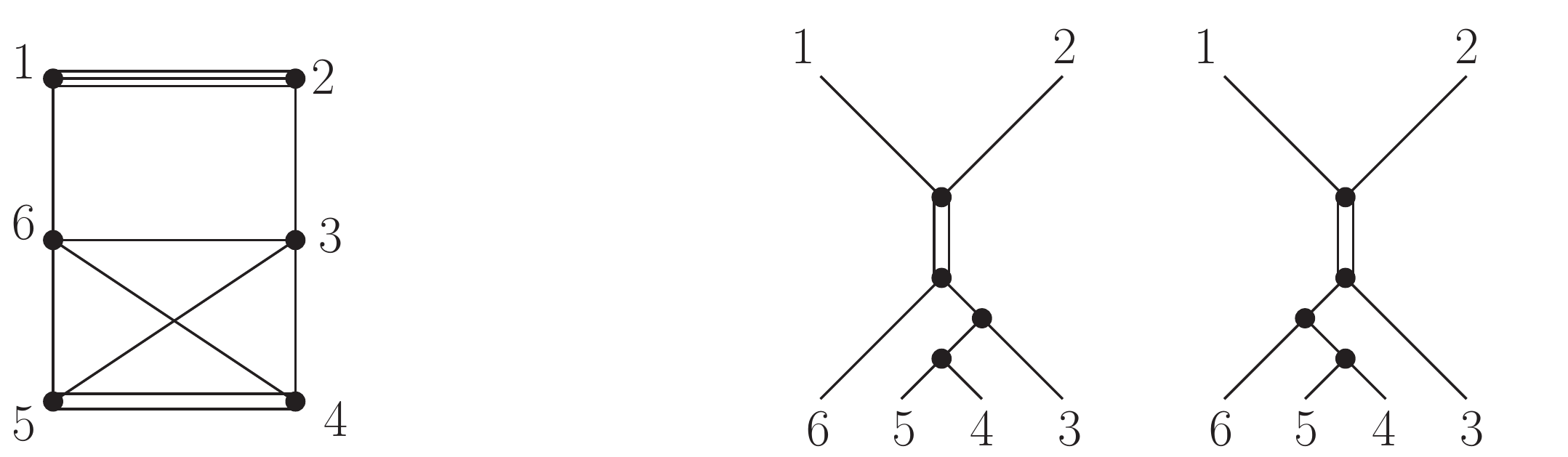}\\
  \caption{The {\sl 4-regular} graph of a given CHY-integrand, and its two
corresponding Feynman diagrams.}\label{FigA62}
\end{figure}
From these combinations we can draw two corresponding Feynman
diagrams as shown in Figure \ref{FigA62}, and using $\rulei$ we can
read out
\bea &&{1\over s_{45}s_{345}}
\ruleI{\{1\},\{2\},\{3,4,5\},\{6\}}+{1\over s_{45}s_{456}}
\ruleI{\{1\},\{2\},\{3\},\{4,5,6\}}\nonumber\\
&=&{2p_1p_{345}+2p_2p_6\over
2s_{12}^2s_{45}s_{345}}+{2p_1p_{3}+2p_2p_{456}\over
2s_{12}^2s_{45}s_{456}}=\frac{s_{13}}{s_{12}^2 s_{45}
s_{456}}+\frac{s_{26}}{s_{12}^2 s_{45} s_{345}}-\frac{1}{s_{12}^2
s_{45}}~.~~~\eea
%

\subsection*{Example Figure \ref{Figsupport1}.b:}

This is a six-point example with CHY-integrand
\bea {1\over z_{12}^2 z_{23}^2 z_{45}^2z_{56}^2z_{13}
z_{34}z_{16}z_{46}}~.~~~\eea
The possible subsets for poles are $\underline{\{1,2,3\}}, \{1,2\},
\{2,3\}, \{4,5\}, \{5,6\}$, so four compatible combinations of
subsets can be found as
\bea &&\{\underline{\{1,2,3\}}, \{1,2\},
 \{4,5\}\}~~~,~~~\{\underline{\{1,2,3\}}, \{1,2\},
\{5,6\}\}~,~~~\nonumber\\
&&\{\underline{\{1,2,3\}},\{2,3\},
\{4,5\}\}~~~,~~~\{\underline{\{1,2,3\}}, \{2,3\},
\{5,6\}\}~.~~~\nonumber\eea
\begin{figure}[ht]
  \centering
  \includegraphics[width=5.5in]{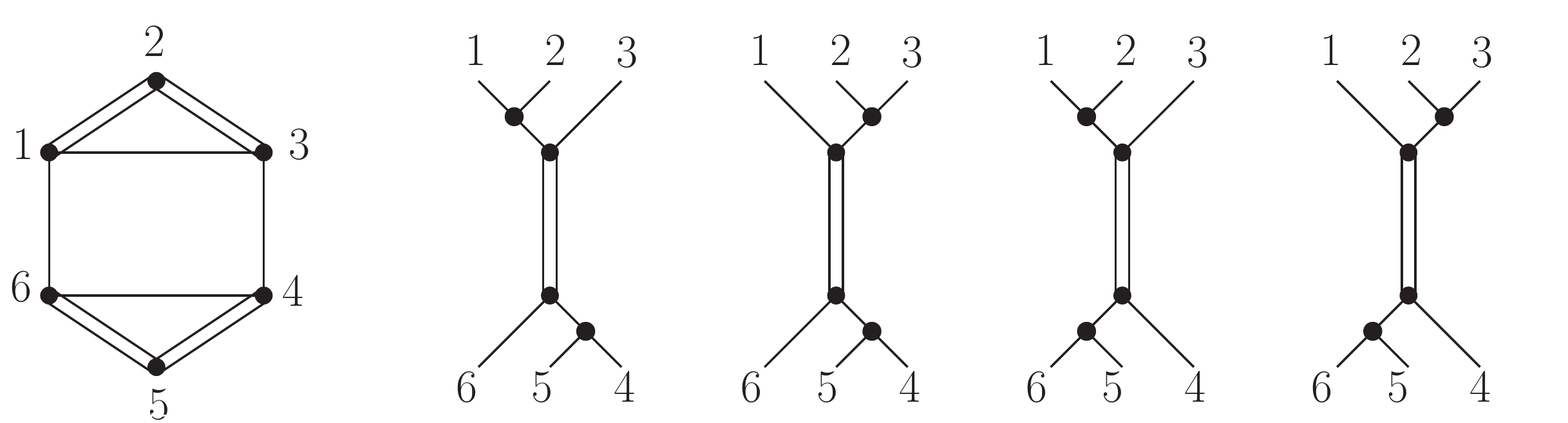}\\
  \caption{The {\sl 4-regular} graph of a given CHY-integrand, and its four
corresponding Feynman diagrams.}\label{FigA66}
\end{figure}
From these combinations we can draw four corresponding Feynman
diagrams as shown in Figure \ref{FigA66}. Using the rule $\rulei$,
we get
\bea &&{1\over s_{12}s_{45}}
\ruleI{\{1,2\},\{3\},\{4,5\},\{6\}}+{1\over s_{12}s_{56}}
\ruleI{\{1,2\},\{3\},\{4\},\{5,6\}}\nonumber\\
&&+{1\over s_{23}s_{45}} \ruleI{\{1\},\{2,3\},\{4,5\},\{6\}}+{1\over
s_{23}s_{56}}
\ruleI{\{1\},\{2,3\},\{4\},\{5,6\}}\nonumber\\
&=&{2p_{12}p_{45}+2p_3p_6\over
2s_{123}^2s_{12}s_{45}}+{2p_{12}p_{4}+2p_3p_{56}\over
2s_{123}^2s_{12}s_{56}}+{2p_{1}p_{45}+2p_{23}p_6\over
2s_{123}^2s_{23}s_{45}}+{2p_{1}p_{4}+2p_{23}p_{56}\over
2s_{123}^2s_{23}s_{56}}\nonumber\\
&=&\frac{s_{14}}{s_{23} s_{56} s_{123}^2}+\frac{s_{124}}{s_{12}
s_{56} s_{123}^2}+\frac{s_{145}}{s_{23} s_{45}
s_{123}^2}-\frac{1}{s_{12} s_{123}^2}-\frac{1}{s_{23}
s_{123}^2}-\frac{1}{s_{45} s_{123}^2}+\frac{s_{36}}{s_{12} s_{45}
s_{123}^2}-\frac{1}{s_{56} s_{123}^2}~.~~~\eea
%

\subsection*{Example Figure \ref{Figsupport1}.c:}

\begin{figure}[h]
  \centering
  \includegraphics[width=6.5in]{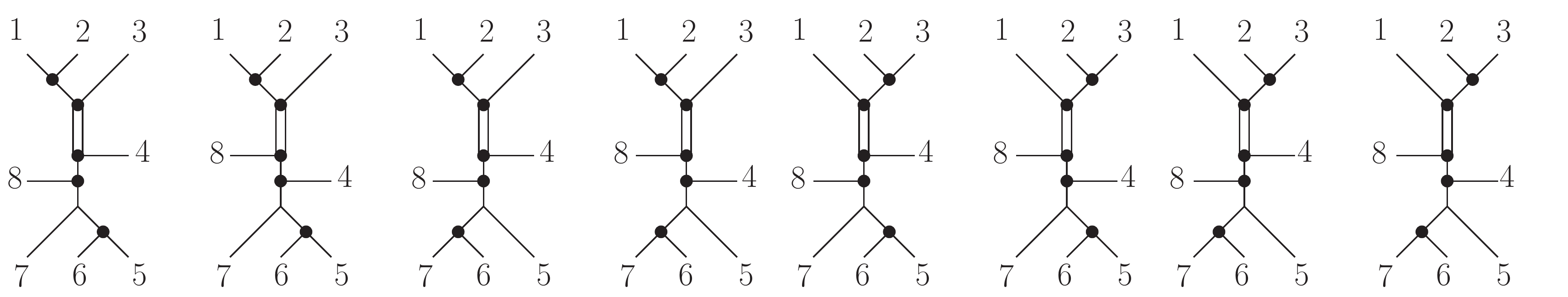}\\
  \caption{The eight Feynman diagrams corresponding to the {\sl 4-regular} graph Figure \ref{Figsupport1}.c.}\label{FigA83}
\end{figure}

This is an eight-point example with CHY-integrand
\bea {1\over z_{12}^2 z_{23}^2z_{56}^2
z_{67}^2z_{13}z_{18}z_{34}z_{48}z_{45}z_{78}z_{47}z_{58}}~.~~~\eea
The possible subsets for poles are $\{\underline{1,2,3}\}$,
$\{4,5,6,7\}$, $\{5,6,7,8\}$, $\{5,6,7\}$, $\{1,2\}$, $\{2,3\}$,
$\{5,6\}$, $\{6,7\}$, so we can construct eight compatible
combinations of subsets as
\bea &&\{\{\underline{1,2,3}\}, \{4,5,6,7\}, \{5,6,7\}, \{5,6\},
\{1,2\}\}~~~,~~~\{\{\underline{1,2,3}\}, \{4,5,6,7\}, \{5,6,7\},
\{6,7\},
\{1,2\}\}~,~~~\nonumber\\
&&\{\{\underline{1,2,3}\}, \{4,5,6,7\}, \{5,6,7\}, \{5,6\},
\{2,3\}\}~~~,~~~\{\{\underline{1,2,3}\}, \{4,5,6,7\}, \{5,6,7\},
\{6,7\},
\{2,3\}\}~,~~~\nonumber\\
&&\{\{\underline{1,2,3}\}, \{5,6,7,8\}, \{5,6,7\}, \{5,6\},
\{1,2\}\}~~~,~~~\{\{\underline{1,2,3}\}, \{5,6,7,8\}, \{5,6,7\},
\{6,7\},
\{1,2\}\}~,~~~\nonumber\\
&&\{\{\underline{1,2,3}\}, \{5,6,7,8\}, \{5,6,7\}, \{5,6\},
\{2,3\}\}~~~,~~~\{\{\underline{1,2,3}\}, \{5,6,7,8\}, \{5,6,7\},
\{6,7\}, \{2,3\}\}~.~~~\nonumber\eea
From them we can draw eight Feynman diagrams as shown in Figure
\ref{FigA83},
According to the rule $\rulei$, the result is given by
{\footnotesize \bea &&{1\over s_{12}s_{567}s_{4567}}\left({1\over
s_{56}}+{1\over s_{67}}\right)
\ruleI{\{1,2\},\{3\},\{4,5,6,7\},\{8\}}+{1\over
s_{12}s_{567}s_{5678}}\left({1\over s_{56}}+{1\over s_{67}}\right)
\ruleI{\{1,2\},\{3\},\{4\},\{5,6,7,8\}}\nonumber\\
&&+{1\over s_{23}s_{567}s_{4567}}\left({1\over s_{56}}+{1\over
s_{67}}\right) \ruleI{\{1\},\{2,3\},\{4,5,6,7\},\{8\}}+{1\over
s_{23}s_{567}s_{5678}}\left({1\over s_{56}}+{1\over s_{67}}\right)
\ruleI{\{1\},\{2,3\},\{4\},\{5,6,7,8\}}~.~~~\nonumber\eea}
%

\subsection*{Example Figure \ref{Figsupport1}.d:}

\begin{figure}[h]
  \centering
  \includegraphics[width=6.5in]{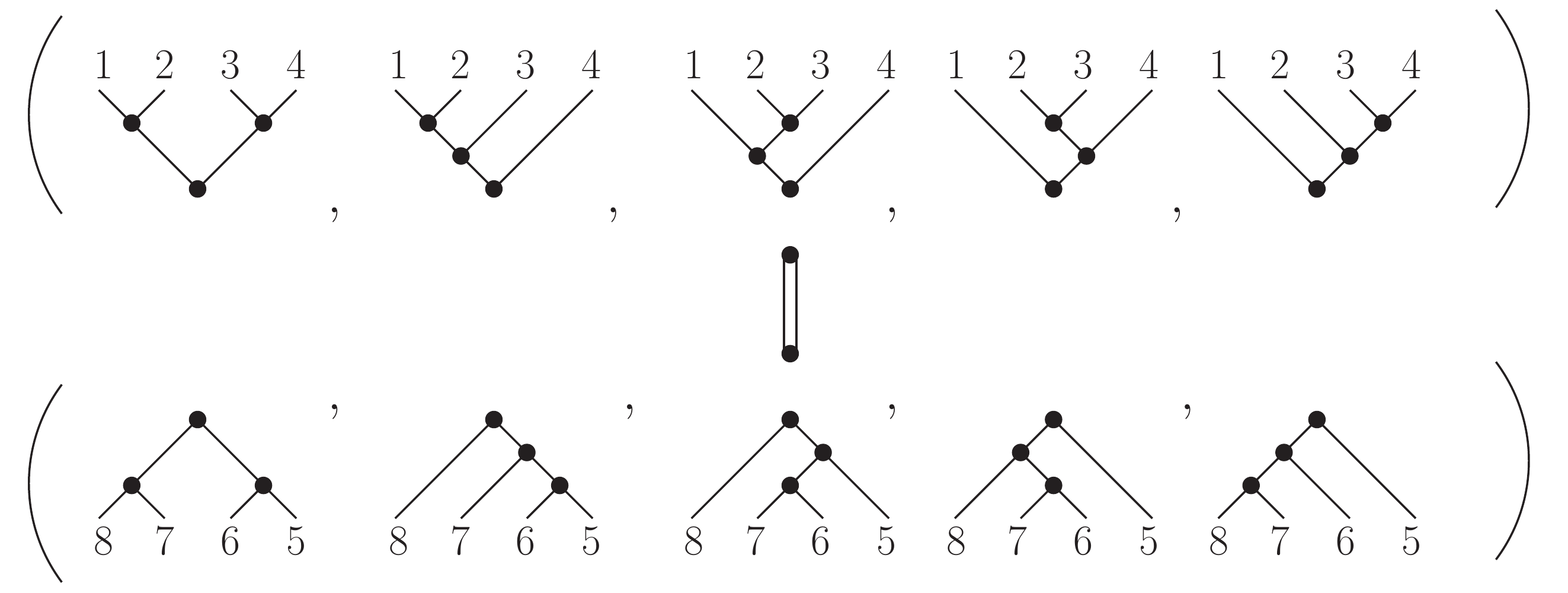}\\
  \caption{The 25 Feynman diagrams corresponding to the {\sl 4-regular} graph Figure \ref{Figsupport1}.d.}\label{FigA81}
\end{figure}

This is an eight-point example with CHY-integrand
\bea {1\over z_{12}^2 z_{23}^2z_{34}^2
z_{56}^2z_{67}^2z_{78}^2z_{45}z_{58}z_{18}z_{14}}~.~~~\eea
The possible subsets for poles are $\{\underline{1,2,3,4}\}$,
$\{1,2,3\}$, $\{2,3,4\}$, $\{5,6,7\}$, $\{6,7,8\}$, $\{1,2\}$,
$\{2,3\}$, $\{3,4\}$, $\{5,6\}$, $\{6,7\}$, $\{7,8\}$, so we can
construct 25 compatible combinations of subsets as
{\small \bea &&\{\{\underline{1,2,3,4}\}, \{1,2\}, \{3,4\}, \{5,6\},
\{7,8\}\}~,~~~\nonumber\\
&&\{\{\underline{1,2,3,4}\}, \{1,2,3\},
\Big[\{1,2\}~\mbox{or}~\{2,3\}\Big], \{5,6\},
\{7,8\}\}~~,~~\{\{\underline{1,2,3,4}\}, \{2,3,4\},
\Big[\{2,3\}~\mbox{or}~\{3,4\}\Big], \{5,6\},
\{7,8\}\}~,~~~\nonumber\\
&&\{\{\underline{1,2,3,4}\}, \{1,2\}, \{3,4\}, \{5,6,7\},
\Big[\{5,6\}~\mbox{or}~\{6,7\}\Big]\}~~,~~\{\{\underline{1,2,3,4}\},
\{1,2\}, \{3,4\}, \{6,7,8\},
\Big[\{6,7\}~\mbox{or}~\{7,8\}\Big]\}~,~~~\nonumber\\
&&\{\{\underline{1,2,3,4}\}, \{1,2,3\}, \{5,6,7\}~,~
\Big[\{1,2\}~\mbox{or}~\{2,3\}\Big]~,~\Big[
\{5,6\}~\mbox{or}~\{6,7\}\Big]\}~,~~~\nonumber\\
&&\{\{\underline{1,2,3,4}\}, \{1,2,3\},
\{6,7,8\}~,~\Big[\{1,2\}~\mbox{or}~\{2,3\}\Big]~,~
\Big[\{6,7\}~\mbox{or}~\{7,8\}\Big]\}~,~~~\nonumber\\
&&\{\{\underline{1,2,3,4}\}, \{2,3,4\},
\{5,6,7\}~,~\Big[\{2,3\}~\mbox{or}~\{3,4\}\Big]~,~
\Big[\{5,6\}~\mbox{or}~\{6,7\}\Big]\}~,~~~\nonumber\\
&&\{\{\underline{1,2,3,4}\}, \{2,3,4\}, \{6,7,8\}~,~
\Big[\{2,3\}~\mbox{or}~\{3,4\}\Big]~,~
\Big[\{6,7\}~\mbox{or}~\{7,8\}\Big]\}~.~~~\nonumber\eea}
From them we can draw 25 Feynman diagrams as shown in Figure
\ref{FigA81},
which immediately leads to the final result by Feynman rule $\rulei$
as
{\footnotesize \bea &&{1\over s_{12}s_{34}s_{56}s_{78}}
\ruleI{\{1,2\},\{3,4\},\{5,6\},\{7,8\}}\nonumber\\
&&+{1\over s_{123}s_{56}s_{78}}\left({1\over s_{12}}+{1\over
s_{23}}\right)\ruleI{\{1,2,3\},\{4\},\{5,6\},\{7,8\}}+{1\over
s_{234}s_{56}s_{78}}\left({1\over s_{23}}+{1\over
s_{34}}\right)\ruleI{\{1\},\{2,3,4\},\{5,6\},\{7,8\}}\nonumber\\
&&+{1\over s_{12}s_{34}s_{567}}\left({1\over s_{56}}+{1\over
s_{67}}\right)\ruleI{\{1,2\},\{3,4\},\{5,6,7\},\{8\}}+{1\over
s_{12}s_{34}s_{678}}\left({1\over s_{67}}+{1\over
s_{78}}\right)\ruleI{\{1,2\},\{3,4\},\{5\},\{6,7,8\}}\nonumber\\
&&+{1\over s_{123}s_{567}}\left({1\over s_{12}}+{1\over
s_{23}}\right)\left({1\over s_{56}}+{1\over
s_{67}}\right)\ruleI{\{1,2,3\},\{4\},\{5,6,7\},\{8\}}\nonumber\\
&&+{1\over s_{123}s_{678}}\left({1\over s_{12}}+{1\over
s_{23}}\right)\left({1\over s_{67}}+{1\over
s_{78}}\right)\ruleI{\{1,2,3\},\{4\},\{5\},\{6,7,8\}}\nonumber\\
&&+{1\over s_{234}s_{567}}\left({1\over s_{23}}+{1\over
s_{34}}\right)\left({1\over s_{56}}+{1\over
s_{67}}\right)\ruleI{\{1\},\{2,3,4\},\{5,6,7\},\{8\}}\nonumber\\
&&+{1\over s_{234}s_{678}}\left({1\over s_{23}}+{1\over
s_{34}}\right)\left({1\over s_{67}}+{1\over
s_{78}}\right)\ruleI{\{1\},\{2,3,4\},\{5\},\{6,7,8\}}~.~~~\eea}
%

\subsection{Examples with only triple poles and simple poles}
\label{secSupportingtriple}

In this subsection, we will examine the Feynman rule $\ruleii$ of
triple pole by eight examples in Figure \ref{Figsupport2}.

\subsection*{Example Figure \ref{Figsupport2}.a:}

\begin{figure}[h]
  \centering
  \includegraphics[width=5.5in]{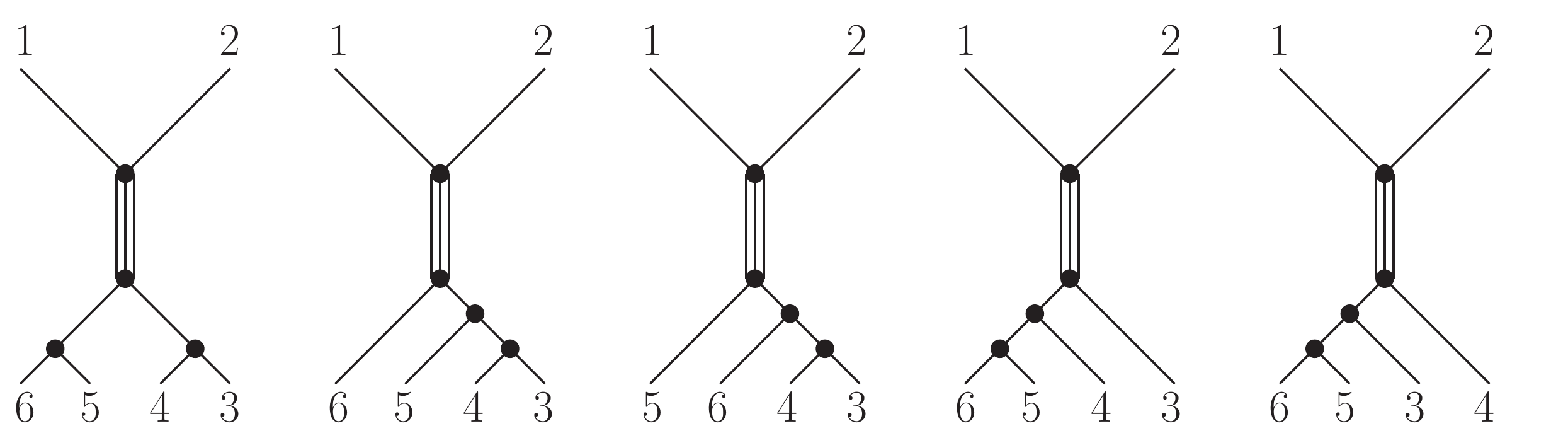}\\
  \caption{The five Feynman diagrams corresponding to the {\sl 4-regular} graph Figure \ref{Figsupport2}.a.}\label{FigA63}
\end{figure}

This is a six-point example with CHY-integrand
\bea {1\over z_{12}^4 z_{34}^2 z_{56}^2z_{36}z_{45}
z_{35}z_{46}}~.~~~\eea
The possible subsets for poles are $\{\dunderline{1,2}\}, \{3,4\},
\{5,6\}, \{3,4,5\}, \{3,4,6\}, \{3,5,6\}, \{4,5,6\}$, so we can
construct five compatible combinations as
\bea &&\{\{\dunderline{1,2}\}, \{3,4\},
\{5,6\}\}~~~,~~~\{\{\dunderline{1,2}\}, \{3,4\},
\{3,4,5\}\}~,~~~\nonumber\\
&&\{\{\dunderline{1,2}\}, \{3,4\},
\{3,4,6\}\}~~~,~~~\{\{\dunderline{1,2}\}, \{5,6\},
\{3,5,6\}\}~~~,~~~\{\{\dunderline{1,2}\}, \{5,6\},
\{4,5,6\}\}~.~~~\nonumber\eea
The five Feynman diagrams then follows in Figure \ref{FigA63}.
By $\ruleii$ we get the result as
{\small \bea &&{1\over s_{34}s_{56}}
\ruleII{\{1\},\{2\},\{3,4\},\{5,6\}}+{1\over s_{34}s_{345}}
\ruleII{\{1\},\{2\},\{3,4,5\},\{6\}}+{1\over s_{34}s_{346}}
\ruleII{\{1\},\{2\},\{3,4,6\},\{5\}}\nonumber\\
&&+{1\over s_{56}s_{356}}
\ruleII{\{1\},\{2\},\{3,5,6\},\{4\}}+{1\over s_{56}s_{456}}
\ruleII{\{1\},\{2\},\{4,5,6\},\{3\}}\nonumber\\
&=&-\frac{s_{15}}{s_{12}^2 s_{34} s_{56}}-\frac{s_{23}
s_{15}}{s_{12}^3 s_{34} s_{56}}-\frac{s_{24} s_{15}}{s_{12}^3 s_{34}
s_{56}}+\frac{1}{s_{12}^2 s_{56}}-\frac{s_{16}}{s_{12}^2 s_{34}
s_{56}}-\frac{s_{16} s_{23}}{s_{12}^3 s_{34}
s_{56}}\nonumber\\
&&~~~~~~~~~~~~~~~~~~~~~~~-\frac{s_{16} s_{24}}{s_{12}^3 s_{34}
s_{56}}+\frac{s_{13} s_{23}}{s_{12}^3 s_{56} s_{456}}+\frac{s_{16}
s_{26}}{s_{12}^3 s_{34} s_{345}}+\frac{s_{15} s_{25}}{s_{12}^3
s_{34} s_{346}}+\frac{s_{14} s_{24}}{s_{12}^3 s_{56}
s_{356}}~.~~~\eea}
%

\subsection*{Example Figure \ref{Figsupport2}.b:}

\begin{figure}[h]
  \centering
  \includegraphics[width=5.5in]{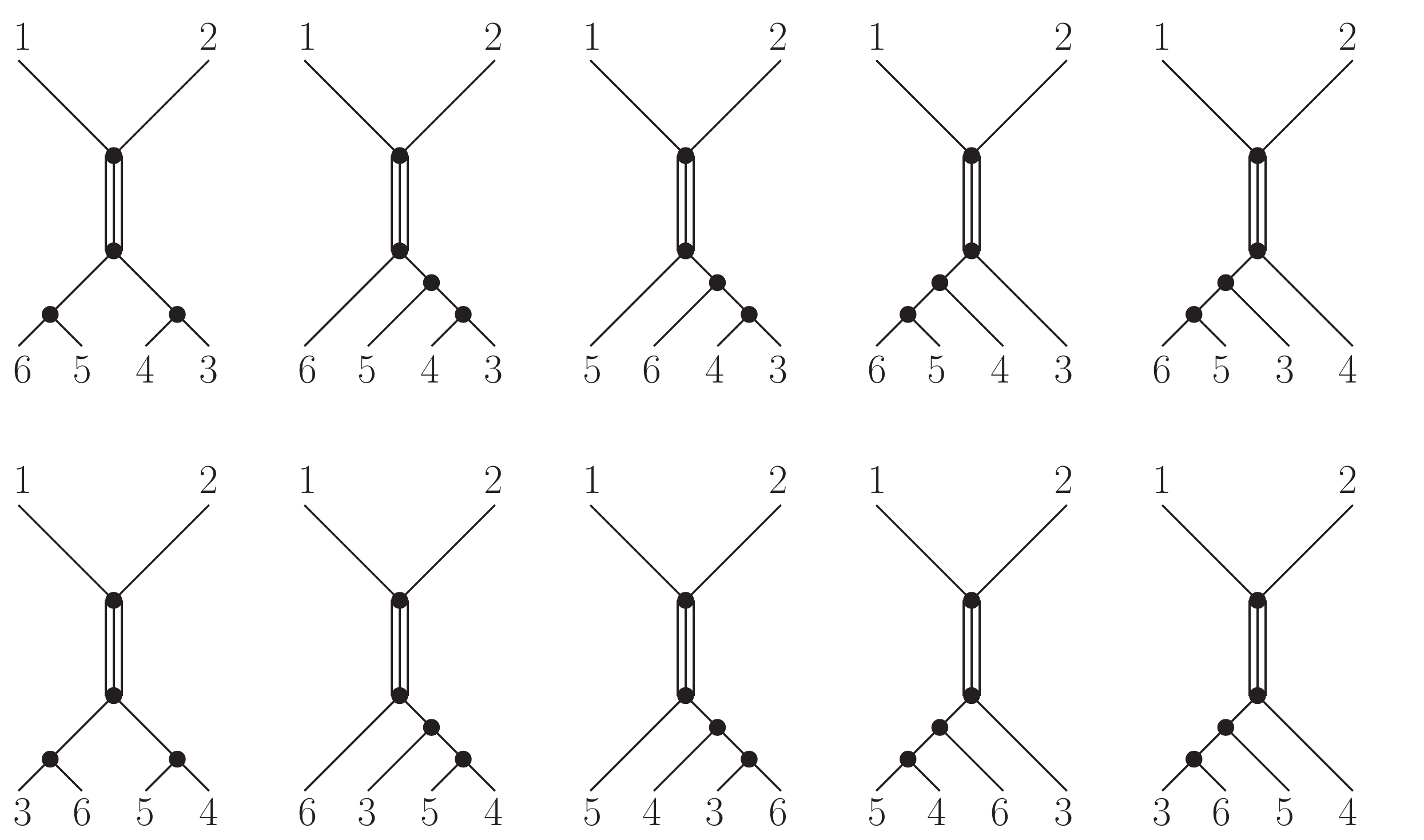}\\
  \caption{The 10 Feynman diagrams corresponding to the {\sl 4-regular} graph Figure \ref{Figsupport2}.b.}\label{FigA65}
\end{figure}

This is a six-point example with CHY-integrand
\bea {1\over z_{12}^4 z_{34}^2z_{45}^2 z_{56}^2z_{16}^2}~.~~~\eea
The possible subsets for poles are $\{\dunderline{1,2}\}, \{3,4\},
\{4,5\}, \{5,6\}, \{3,6\}, \{3,4,5\}, \{3,4,6\}, \{3,5,6\},
\{4,5,6\}$, so the 10 compatible combinations can be given as
\bea &&\{\{\dunderline{1,2}\}, \{3,4\},
\{5,6\}\}~~,~~\{\{\dunderline{1,2}\}, \{4,5\},
\{3,6\}\}~~,~~\{\{\dunderline{1,2}\}, \{3,4\},
\{3,4,5\}\}~~,~~\{\{\dunderline{1,2}\}, \{3,4\},
\{3,4,6\}\}~,~~\nonumber\\
&&\{\{\dunderline{1,2}\}, \{5,6\},
\{3,5,6\}\}~~,~~\{\{\dunderline{1,2}\}, \{5,6\},
\{4,5,6\}\}~~,~~\{\{\dunderline{1,2}\}, \{4,5\},
\{3,4,5\}\}~~,~~\{\{\dunderline{1,2}\}, \{3,6\},
\{3,4,6\}\}~,~~~\nonumber\\
&&\{\{\dunderline{1,2}\}, \{3,6\},
\{3,5,6\}\}~~,~~\{\{\dunderline{1,2}\}, \{4,5\},
\{4,5,6\}\}~,~~~\nonumber\eea
and the corresponding 10 Feynman diagrams are presented in Figure
\ref{FigA65}.
Accordingly, we write down the result by $\ruleii$ as
{\small \bea &&{1\over s_{34}s_{56}}
\ruleII{\{1\},\{2\},\{3,4\},\{5,6\}}+{1\over s_{36}s_{45}}
\ruleII{\{1\},\{2\},\{3,6\},\{4,5\}}\nonumber\\
&&+\left({1\over s_{34}s_{345}}+{1\over s_{45}s_{345}}\right)
\ruleII{\{1\},\{2\},\{3,4,5\},\{6\}}+\left({1\over
s_{34}s_{346}}+{1\over s_{36}s_{346}}\right)
\ruleII{\{1\},\{2\},\{3,4,6\},\{5\}}\nonumber\\
&&+\left({1\over s_{56}s_{356}}+{1\over s_{36}s_{356}}\right)
\ruleII{\{1\},\{2\},\{3,5,6\},\{4\}}+\left({1\over
s_{56}s_{456}}+{1\over s_{45}s_{456}}\right)
\ruleII{\{1\},\{2\},\{4,5,6\},\{3\}}~.~~~\eea}
%

\subsection*{Example Figure \ref{Figsupport2}.c:}

\begin{figure}[h]
  \centering
  \includegraphics[width=1in]{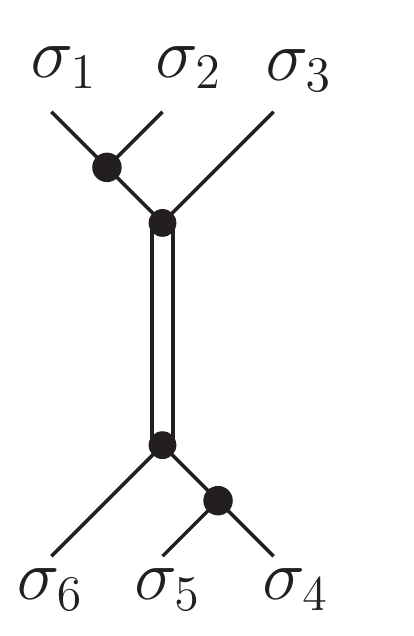}\\
  \caption{The nine Feynman diagrams corresponding to the {\sl 4-regular} graph Figure \ref{Figsupport2}.c. $\{\sigma_1,\sigma_2,\sigma_3\}$ takes the
  cyclic permutations of $\{1,2,3\}$, and $\{\sigma_4,\sigma_5,\sigma_6\}$ takes the cyclic permutation of $\{4,5,6\}$.}\label{FigA67}
\end{figure}

This is a six-point example with CHY-integrand
\bea {1\over z_{12}^2 z_{23}^2z_{31}^2
z_{45}^2z_{56}^2z_{64}^2}~.~~~\eea
The possible subsets for poles are $\{\dunderline{1,2,3}\}, \{1,2\},
\{2,3\}, \{1,2\}, \{4,5\}, \{5,6\}, \{4,6\}$, so we can construct
nine compatible combinations as
\bea &&\{\{\dunderline{1,2,3}\}, \{1,2\},
\{4,5\}\}~~~,~~~\{\{\dunderline{1,2,3}\}, \{1,2\},
\{5,6\}\}~~~,~~~\{\{\dunderline{1,2,3}\}, \{1,2\},
\{4,6\}\}~,~~~\nonumber\\
&&\{\{\dunderline{1,2,3}\}, \{2,3\},
\{4,5\}\}~~~,~~~\{\{\dunderline{1,2,3}\}, \{2,3\},
\{5,6\}\}~~~,~~~\{\{\dunderline{1,2,3}\}, \{2,3\},
\{4,6\}\}\nonumber\\
&&\{\{\dunderline{1,2,3}\}, \{1,3\},
\{4,5\}\}~,~~~\{\{\dunderline{1,2,3}\}, \{1,3\},
\{5,6\}\}~~~,~~~\{\{\dunderline{1,2,3}\}, \{1,3\},
\{4,6\}\}~,~~~\nonumber\eea
with nine Feynman diagrams as shown in Figure \ref{FigA67}.
Then the answer is trivially given by
{\small \bea &&{1\over s_{12}s_{45}}
\ruleII{\{1,2\},\{3\},\{4,5\},\{6\}}+{1\over s_{12}s_{56}}
\ruleII{\{1,2\},\{3\},\{5,6\},\{4\}}+{1\over s_{12}s_{46}}
\ruleII{\{1,2\},\{3\},\{4,6\},\{5\}}\nonumber\\
&&+{1\over s_{23}s_{45}}
\ruleII{\{2,3\},\{1\},\{4,5\},\{6\}}+{1\over s_{23}s_{56}}
\ruleII{\{2,3\},\{1\},\{5,6\},\{4\}}+{1\over s_{23}s_{46}}
\ruleII{\{2,3\},\{1\},\{4,6\},\{5\}}\nonumber\\
&&+{1\over s_{13}s_{45}}
\ruleII{\{1,3\},\{2\},\{4,5\},\{6\}}+{1\over s_{13}s_{56}}
\ruleII{\{1,3\},\{2\},\{5,6\},\{4\}}+{1\over s_{13}s_{46}}
\ruleII{\{1,3\},\{2\},\{4,6\},\{5\}}~.~~~\nonumber\eea}
%

\subsection*{Example Figure \ref{Figsupport2}.d:}

\begin{figure}[h]
  \centering
  \includegraphics[width=6in]{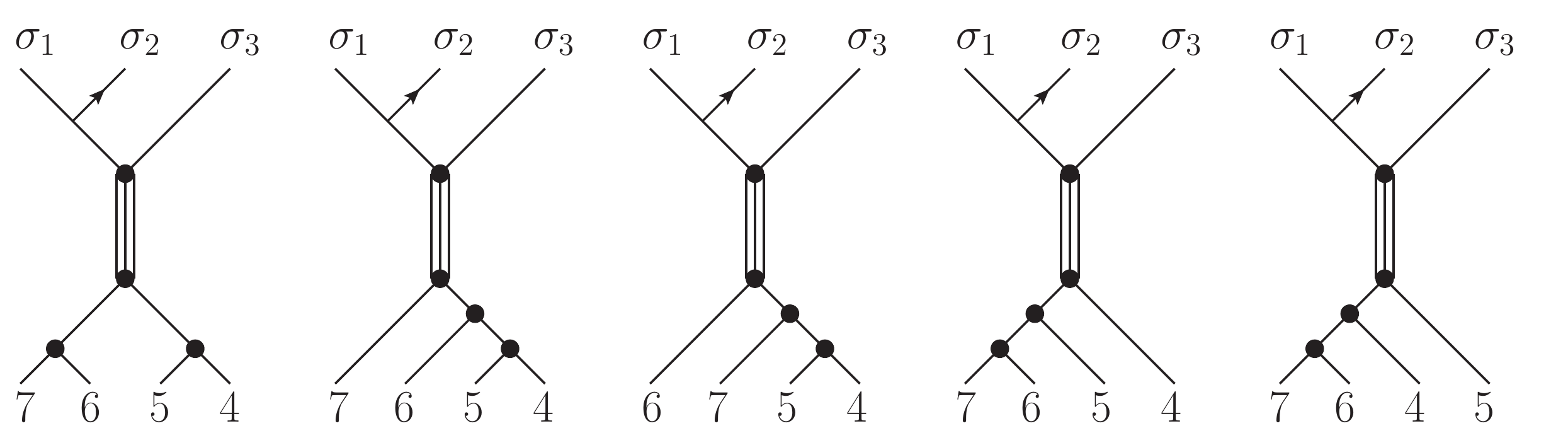}\\
  \caption{The 15 Feynman diagrams corresponding to the {\sl 4-regular} graph Figure \ref{Figsupport2}.d. $\{\sigma_1,\sigma_2,\sigma_3\}$ takes the
  cyclic permutations of $\{1,2,3\}$.}\label{FigA71}
\end{figure}

This is a seven-point example with CHY-integrand
\bea -{1\over z_{12}^2 z_{23}^2z_{31}^2
z_{45}^2z_{67}^2z_{56}z_{64}z_{47}z_{75}}~.~~~\eea
The possible subsets for poles are $\{\dunderline{1,2,3}\},
\{4,5,6\}, \{4,5,7\}, \{4,6,7\}, \{5,6,7\}, \{1,2\}, \{2,3\},
\{1,3\}, \{4,5\}, \{6,7\}$, so we can construct 15 compatible
combinations of subsets as
{\small \bea &&\{\{\dunderline{1,2,3}\}, \{1,2\}, \{4,5\},
\{6,7\}\}~~~,~~~\{\{\dunderline{1,2,3}\}, \{1,2\}, \{4,5\},
\{4,5,6\}\}~~~,~~~\{\{\dunderline{1,2,3}\}, \{1,2\},
\{4,5\}, \{4,5,7\}\}~,~~~\nonumber\\
&&\{\{\dunderline{1,2,3}\}, \{1,2\}, \{6,7\},
\{4,6,7\}\}~~~,~~~\{\{\dunderline{1,2,3}\}, \{1,2\}, \{6,7\},
\{5,6,7\}\}~~~,~~~\nonumber\\
&&\{\{\dunderline{1,2,3}\}, \{1,3\}, \{4,5\},
\{6,7\}\}~~~,~~~\{\{\dunderline{1,2,3}\}, \{1,3\}, \{4,5\},
\{4,5,6\}\}~~~,~~~\{\{\dunderline{1,2,3}\}, \{1,3\},
\{4,5\}, \{4,5,7\}\}~,~~~\nonumber\\
&&\{\{\dunderline{1,2,3}\}, \{1,3\}, \{6,7\},
\{4,6,7\}\}~~~,~~~\{\{\dunderline{1,2,3}\}, \{1,3\}, \{6,7\},
\{5,6,7\}\}~~~,~~~\nonumber\\
&&\{\{\dunderline{1,2,3}\}, \{2,3\}, \{4,5\},
\{6,7\}\}~~~,~~~\{\{\dunderline{1,2,3}\}, \{2,3\}, \{4,5\},
\{4,5,6\}\}~~~,~~~\{\{\dunderline{1,2,3}\}, \{2,3\},
\{4,5\}, \{4,5,7\}\}~,~~~\nonumber\\
&&\{\{\dunderline{1,2,3}\}, \{2,3\}, \{6,7\},
\{4,6,7\}\}~~~,~~~\{\{\dunderline{1,2,3}\}, \{2,3\}, \{6,7\},
\{5,6,7\}\}~.~~~\nonumber\eea}
With them we can draw 15 Feynman diagrams as presented in Figure
\ref{FigA71}.
According to rule $\ruleii$, the result is given by
{\small \bea &&{1\over s_{12}s_{45}s_{67}}
\ruleII{\{1,2\},\{3\},\{4,5\},\{6,7\}}+{1\over s_{12}s_{45}s_{456}}
\ruleII{\{1,2\},\{3\},\{4,5,6\},\{7\}}\nonumber\\
&&+{1\over s_{12}s_{45}s_{457}}
\ruleII{\{1,2\},\{3\},\{4,5,7\},\{6\}}+{1\over s_{12}s_{67}s_{467}}
\ruleII{\{1,2\},\{3\},\{4,6,7\},\{5\}}\nonumber\\
&&+{1\over s_{12}s_{67}s_{567}}
\ruleII{\{1,2\},\{3\},\{5,6,7\},\{4\}}+\Big(~~\{1,2,3\}\to
\{2,3,1\}~~,~~\{3,1,2\}~~\Big)~.~~~\nonumber\eea}
%

\subsection*{Example Figure \ref{Figsupport2}.e:}

\begin{figure}[h]
  \centering
  \includegraphics[width=5.5in]{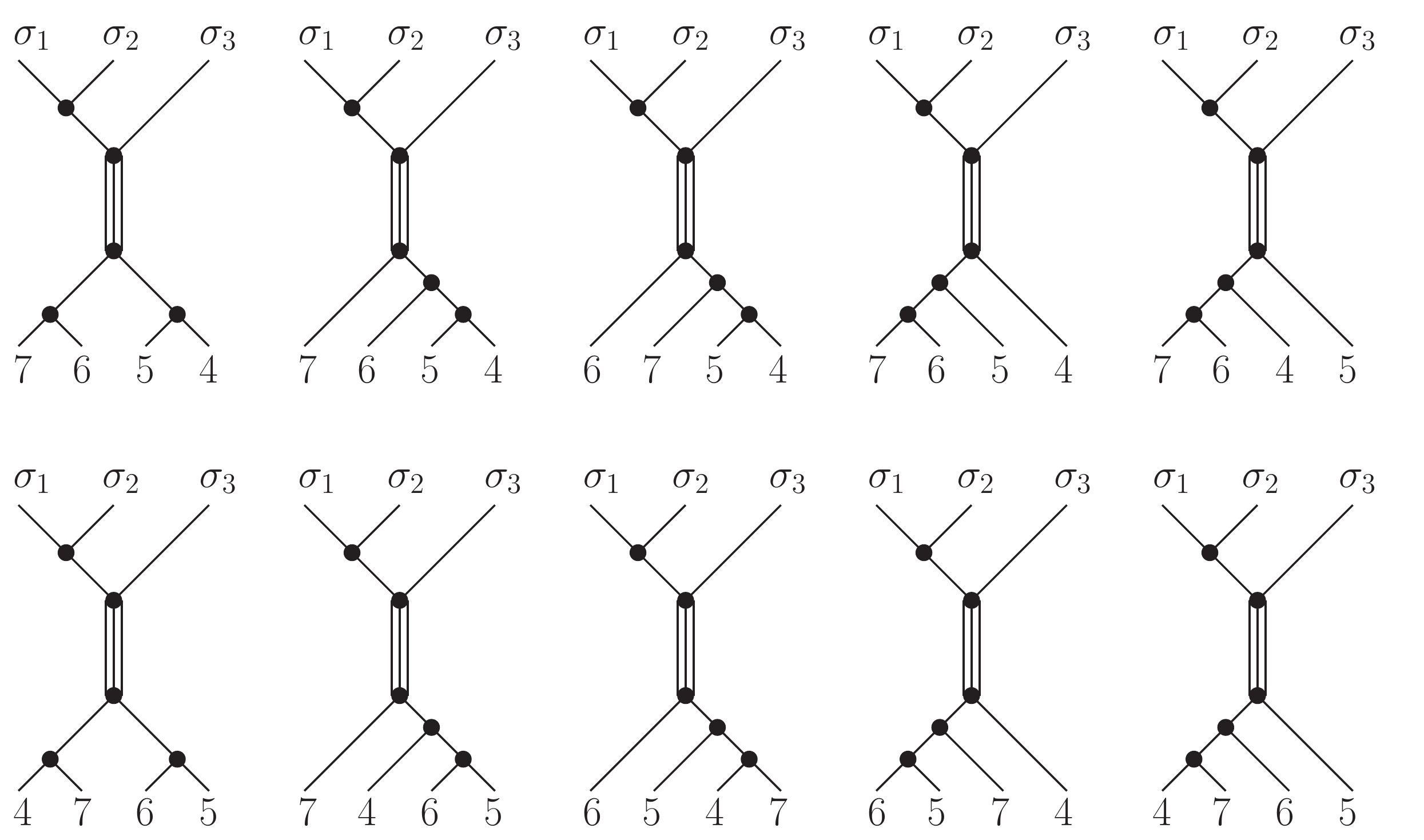}\\
  \caption{The 30 Feynman diagrams corresponding to the {\sl 4-regular} graph Figure \ref{Figsupport2}.e. $\{\sigma_1,\sigma_2,\sigma_3\}$ takes the
  cyclic permutations of $\{1,2,3\}$.}\label{FigA72}
\end{figure}

This is a seven-point example with CHY-integrand
\bea -{1\over z_{12}^2 z_{23}^2z_{31}^2
z_{45}^2z_{56}^2z_{67}^2z_{74}^2}~.~~~\eea
The possible subsets for poles are $\{\dunderline{1,2,3}\}$,
$\{4,5,6\}$, $\{4,5,7\}$, $\{4,6,7\}$, $\{5,6,7\}$, $\{1,2\}$,
$\{2,3\}$, $\{1,3\}$, $\{4,5\}$, $\{5,6\}$, $\{6,7\}$, $\{4,7\}$, so
we can construct 30 compatible combinations as
{\small \bea &&\{\{\dunderline{1,2,3}\}, \{1,2\}, \{4,5\},
\{6,7\}\}~~~,~~~\{\{\dunderline{1,2,3}\}, \{1,2\}, \{4,7\},
\{5,6\}\}~~~,~~~\{\{\dunderline{1,2,3}\}, \{1,2\}, \{4,5\},
\{4,5,6\}\}~,~~~\nonumber\\
&&\{\{\dunderline{1,2,3}\}, \{1,2\}, \{4,5\},
\{4,5,7\}\}~~~,~~~\{\{\dunderline{1,2,3}\}, \{1,2\}, \{6,7\},
\{4,6,7\}\}~~~,~~~\{\{\dunderline{1,2,3}\}, \{1,2\}, \{6,7\},
\{5,6,7\}\}~,~~~\nonumber\\
&&\{\{\dunderline{1,2,3}\}, \{1,2\}, \{5,6\},
\{4,5,6\}\}~~~,~~~\{\{\dunderline{1,2,3}\}, \{1,2\}, \{4,7\},
\{4,5,7\}\}~~~,~~~\{\{\dunderline{1,2,3}\}, \{1,2\}, \{4,7\},
\{4,6,7\}\}~,~~~\nonumber\\
&&\{\{\dunderline{1,2,3}\}, \{1,2\}, \{5,6\},
\{5,6,7\}\}~,~~~\nonumber\\
&&\{\{\dunderline{1,2,3}\}, \{2,3\}, \{4,5\},
\{6,7\}\}~~~,~~~\{\{\dunderline{1,2,3}\}, \{2,3\}, \{4,7\},
\{5,6\}\}~~~,~~~\{\{\dunderline{1,2,3}\}, \{2,3\}, \{4,5\},
\{4,5,6\}\}~,~~~\nonumber\\
&&\{\{\dunderline{1,2,3}\}, \{2,3\}, \{4,5\},
\{4,5,7\}\}~~~,~~~\{\{\dunderline{1,2,3}\}, \{2,3\}, \{6,7\},
\{4,6,7\}\}~~~,~~~\{\{\dunderline{1,2,3}\}, \{2,3\}, \{6,7\},
\{5,6,7\}\}~,~~~\nonumber\\
&&\{\{\dunderline{1,2,3}\}, \{2,3\}, \{5,6\},
\{4,5,6\}\}~~~,~~~\{\{\dunderline{1,2,3}\}, \{2,3\}, \{4,7\},
\{4,5,7\}\}~~~,~~~\{\{\dunderline{1,2,3}\}, \{2,3\}, \{4,7\},
\{4,6,7\}\}~,~~~\nonumber\\
&&\{\{\dunderline{1,2,3}\}, \{2,3\}, \{5,6\},
\{5,6,7\}\}~,~~~\nonumber\\
&&\{\{\dunderline{1,2,3}\}, \{1,3\}, \{4,5\},
\{6,7\}\}~~~,~~~\{\{\dunderline{1,2,3}\}, \{1,3\}, \{4,7\},
\{5,6\}\}~~~,~~~\{\{\dunderline{1,2,3}\}, \{1,3\}, \{4,5\},
\{4,5,6\}\}~,~~~\nonumber\\
&&\{\{\dunderline{1,2,3}\}, \{1,3\}, \{4,5\},
\{4,5,7\}\}~~~,~~~\{\{\dunderline{1,2,3}\}, \{1,3\}, \{6,7\},
\{4,6,7\}\}~~~,~~~\{\{\dunderline{1,2,3}\}, \{1,3\}, \{6,7\},
\{5,6,7\}\}~,~~~\nonumber\\
&&\{\{\dunderline{1,2,3}\}, \{1,3\}, \{5,6\},
\{4,5,6\}\}~~~,~~~\{\{\dunderline{1,2,3}\}, \{1,3\}, \{4,7\},
\{4,5,7\}\}~~~,~~~\{\{\dunderline{1,2,3}\}, \{1,3\}, \{4,7\},
\{4,6,7\}\}~,~~~\nonumber\\
&&\{\{\dunderline{1,2,3}\}, \{1,3\}, \{5,6\},
\{5,6,7\}\}~.~~~\nonumber\eea}
The 30 Feynman diagrams corresponding to this CHY-integrand is shown
in Figure \ref{FigA72}.
Then the answer is given by
{\small \bea &&{1\over s_{12}s_{45}s_{67}}
\ruleII{\{1,2\},\{3\},\{4,5\},\{6,7\}}+{1\over s_{12}s_{47}s_{56}}
\ruleII{\{1,2\},\{3\},\{4,7\},\{5,6\}}\nonumber\\
&&+{1\over s_{12}}\left({1\over s_{45}s_{456}}+{1\over
s_{56}s_{456}}\right) \ruleII{\{1,2\},\{3\},\{4,5,6\},\{7\}}+{1\over
s_{12}}\left({1\over s_{45}s_{457}}+{1\over s_{47}s_{457}}\right)
\ruleII{\{1,2\},\{3\},\{4,5,7\},\{6\}}\nonumber\\
&&+{1\over s_{12}}\left({1\over s_{47}s_{467}}+{1\over
s_{67}s_{467}}\right)\ruleII{\{1,2\},\{3\},\{4,6,7\},\{5\}}+{1\over
s_{12}}\left({1\over s_{56}s_{567}}+{1\over s_{67}s_{567}}\right)
\ruleII{\{1,2\},\{3\},\{5,6,7\},\{4\}}\nonumber\\
&&~~~~~~~~~+\Big(\{1,2,3\}\to
\{2,3,1\}~~,~~\{3,1,2\}\Big)~.~~~\nonumber\eea}
%

\subsection*{Example Figure \ref{Figsupport2}.f:}

\begin{figure}[h]
  \centering
  \includegraphics[width=6in]{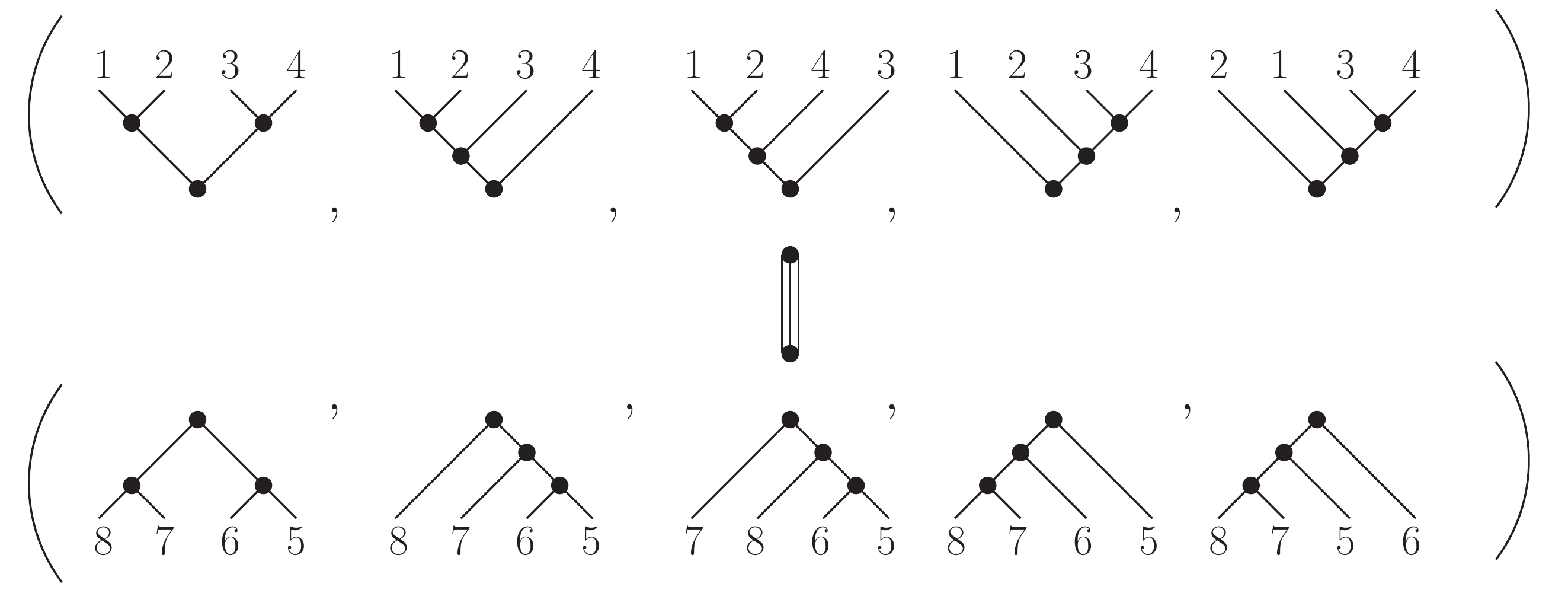}\\
  \caption{The 25 Feynman diagrams corresponding to the {\sl 4-regular} graph Figure \ref{Figsupport2}.f.}\label{FigA82}
\end{figure}

This is an eight-point example with CHY-integrand
\bea {1\over z_{12}^2 z_{34}^2
z_{56}^2z_{78}^2z_{23}z_{31}z_{14}z_{42}z_{67}z_{75}z_{58}z_{86}}~.~~~\eea
The possible subsets for poles are $\{\dunderline{1,2,3,4}\}$,
$\{1,2,3\}$, $\{2,3,4\}$, $\{1,3,4\}$, $\{1,2,4\}$, $\{5,6,7\}$,
$\{6,7,8\}$, $\{5,7,8\}$, $\{5,6,8\}$, $\{1,2\}$, $\{3,4\}$,
$\{5,6\}$, $\{7,8\}$, so we can construct 25 compatible combinations
of subsets as
\bea
&&\{\{\dunderline{1,2,3,4}\},\{1,2\},\{3,4\},\{5,6\},\{7,8\}\}~,~~~\nonumber\\
&&\{\{\dunderline{1,2,3,4}\},\Big[\{1,2,3\},\{1,2\}~\mbox{or}~\{1,2,4\},\{1,2\}~\mbox{or}~\{1,3,4\},\{3,4\}~\mbox{or}~\{2,3,4\},\{3,4\}\Big],\{5,6\},\{7,8\}\}~,~~~\nonumber\\
&&\{\{\dunderline{1,2,3,4}\},\{1,2\}, \{3,4\},
\Big[\{5,6,7\},\{5,6\}~\mbox{or}~\{5,6,8\},\{5,6\}~\mbox{or}~\{5,7,8\},\{7,8\}~\mbox{or}~\{6,7,8\},\{7,8\}\Big]\}~,~~~\nonumber\\
&&\{\{\dunderline{1,2,3,4}\},\Big[\{1,2,3\},\{1,2\}~\mbox{or}~\{1,2,4\},\{1,2\}~\mbox{or}~\{1,3,4\},\{3,4\}~\mbox{or}~\{2,3,4\},\{3,4\}\Big],\nonumber\\
&&~~~~~~~~~~~~~~~~~~~~~~~~~~~~~~~~~\Big[\{5,6,7\},\{5,6\}~\mbox{or}~\{5,6,8\},\{5,6\}~\mbox{or}~\{5,7,8\},\{7,8\}~\mbox{or}~\{6,7,8\},\{7,8\}\Big]\}~.~~~\nonumber\eea
The Feynman diagrams for them are presented in Figure \ref{FigA82}.
Then the answer is given by
{\small \bea &&{1\over s_{12}s_{34}s_{56}s_{78}}
\ruleII{\{1,2\},\{3,4\},\{5,6\},\{7,8\}}\nonumber\\
&&+{1\over s_{56}s_{78}}\left({1\over
s_{12}s_{123}}\ruleII{\{1,2,3\},\{4\},\{5,6\},\{7,8\}}+{1\over
s_{12}s_{124}}\ruleII{\{1,2,4\},\{3\},\{5,6\},\{7,8\}}\right.\nonumber\\
&&~~~~~~~~~~~~~~\left.+{1\over
s_{34}s_{134}}\ruleII{\{1,3,4\},\{2\},\{5,6\},\{7,8\}}+{1\over
s_{34}s_{234}}\ruleII{\{2,3,4\},\{1\},\{5,6\},\{7,8\}}\right)\nonumber\\
&&+{1\over s_{12}s_{34}}\left({1\over
s_{56}s_{567}}\ruleII{\{1,2,3\},\{4\},\{5,6\},\{7,8\}}+{1\over
s_{12}s_{124}}\ruleII{\{1,2,4\},\{3\},\{5,6\},\{7,8\}}\right.\nonumber\\
&&~~~~~~~~~~~~~~\left.+{1\over
s_{34}s_{134}}\ruleII{\{1,3,4\},\{2\},\{5,6\},\{7,8\}}+{1\over
s_{34}s_{234}}\ruleII{\{2,3,4\},\{1\},\{5,6\},\{7,8\}}\right)\nonumber\\
&&+\left({1\over
s_{12}s_{123}}\ruleII{\{1,2,3\},\{4\},\{\bullet\},\{\bullet\}}+{1\over
s_{12}s_{124}}\ruleII{\{1,2,4\},\{3\},\{\bullet\},\{\bullet\}}\right.\nonumber\\
&&~~~~~~~~~~~~~~\left.+{1\over
s_{34}s_{134}}\ruleII{\{1,3,4\},\{2\},\{\bullet\},\{\bullet\}}+{1\over
s_{34}s_{234}}\ruleII{\{2,3,4\},\{1\},\{\bullet\},\{\bullet\}}\right)\nonumber\\
&&~~~~\bigotimes\left({1\over
s_{56}s_{567}}\ruleII{\{\bullet\},\{\bullet\},\{5,6,7\},\{8\}}+{1\over
s_{56}s_{568}}\ruleII{\{\bullet\},\{\bullet\},\{5,6,8\},\{7\}}\right.\nonumber\\
&&~~~~~~~~~~~~~~\left.+{1\over
s_{78}s_{578}}\ruleII{\{\bullet\},\{\bullet\},\{5,7,8\},\{6\}}+{1\over
s_{78}s_{678}}\ruleII{\{\bullet\},\{\bullet\},\{6,7,8\},\{5\}}\right)~,~~~\eea}
where in order to simplify the presentation we use product
$\bigotimes$ to denote
\bea
\ruleII{P_A,P_B,\bullet,\bullet}\bigotimes\ruleII{\bullet,\bullet,P_C,P_D}:=\ruleII{P_A,P_B,P_C,P_D}~.~~~
\eea
%

\subsection*{Example Figure \ref{Figsupport2}.g:}

\begin{figure}[h]
  \centering
  \includegraphics[width=7in]{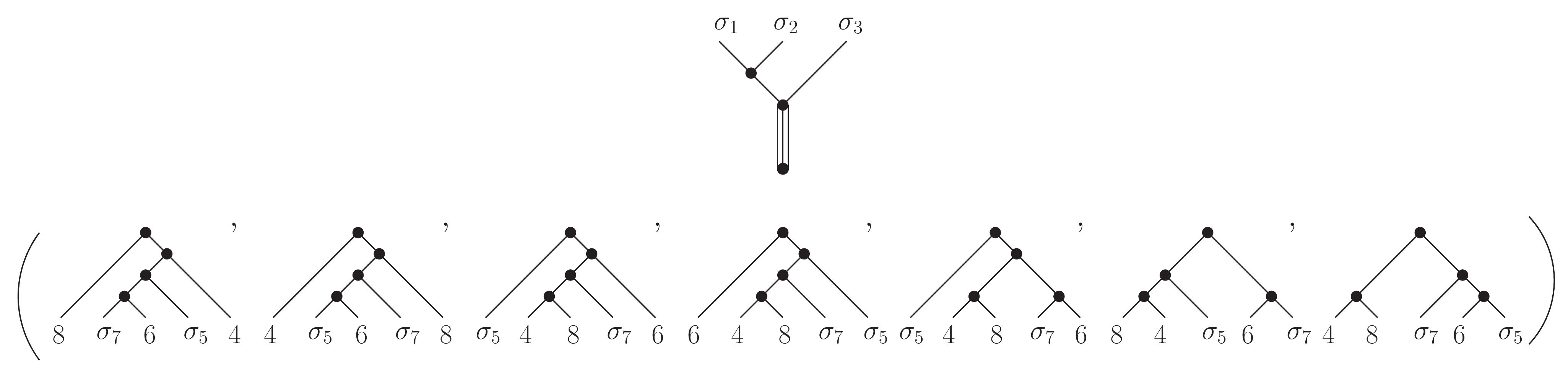}\\
  \caption{The 42 Feynman diagrams corresponding to the {\sl 4-regular} graph Figure \ref{Figsupport2}.g. $\{\sigma_1,\sigma_2,\sigma_3\}$ takes
  the cyclic permutations of $\{1,2,3\}$, while $\{\sigma_5,\sigma_7\}$ takes the permutations of $\{5,7\}$.}\label{FigA84}
\end{figure}

This is an eight-point example with CHY-integrand
\bea {1\over z_{12}^2 z_{23}^2
z_{31}^2z_{84}^2z_{56}^2z_{67}^2z_{45}z_{58}z_{87}z_{74}}~.~~~\eea
The possible subsets for poles are $\{\dunderline{1,2,3}\}$,
$\{4,5,6,7\}$, $\{5,6,7,8\}$, $\{6,7,8,4\}$, $\{7,8,4,5\}$,
$\{8,4,5,6\}$, $\{4,5,8\}$, $\{4,7,8\}$, $\{5,6,7\}$, $\{1,2\}$,
$\{2,3\}$, $\{1,3\}$, $\{4,8\}$, $\{5,6\}$, $\{6,7\}$, so we can
construct 42 compatible combinations of subsets as
{\small \bea
&&\{\{\dunderline{1,2,3}\},\{4,5,6,7\},\{5,6,7\},\Big[\{5,6\}~\mbox{or}~\{6,7\}\Big], \Big[\{1,2\}~\mbox{or}~\{2,3\}~\mbox{or}~\{1,3\}\Big]\}~,~~~\nonumber\\
&&\{\{\dunderline{1,2,3}\},\{5,6,7,8\},\{5,6,7\},\Big[\{5,6\}~\mbox{or}~\{6,7\}\Big],
\Big[\{1,2\}~\mbox{or}~\{2,3\}~\mbox{or}~\{1,3\}\Big]\}~,~~~\nonumber\\
&&\{\{\dunderline{1,2,3}\},\{6,7,8,4\},\Big[\{6,7\},
\{4,8\}~\mbox{or}~\{4,7,8\}, \{4,8\}\Big],
\Big[\{1,2\}~\mbox{or}~\{2,3\}~\mbox{or}~\{1,3\}\Big]\}~,~~~\nonumber\\
&&\{\{\dunderline{1,2,3}\},\{7,8,4,5\},\Big[\{4,5,8\},
\{4,8\}~\mbox{or}~\{4,7,8\}, \{4,8\}\Big],
\Big[\{1,2\}~\mbox{or}~\{2,3\}~\mbox{or}~\{1,3\}\Big]\}~,~~~\nonumber\\
&&\{\{\dunderline{1,2,3}\},\{8,4,5,6\},\Big[\{4,5,8\},
\{4,8\}~\mbox{or}~\{5,6\}, \{4,8\}\Big],
\Big[\{1,2\}~\mbox{or}~\{2,3\}~\mbox{or}~\{1,3\}\Big]\}~,~~~\nonumber\\
&&\{\{\dunderline{1,2,3}\},\{5,6,7\}, \{4,8\},
\Big[\{5,6\}~\mbox{or}~\{6,7\}\Big],
\Big[\{1,2\}~\mbox{or}~\{2,3\}~\mbox{or}~\{1,3\}\Big]\}~,~~~\nonumber\\
&&\{\{\dunderline{1,2,3}\},\{4,5,8\}, \{6,7\}, \{4,8\},
\Big[\{1,2\}~\mbox{or}~\{2,3\}~\mbox{or}~\{1,3\}\Big]\}~,~~~\nonumber\\
&&\{\{\dunderline{1,2,3}\},\{4,7,8\}, \{5,6\}, \{4,8\},
\Big[\{1,2\}~\mbox{or}~\{2,3\}~\mbox{or}~\{1,3\}\Big]\}~.~~~\nonumber\eea}
The corresponding 42 Feynman diagrams are presented in Figure
\ref{FigA84}.
By $\ruleii$ we immediately arrive at the result
{\small\bea &&\left({1\over
s_{12}}\ruleII{\{1,2\},\{3\},\{\bullet\},\{\bullet\}}+{1\over
s_{23}}\ruleII{\{2,3\},\{1\},\{\bullet\},\{\bullet\}}+{1\over
s_{13}}\ruleII{\{1,3\},\{2\},\{\bullet\},\{\bullet\}}\right)\nonumber\\
&&\bigotimes\left({1\over s_{4567}s_{567}}\left({1\over
s_{56}}+{1\over
s_{67}}\right)\ruleII{\{\bullet\},\{\bullet\},\{4,5,6,7\},\{8\}}+{1\over
s_{5678}s_{567}}\left({1\over s_{56}}+{1\over
s_{67}}\right)\ruleII{\{\bullet\},\{\bullet\},\{5,6,7,8\},\{4\}}\right.\nonumber\\
&&~~~~+{1\over s_{6784}s_{48}}\left({1\over s_{478}}+{1\over
s_{67}}\right)\ruleII{\{\bullet\},\{\bullet\},\{6,7,8,4\},\{5\}}+{1\over
s_{7845}s_{48}}\left({1\over s_{458}}+{1\over
s_{478}}\right)\ruleII{\{\bullet\},\{\bullet\},\{7,8,4,5\},\{6\}}\nonumber\\
&&~~~~+{1\over s_{8456}s_{48}}\left({1\over s_{458}}+{1\over
s_{56}}\right)\ruleII{\{\bullet\},\{\bullet\},\{8,4,5,6\},\{7\}}+{1\over
s_{567}s_{48}}\left({1\over s_{56}}+{1\over
s_{67}}\right)\ruleII{\{\bullet\},\{\bullet\},\{5,6,7\},\{4,8\}}\nonumber\\
&&~~~~~~~~~~~~~~\left.+{1\over
s_{458}s_{67}s_{48}}\ruleII{\{\bullet\},\{\bullet\},\{4,5,8\},\{6,7\}}+{1\over
s_{478}s_{56}s_{48}}\ruleII{\{\bullet\},\{\bullet\},\{4,7,8\},\{5,6\}}\right)~.~~~\eea}
%

\subsection*{Example Figure \ref{Figsupport2}.h:}

\begin{figure}[h]
  \centering
  \includegraphics[width=7in]{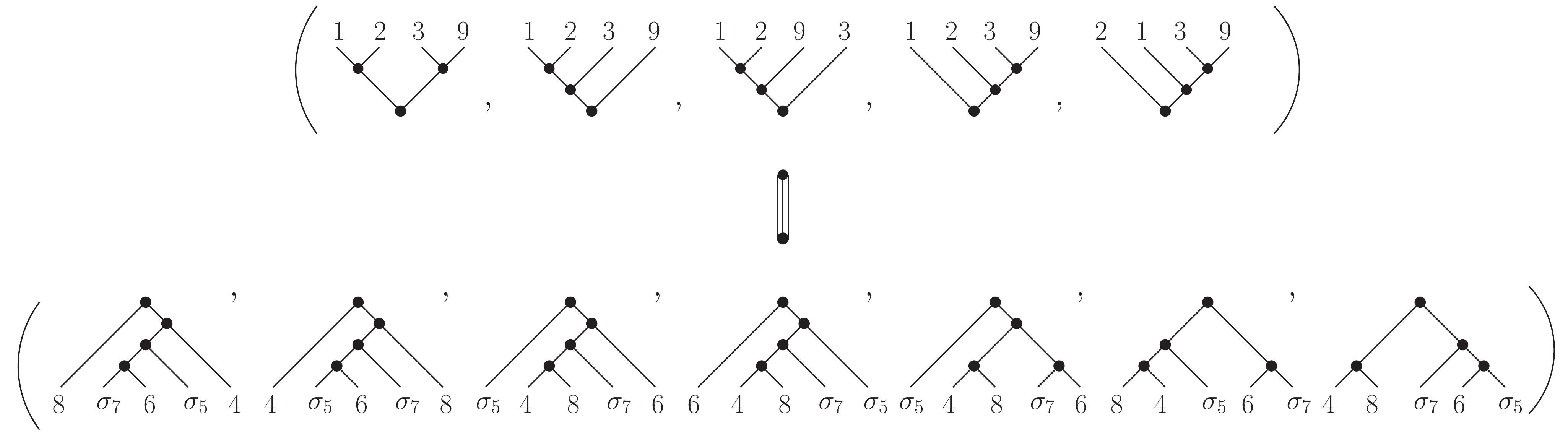}\\
  \caption{The 70 Feynman diagrams corresponding to the {\sl 4-regular} graph Figure \ref{Figsupport2}.h. $\{\sigma_5,\sigma_7\}$ takes the permutations of $\{5,7\}$.}\label{FigA91}
\end{figure}

This is a nine-point example with CHY-integrand
\bea -{1\over z_{12}^2 z_{39}^2
z_{84}^2z_{56}^2z_{67}^2z_{28}z_{91}z_{13}z_{32}z_{47}z_{78}z_{85}z_{54}}~.~~~\eea
The possible subsets for poles are
$\{\dunderline{4,5,6,7,8}\}=\{\dunderline{1,2,3,9}\}$,
$\{5,6,7,8\}$, $\{6,7,8,4\}$, $\{7,8,4,5\}$, $\{8,4,5,6\}$,
$\{4,5,6,7\}$, $\{5,6,7\}$, $\{7,8,4\}$, $\{8,4,5\}$, $\{3,9,1\}$,
$\{3,9,2\}$, $\{3,1,2\}$, $\{9,1,2\}$, $\{5,6\}$, $\{6,7\}$,
$\{8,4\}$, $\{3,9\}$, $\{1,2\}$, so we can construct 70 compatible
combinations of subsets as
{\footnotesize \bea &&\{\{\dunderline{1,2,3,9}\}, \{5,6,7\},
\{5,6\}, \{8,4\},
\Big[\{3,9\},\{1,2\}~\mbox{or}~\{3,9\},\{3,9,1\}~\mbox{or}~\{3,9\},\{3,9,2\}~\mbox{or}~\{1,2\},\{3,1,2\}~\mbox{or}~\{1,2\},
\{9,1,2\}\Big]\}~,~~~\nonumber\\
&&\{\{\dunderline{1,2,3,9}\}, \{5,6,7\}, \{6,7\}, \{8,4\},
\Big[\{3,9\},\{1,2\}~\mbox{or}~\{3,9\},\{3,9,1\}~\mbox{or}~\{3,9\},\{3,9,2\}~\mbox{or}~\{1,2\},\{3,1,2\}~\mbox{or}~\{1,2\},
\{9,1,2\}\Big]\}~,~~~\nonumber\\
&&\{\{\dunderline{1,2,3,9}\}, \{7,8,4\}, \{8,4\}, \{5,6\},
\Big[\{3,9\},\{1,2\}~\mbox{or}~\{3,9\},\{3,9,1\}~\mbox{or}~\{3,9\},\{3,9,2\}~\mbox{or}~\{1,2\},\{3,1,2\}~\mbox{or}~\{1,2\},
\{9,1,2\}\Big]\}~,~~~\nonumber\\
&&\{\{\dunderline{1,2,3,9}\}, \{8,4,5\}, \{8,4\}, \{6,7\},
\Big[\{3,9\},\{1,2\}~\mbox{or}~\{3,9\},\{3,9,1\}~\mbox{or}~\{3,9\},\{3,9,2\}~\mbox{or}~\{1,2\},\{3,1,2\}~\mbox{or}~\{1,2\},
\{9,1,2\}\Big]\}~,~~~\nonumber\\
&&\{\{\dunderline{1,2,3,9}\}, \{5,6,7,8\}, \{5,6,7\}, \{5,6\},
\Big[\{3,9\},\{1,2\}~\mbox{or}~\{3,9\},\{3,9,1\}~\mbox{or}~\{3,9\},\{3,9,2\}~\mbox{or}~\{1,2\},\{3,1,2\}~\mbox{or}~\{1,2\},
\{9,1,2\}\Big]\}~,~~~\nonumber\\
&&\{\{\dunderline{1,2,3,9}\}, \{5,6,7,8\}, \{5,6,7\}, \{6,7\},
\Big[\{3,9\},\{1,2\}~\mbox{or}~\{3,9\},\{3,9,1\}~\mbox{or}~\{3,9\},\{3,9,2\}~\mbox{or}~\{1,2\},\{3,1,2\}~\mbox{or}~\{1,2\},
\{9,1,2\}\Big]\}~,~~~\nonumber\\
&&\{\{\dunderline{1,2,3,9}\}, \{6,7,8,4\}, \{6,7\}, \{8,4\},
\Big[\{3,9\},\{1,2\}~\mbox{or}~\{3,9\},\{3,9,1\}~\mbox{or}~\{3,9\},\{3,9,2\}~\mbox{or}~\{1,2\},\{3,1,2\}~\mbox{or}~\{1,2\},
\{9,1,2\}\Big]\}~,~~~\nonumber\\
&&\{\{\dunderline{1,2,3,9}\}, \{6,7,8,4\}, \{7,8,4\}, \{8,4\},
\Big[\{3,9\},\{1,2\}~\mbox{or}~\{3,9\},\{3,9,1\}~\mbox{or}~\{3,9\},\{3,9,2\}~\mbox{or}~\{1,2\},\{3,1,2\}~\mbox{or}~\{1,2\},
\{9,1,2\}\Big]\}~,~~~\nonumber\\
&&\{\{\dunderline{1,2,3,9}\}, \{7,8,4,5\}, \{7,8,4\}, \{8,4\},
\Big[\{3,9\},\{1,2\}~\mbox{or}~\{3,9\},\{3,9,1\}~\mbox{or}~\{3,9\},\{3,9,2\}~\mbox{or}~\{1,2\},\{3,1,2\}~\mbox{or}~\{1,2\},
\{9,1,2\}\Big]\}~,~~~\nonumber\\
&&\{\{\dunderline{1,2,3,9}\}, \{7,8,4,5\}, \{8,4,5\}, \{8,4\},
\Big[\{3,9\},\{1,2\}~\mbox{or}~\{3,9\},\{3,9,1\}~\mbox{or}~\{3,9\},\{3,9,2\}~\mbox{or}~\{1,2\},\{3,1,2\}~\mbox{or}~\{1,2\},
\{9,1,2\}\Big]\}~,~~~\nonumber \\
&&\{\{\dunderline{1,2,3,9}\}, \{8,4,5,6\}, \{8,4,5\}, \{8,4\},
\Big[\{3,9\},\{1,2\}~\mbox{or}~\{3,9\},\{3,9,1\}~\mbox{or}~\{3,9\},\{3,9,2\}~\mbox{or}~\{1,2\},\{3,1,2\}~\mbox{or}~\{1,2\},
\{9,1,2\}\Big]\}~,~~~\nonumber\\
&&\{\{\dunderline{1,2,3,9}\}, \{8,4,5,6\}, \{8,4\}, \{5,6\},
\Big[\{3,9\},\{1,2\}~\mbox{or}~\{3,9\},\{3,9,1\}~\mbox{or}~\{3,9\},\{3,9,2\}~\mbox{or}~\{1,2\},\{3,1,2\}~\mbox{or}~\{1,2\},
\{9,1,2\}\Big]\}~,~~~\nonumber\\
&&\{\{\dunderline{1,2,3,9}\}, \{4,5,6,7\}, \{5,6,7\}, \{5,6\},
\Big[\{3,9\},\{1,2\}~\mbox{or}~\{3,9\},\{3,9,1\}~\mbox{or}~\{3,9\},\{3,9,2\}~\mbox{or}~\{1,2\},\{3,1,2\}~\mbox{or}~\{1,2\},
\{9,1,2\}\Big]\}~,~~~\nonumber\\
&&\{\{\dunderline{1,2,3,9}\}, \{4,5,6,7\}, \{5,6,7\}, \{6,7\},
\Big[\{3,9\},\{1,2\}~\mbox{or}~\{3,9\},\{3,9,1\}~\mbox{or}~\{3,9\},\{3,9,2\}~\mbox{or}~\{1,2\},\{3,1,2\}~\mbox{or}~\{1,2\},
\{9,1,2\}\Big]\}~.~~~\nonumber\eea
} The corresponding 70 Feynman diagrams for them are shown in Figure
\ref{FigA91}.
According to the rule $\ruleii$, the result is given by
{\small \bea &&\left({1\over s_{567}s_{84}}\left({1\over
s_{56}}+{1\over
s_{67}}\right)\ruleII{\{5,6,7\},\{8,4\},\{\bullet\},\{\bullet\}}+{1\over
s_{784}s_{56}s_{84}}\ruleII{\{5,6\},\{7,8,4\},\{\bullet\},\{\bullet\}}\right.\nonumber\\
&&~+{1\over
s_{845}s_{84}s_{67}}\ruleII{\{8,4,5\},\{6,7\},\{\bullet\},\{\bullet\}}+{1\over
s_{5678}s_{567}}\left({1\over s_{56}}+{1\over
s_{67}}\right)\ruleII{\{5,6,7,8\},\{4\},\{\bullet\},\{\bullet\}}\nonumber\\
&&~+{1\over s_{6784}s_{84}}\left({1\over s_{67}}+{1\over
s_{784}}\right)\ruleII{\{6,7,8,4\},\{5\},\{\bullet\},\{\bullet\}}+{1\over
s_{7845}s_{84}}\left({1\over s_{784}}+{1\over
s_{845}}\right)\ruleII{\{7,8,4,5\},\{6\},\{\bullet\},\{\bullet\}}\nonumber\\
&&~\left.+{1\over s_{8456}s_{84}}\left({1\over s_{845}}+{1\over
s_{56}}\right)\ruleII{\{8,4,5,6\},\{7\},\{\bullet\},\{\bullet\}}+{1\over
s_{4567}s_{567}}\left({1\over s_{56}}+{1\over
s_{67}}\right)\ruleII{\{4,5,6,7\},\{8\},\{\bullet\},\{\bullet\}}\right)\nonumber\\
&&\bigotimes\left({1\over
s_{39}s_{12}}\ruleII{\{\bullet\},\{\bullet\},\{3,9\},\{1,2\}}+{1\over
s_{39}s_{391}}\ruleII{\{\bullet\},\{\bullet\},\{3,9,1\},\{2\}}+{1\over
s_{39}s_{392}}\ruleII{\{\bullet\},\{\bullet\},\{3,9,2\},\{1\}}\right.\nonumber\\
&&~~~~~~~~~~~~~~\left.+{1\over
s_{12}s_{312}}\ruleII{\{\bullet\},\{\bullet\},\{3,1,2\},\{9\}}+{1\over
s_{12}s_{912}}\ruleII{\{\bullet\},\{\bullet\},\{9,1,2\},\{3\}}\right)~.~~~\eea}
%

\subsection{Examples with only duplex-double poles and simple poles}
\label{secSupportingduplex}

In this subsection, we examine Feynman rule $\ruleiii$ of
duplex-double pole by four examples in Figure \ref{Figsupport3}.

\subsection*{Example Figure \ref{Figsupport3}.a:}

\begin{figure}[h]
  \centering
  \includegraphics[width=4in]{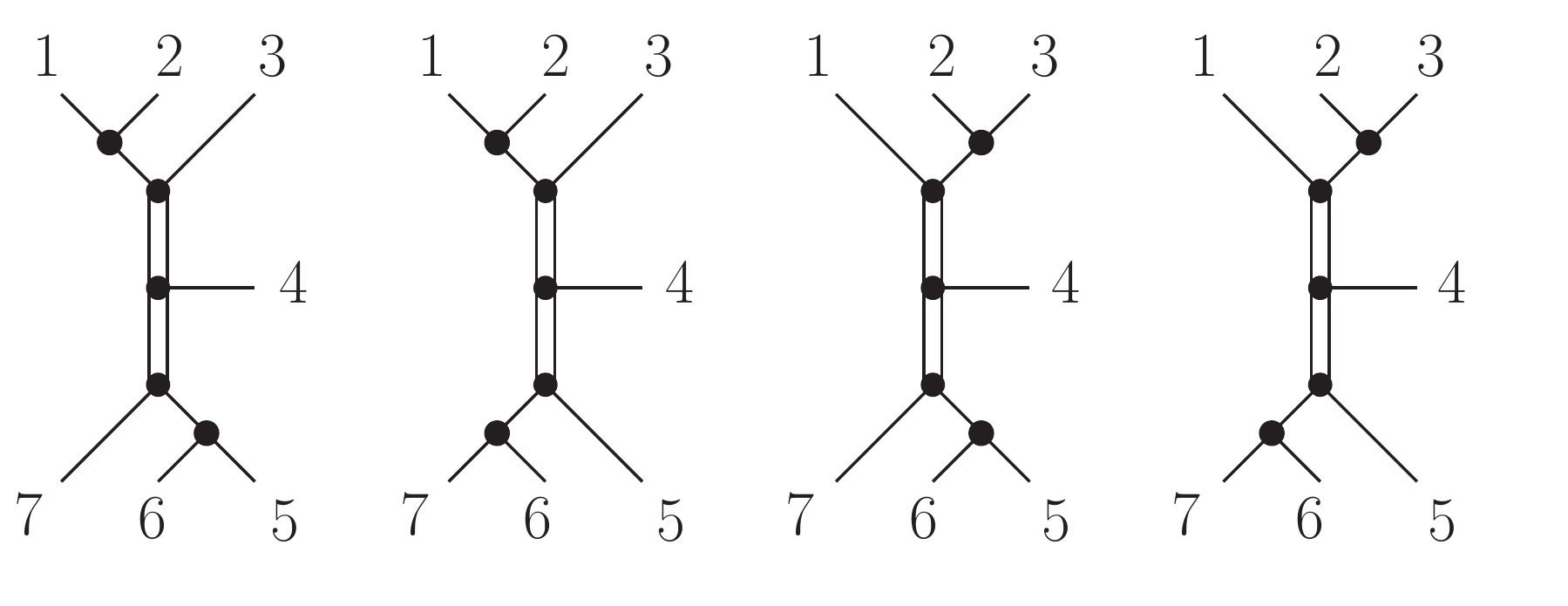}\\
  \caption{The four Feynman diagrams corresponding to the {\sl 4-regular} graph Figure \ref{Figsupport3}.a.}\label{FigA74}
\end{figure}

This is a seven-point example with CHY-integrand
\bea -{1\over z_{12}^2 z_{23}^2z_{56}^2
z_{67}^2z_{34}z_{45}z_{57}z_{74}z_{41}z_{13}}~.~~~\eea
The possible subsets for poles are $\{\underline{1,2,3}\}$,
$\{\underline{5,6,7}\}$, $\{1,2\}$, $\{2,3\}$, $\{5,6\}$, $\{6,7\}$,
so we can construct four compatible combinations of subsets as
\bea &&\{\{\underline{1,2,3}\}, \{\underline{5,6,7}\}, \{1,2\},
\{5,6\}\}~~~,~~~\{\{\underline{1,2,3}\}, \{\underline{5,6,7}\},
\{1,2\},
\{6,7\}\}~,~~~\nonumber\\
&&\{\{\underline{1,2,3}\}, \{\underline{5,6,7}\}, \{2,3\},
\{5,6\}\}~~~,~~~\{\{\underline{1,2,3}\}, \{\underline{5,6,7}\},
\{2,3\}, \{6,7\}\}~.~~~\nonumber\eea
From them we can draw four Feynman diagrams as shown in Figure
\ref{FigA74}.
Immediately we obtain the result by using $\ruleiii$,
\bea &&{1\over
s_{12}s_{56}}\ruleIII{\{1,2\},\{3\},\{4\},\{5,6\},\{7\}}+{1\over
s_{12}s_{67}}
\ruleIII{\{1,2\},\{3\},\{4\},\{5\},\{6,7\}}\nonumber\\
&&+{1\over
s_{23}s_{56}}\ruleIII{\{1\},\{2,3\},\{4\},\{5,6\},\{7\}}+{1\over
s_{23}s_{67}}\ruleIII{\{1\},\{2,3\},\{4\},\{5\},\{6,7\}}~.~~~\eea
%

\subsection*{Example Figure \ref{Figsupport3}.b:}

\begin{figure}[h]
  \centering
  \includegraphics[width=4in]{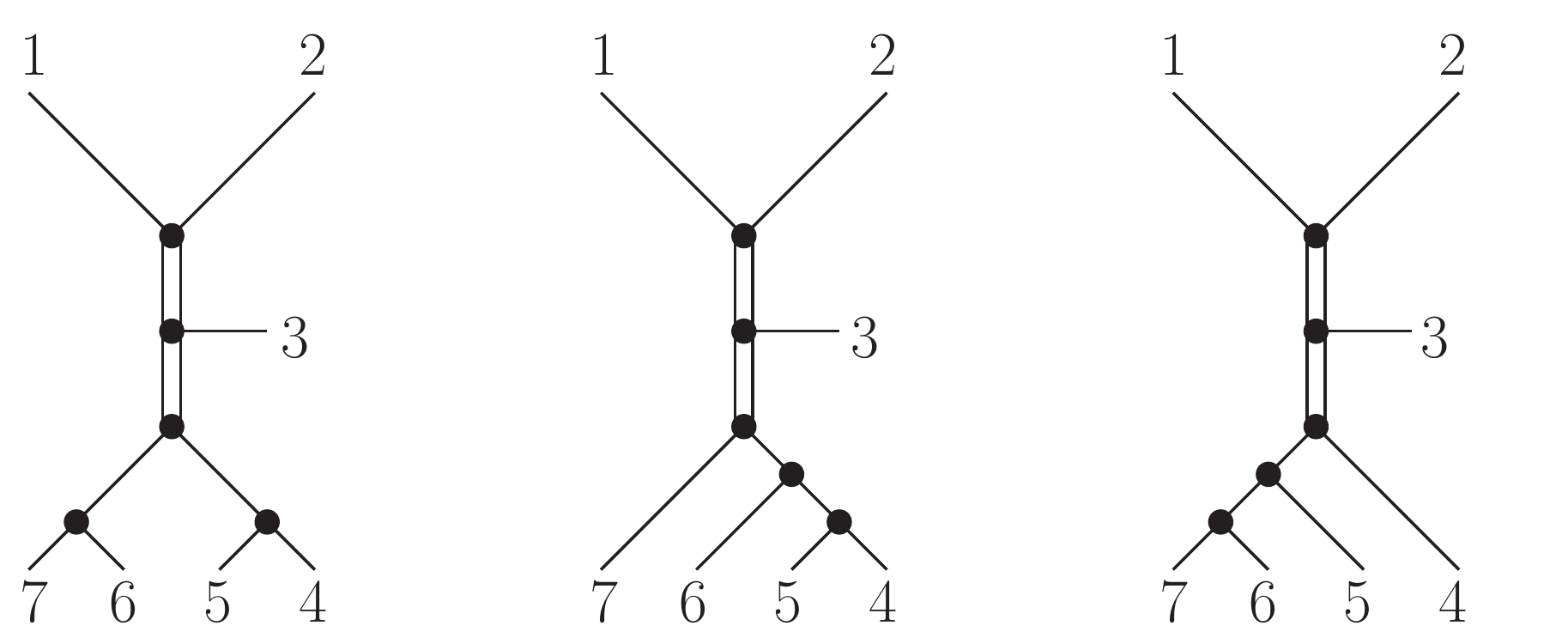}\\
  \caption{The three Feynman diagrams corresponding to the {\sl 4-regular} graph Figure \ref{Figsupport3}.b.}\label{FigA75}
\end{figure}

This is a seven-point example with CHY-integrand
\bea {1\over z_{12}^3 z_{45}^2z_{67}^2
z_{23}z_{34}z_{46}z_{65}z_{57}z_{73}z_{31}}~.~~~\eea
The possible subsets for poles are $\{\underline{1,2}\}$,
$\{\underline{1,2,3}\}$, $\{4,5,6\}$, $\{5,6,7\}$, $\{4,5\}$,
$\{6,7\}$, so we can construct three compatible combinations of
subsets as
\bea &&\{\{\underline{1,2}\}, \{\underline{1,2,3}\}, \{4,5\},
\{6,7\}\}~~,~~\{\{\underline{1,2}\}, \{\underline{1,2,3}\}, \{4,5\},
\{4,5,6\}\}~~,~~\{\{\underline{1,2}\}, \{\underline{1,2,3}\},
\{6,7\}, \{5,6,7\}\}~.~~~\nonumber\eea The corresponding three
Feynman diagrams are shown in Figure \ref{FigA75}.
So we can write the answer as
\bea &&{1\over
s_{45}s_{67}}\ruleIII{\{1\},\{2\},\{3\},\{4,5\},\{6,7\}}+{1\over
s_{45}s_{456}}
\ruleIII{\{1\},\{2\},\{3\},\{4,5,6\},\{7\}}\nonumber\\
&&+{1\over
s_{67}s_{567}}\ruleIII{\{1\},\{2\},\{3\},\{4\},\{5,6,7\}}~.~~~\eea
%

\subsection*{Example Figure \ref{Figsupport3}.c:}

\begin{figure}[h]
  \centering
  \includegraphics[width=5in]{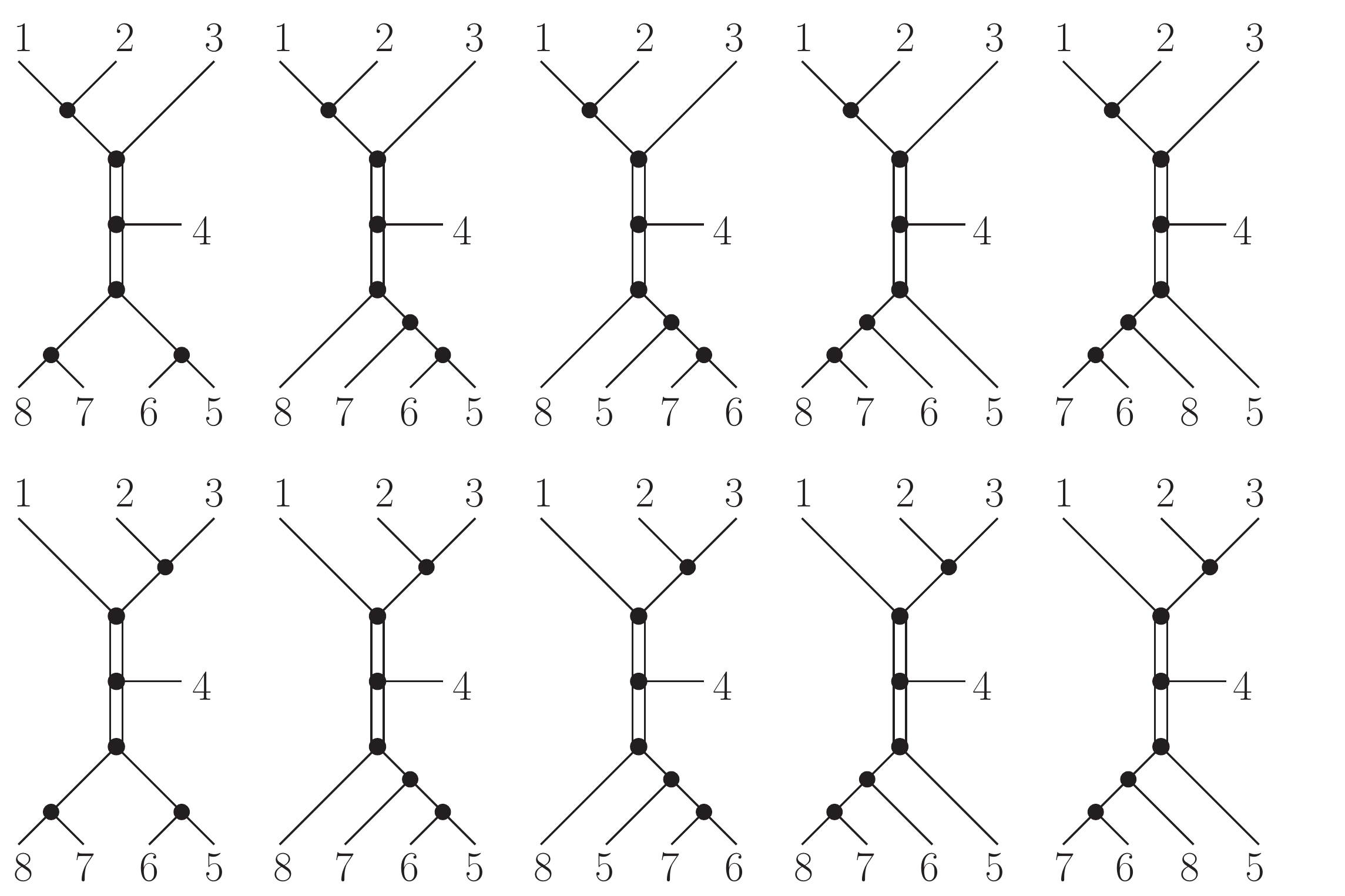}\\
  \caption{The 10 Feynman diagrams corresponding to the {\sl 4-regular} graph Figure \ref{Figsupport3}.c.}\label{FigA85}
\end{figure}

This is an eight-point example with CHY-integrand
\bea {1\over z_{12}^2 z_{23}^2
z_{56}^2z_{67}^2z_{78}^2z_{13}z_{14}z_{34}z_{45}z_{48}z_{58}}~.~~~\eea
The possible subsets for poles are $\{\underline{1,2,3}\}$,
$\{\underline{5,6,7,8}\}$, $\{5,6,7\}$, $\{6,7,8\}$, $\{1,2\}$,
$\{2,3\}$, $\{5,6\}$, $\{6,7\}$, $\{7,8\}$, so we can construct 10
compatible combinations of subsets as
\bea
&&\{\{\underline{1,2,3}\},\{\underline{5,6,7,8}\},\Big[\{1,2\}~\mbox{or}~\{2,3\}\Big], \{5,6,7\}, \Big[\{5,6\}~\mbox{or}~\{6,7\}\Big]\}~,~~~\nonumber\\
&&\{\{\underline{1,2,3}\},\{\underline{5,6,7,8}\},\Big[\{1,2\}~\mbox{or}~\{2,3\}\Big],
\{6,7,8\},
\Big[\{6,7\}~\mbox{or}~\{7,8\}\Big]\}~,~~~\nonumber\\
&&\{\{\underline{1,2,3}\},\{\underline{5,6,7,8}\},\Big[\{1,2\}~\mbox{or}~\{2,3\}\Big],
\{5,6\}, \{7,8\}\}~.~~~\nonumber\eea
The 10 Feynman diagrams can be accordingly drawn as in Figure
\ref{FigA85}.
So according to the rule $\ruleiii$, we can write down the answer as
{\small \bea &&{1\over s_{12}s_{567}}\left({1\over s_{56}}+{1\over
s_{67}}\right)\ruleIII{\{1,2\},\{3\},\{4\},\{5,6,7\},\{8\}}+{1\over
s_{12}s_{678}}\left({1\over s_{67}}+{1\over
s_{78}}\right)\ruleIII{\{1,2\},\{3\},\{4\},\{5\},\{6,7,8\}}\nonumber\\
&&+{1\over
s_{12}s_{56}s_{78}}\ruleIII{\{1,2\},\{3\},\{4\},\{5,6\},\{7,8\}}+{1\over
s_{23}s_{567}}\left({1\over s_{56}}+{1\over
s_{67}}\right)\ruleIII{\{1\},\{2,3\},\{4\},\{5,6,7\},\{8\}}\nonumber\\
&&+{1\over s_{23}s_{678}}\left({1\over s_{67}}+{1\over
s_{78}}\right)\ruleIII{\{1\},\{2,3\},\{4\},\{5\},\{6,7,8\}}+{1\over
s_{23}s_{56}s_{78}}\ruleIII{\{1\},\{2,3\},\{4\},\{5,6\},\{7,8\}}~.~~~\eea}
%

\subsection*{Example Figure \ref{Figsupport3}.d:}

\begin{figure}[h]
  \centering
  \includegraphics[width=6in]{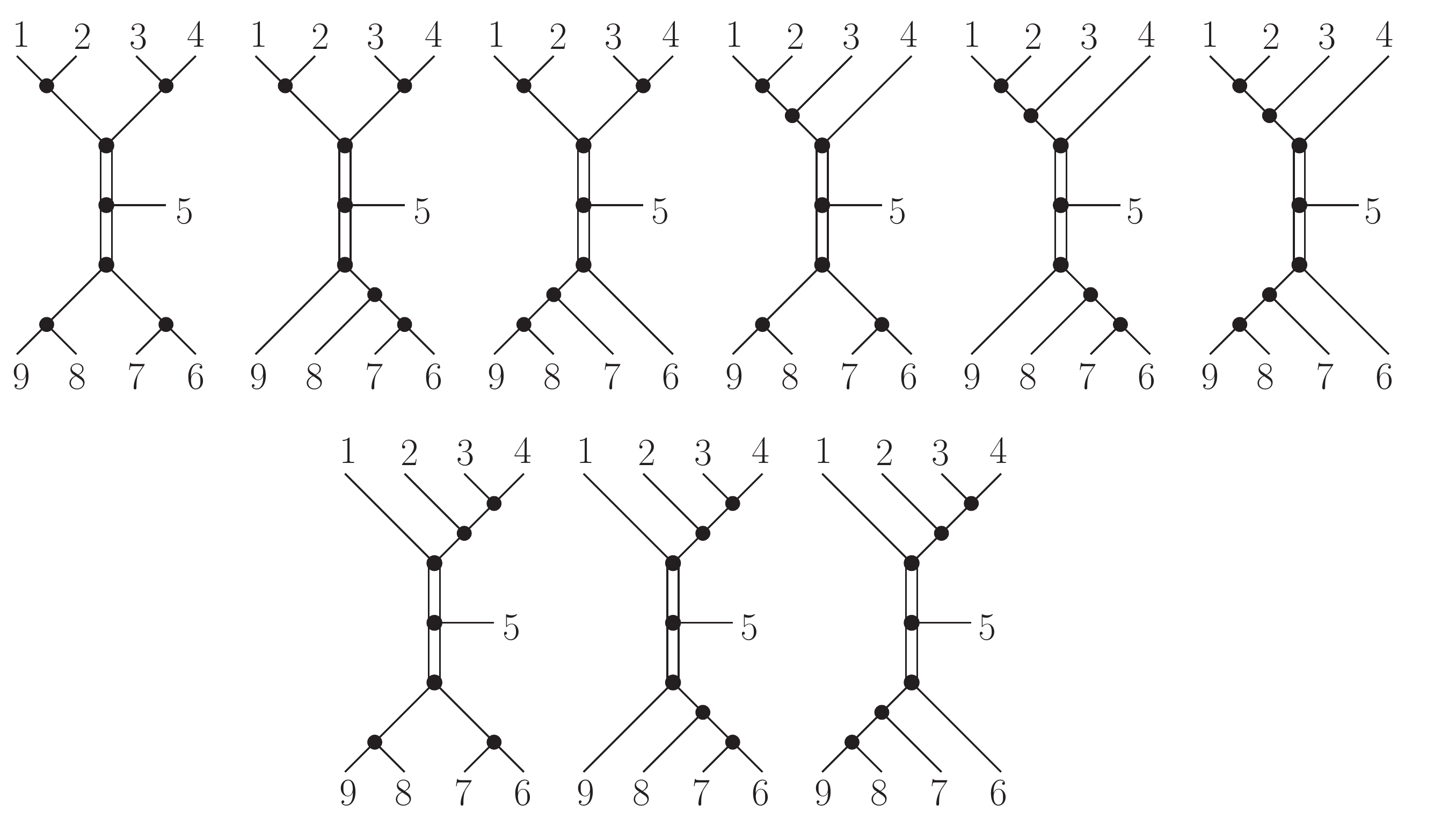}\\
  \caption{The nine Feynman diagrams corresponding to the {\sl 4-regular} graph Figure \ref{Figsupport3}.d.}\label{FigA92}
\end{figure}

This is a nine-point example with CHY-integrand
\bea {1\over z_{12}^2 z_{34}^2
z_{67}^2z_{89}^2z_{23}z_{31}z_{15}z_{56}z_{68}z_{87}z_{79}z_{95}z_{54}z_{42}}~.~~~\eea
The possible groups for poles are $\{\underline{1,2,3,4}\}$,
$\{\underline{6,7,8,9}\}$, $\{1,2,3\}$, $\{2,3,4\}$, $\{6,7,8\}$,
$\{7,8,9\}$, $\{1,2\}$, $\{3,4\}$, $\{6,7\}$, $\{8,9\}$, so we can
construct nine compatible combinations of subsets as
\bea &&\{\{\underline{1,2,3,4}\}, \{\underline{6,7,8,9}\}, \{1,2\},
\{3,4\},
\Big[\{6,7\},\{8,9\}~\mbox{or}~\{6,7\},\{6,7,8\}~\mbox{or}~\{8,9\},\{7,8,9\}\Big]\}~,~~~\nonumber\\
&&\{\{\underline{1,2,3,4}\}, \{\underline{6,7,8,9}\}, \{1,2,3\},
\{1,2\},
\Big[\{6,7\},\{8,9\}~\mbox{or}~\{6,7\},\{6,7,8\}~\mbox{or}~\{8,9\},\{7,8,9\}\Big]\}~,~~~\nonumber\\
&&\{\{\underline{1,2,3,4}\}, \{\underline{6,7,8,9}\}, \{2,3,4\},
\{3,4\},
\Big[\{6,7\},\{8,9\}~\mbox{or}~\{6,7\},\{6,7,8\}~\mbox{or}~\{8,9\},\{7,8,9\}\Big]\}~.~~~\nonumber\eea
From them we can draw nine Feynman diagrams as in Figure
\ref{FigA92}.
According to the rule $\ruleiii$, the result is given by
{\small \bea &&{1\over
s_{12}s_{34}s_{67}s_{89}}\ruleIII{\{1,2\},\{3,4\},\{5\},\{6,7\},\{8,9\}}+{1\over
s_{12}s_{34}s_{67}s_{678}}\ruleIII{\{1,2\},\{3,4\},\{5\},\{6,7,8\},\{9\}}\nonumber\\
&&+{1\over
s_{12}s_{34}s_{89}s_{789}}\ruleIII{\{1,2\},\{3,4\},\{5\},\{6\},\{7,8,9\}}+{1\over
s_{12}s_{123}s_{67}s_{89}}\ruleIII{\{1,2,3\},\{4\},\{5\},\{6,7\},\{8,9\}}\nonumber\\
&&+{1\over
s_{12}s_{123}s_{67}s_{678}}\ruleIII{\{1,2,3\},\{4\},\{5\},\{6,7,8\},\{9\}}+{1\over
s_{12}s_{123}s_{89}s_{789}}\ruleIII{\{1,2,3\},\{4\},\{5\},\{6\},\{7,8,9\}}\nonumber\\
&&+{1\over
s_{34}s_{234}s_{67}s_{89}}\ruleIII{\{1\},\{2,3,4\},\{5\},\{6,7\},\{8,9\}}+{1\over
s_{34}s_{234}s_{67}s_{678}}\ruleIII{\{1\},\{2,3,4\},\{5\},\{6,7,8\},\{9\}}\nonumber\\
&&+{1\over
s_{34}s_{234}s_{89}s_{789}}\ruleIII{\{1\},\{2,3,4\},\{5\},\{6\},\{7,8,9\}}~.~~~\eea}
%

\subsection{Examples with mixed types of higher-order poles}
\label{secSupportingmixed}

In this subsection, we will examine Feynman rules for higher-order
poles by examples in Figure \ref{Figsupportmix}. They all contain
Feynman diagrams with mixed types of higher-order poles, thus really
serve as extremely non-trivial supporting for the validation of
rules.

\subsection*{Example Figure \ref{Figsupportmix}.a:}

\begin{figure}[h]
  \centering
  \includegraphics[width=4in]{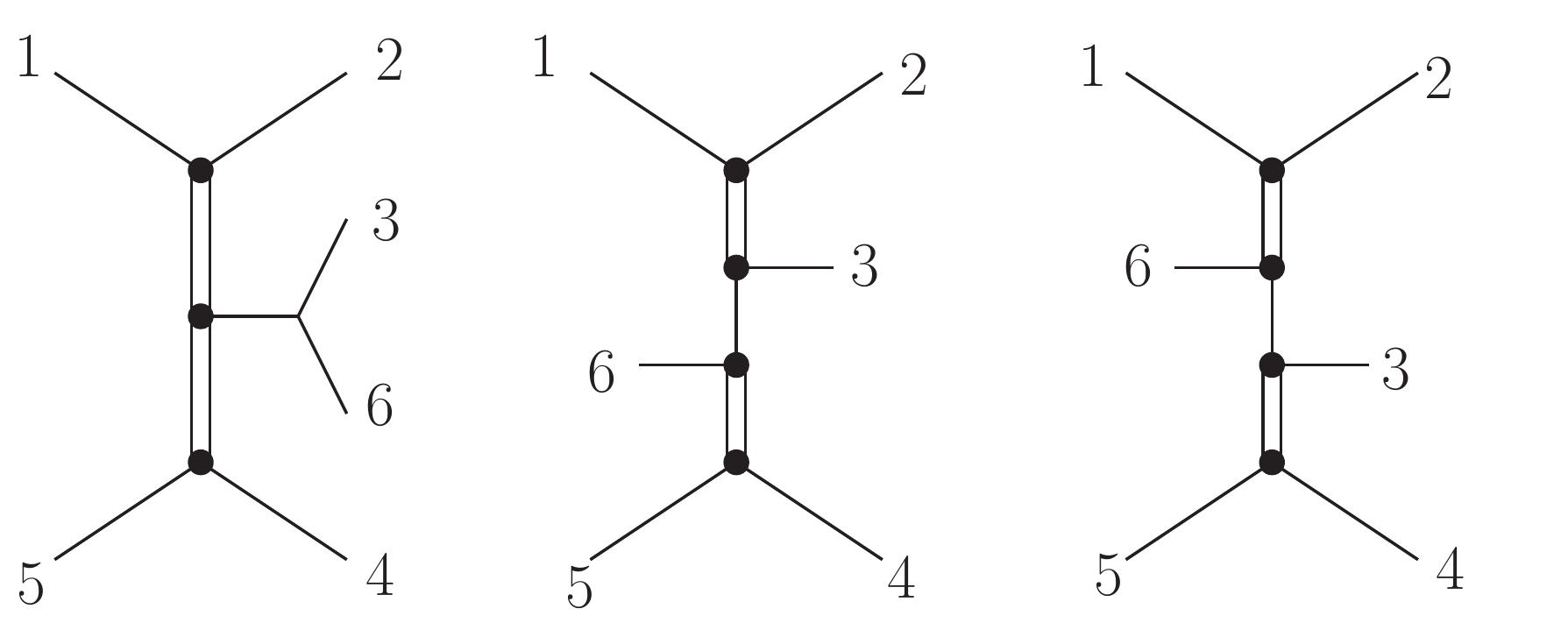}\\
  \caption{The three Feynman diagrams corresponding to the {\sl 4-regular} graph Figure \ref{Figsupportmix}.a.}\label{FigA69}
\end{figure}

This is a six-point example with CHY-integrand
\bea {1\over z_{12}^3 z_{45}^3z_{63}^2
z_{23}z_{34}z_{56}z_{61}}~.~~~\eea
It is simple to find the possible subsets for poles as
$\{\underline{1,2}\}$, $\{\underline{4,5}\}$, $\{1,2,3\}$,
$\{1,2,6\}$, $\{3,6\}$. So we can construct three compatible
combinations of subsets as
\bea &&\{\{\underline{1,2}\}, \{\underline{4,5}\},
\{3,6\}\}~~~,~~~\{\{\underline{1,2}\}, \{\underline{4,5}\},
\{1,2,3\}\}~~~,~~~\{\{\underline{1,2}\}, \{\underline{4,5}\},
\{1,2,6\}\}~.~~~\nonumber\eea
From them we draw three Feynman diagrams as presented in Figure
\ref{FigA69}.
The first Feynman diagram contains a duplex-double pole while the
remaining two contain two double poles. Remind that for the rule
$\ruleiii$ defined in (\ref{rule3}), we mentioned that the terms
with $P_E^2$ can not be inferred from result of five-point
CHY-integrand since therein $P_E^2=0$. But it is crucial for generic
CHY-integrands whose Feynman diagrams could have massive $P_E$. This
example is indeed the one that motives us to think about adding
$P_E^2$ terms in the Feynman rule.

Let us apply rules $\rulei$ and $\ruleiii$ for corresponding
propagators in Feynman diagrams, which gives
\bea &&{1\over s_{36}}
\ruleIII{\{1\},\{2\},\{3,6\},\{4\},\{5\}}+{1\over s_{123}}
\ruleI{\{1\},\{2\},\{3\},\{4,5,6\}}\ruleI{\{6\},\{1,2,3\},\{4\},\{5\}}\nonumber\\
&&+{1\over s_{126}}
\ruleI{\{1\},\{2\},\{3,4,5,\},\{6\}}\ruleI{\{6,1,2\},\{3\},\{4\},\{5\}}\nonumber\\
&=&\frac{s_{15} s_{24}-s_{14} s_{25}}{s_{12}^2
s_{45}^2s_{36}}+\frac{s_{13} s_{46}}{s_{12}^2 s_{45}^2
s_{123}}+\frac{s_{26} s_{35}}{s_{12}^2 s_{45}^2
s_{126}}-\frac{s_{135}}{s_{12}^2 s_{45}^2}~.~~~\eea
%

\subsection*{Example Figure \ref{Figsupportmix}.b:}

\begin{figure}[h]
  \centering
  \includegraphics[width=5in]{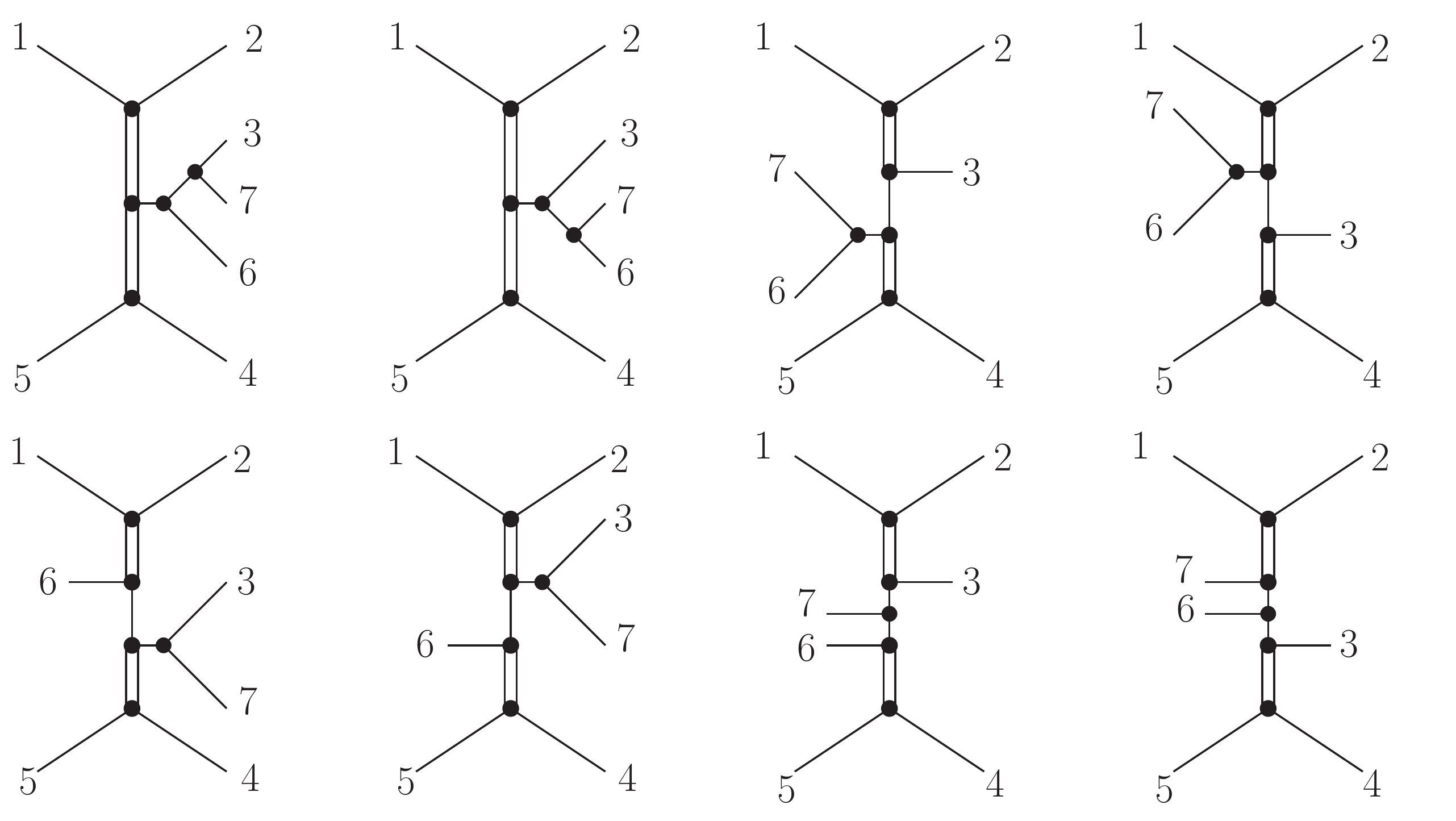}\\
  \caption{The eight Feynman diagrams corresponding to the {\sl 4-regular} graph Figure \ref{Figsupportmix}.b.}\label{FigA77}
\end{figure}

This is a seven-point example with CHY-integrand
\bea -{1\over z_{12}^3 z_{45}^3z_{67}^2
z_{73}^2z_{23}z_{34}z_{56}z_{61}}~.~~~\eea
The possible subsets for poles are $\{\underline{1,2}\}$,
$\{\underline{4,5}\}$, $\{1,2,3\}$, $\{1,2,6\}$, $\{3,4,5\}$,
$\{4,5,6\}$, $\{3,6,7\}$, $\{3,7\}$, $\{6,7\}$. From them we can
construct eight compatible combinations of subsets as
\bea &&\{\{\underline{1,2}\}, \{\underline{4,5}\}, \{3,6,7\},
\{3,7\}\}~~,~~\{\{\underline{1,2}\}, \{\underline{4,5}\}, \{3,6,7\},
\{6,7\}\}~~,~~\{\{\underline{1,2}\}, \{\underline{4,5}\}, \{1,2,3\},
\{6,7\}\}~,~~~\nonumber\\
&&\{\{\underline{1,2}\}, \{\underline{4,5}\}, \{3,4,5\},
\{6,7\}\}~~,~~\{\{\underline{1,2}\}, \{\underline{4,5}\}, \{1,2,6\},
\{3,7\}\}~~,~~\{\{\underline{1,2}\}, \{\underline{4,5}\}, \{4,5,6\},
\{3,7\}\}~,~~~\nonumber\\
&&\{\{\underline{1,2}\}, \{\underline{4,5}\}, \{1,2,3\},
\{4,5,6\}\}~~,~~\{\{\underline{1,2}\}, \{\underline{4,5}\},
\{1,2,6\}, \{3,4,5\}\}~.~~~\nonumber\eea
The corresponding eight Feynman diagrams are shown in Figure
\ref{FigA77}.
So applying the rules we get the result{
\bea &&{1\over s_{367}}\left({1\over s_{37}}+{1\over
s_{67}}\right)\ruleIII{\{1\},\{2\},\{3,6,7\},\{4\},\{5\}}\nonumber\\
&&+{1\over
s_{123}s_{67}}\ruleI{\{1\},\{2\},\{3\},\{4,5,6,7\}}\ruleI{\{6,7\},\{1,2,3\},\{4\},\{5\}}\nonumber\\
&&+{1\over
s_{345}s_{67}}\ruleI{\{1\},\{2\},\{3,4,5\},\{6,7\}}\ruleI{\{6,7,1,2\},\{3\},\{4\},\{5\}}\nonumber\\
&&+{1\over
s_{126}s_{37}}\ruleI{\{1\},\{2\},\{3,4,5,7\},\{6\}}\ruleI{\{6,1,2\},\{3,7\},\{4\},\{5\}}\nonumber\\
&&+{1\over
s_{456}s_{37}}\ruleI{\{1\},\{2\},\{3,7\},\{4,5,6\}}\ruleI{\{6\},\{1,2,3,7\},\{4\},\{5\}}\nonumber\\
&&+{1\over
s_{123}s_{456}}\ruleI{\{1\},\{2\},\{3\},\{4,5,6,7\}}\ruleI{\{6\},\{7,1,2,3\},\{4\},\{5\}}\nonumber\\
&&+{1\over
s_{126}s_{345}}\ruleI{\{1\},\{2\},\{7,3,4,5\},\{6\}}\ruleI{\{6,7,1,2\},\{3\},\{4\},\{5\}}~.~~~\eea}
%

\subsection*{Example Figure \ref{Figsupportmix}.c:}

\begin{figure}[h]
  \centering
  \includegraphics[width=6.5in]{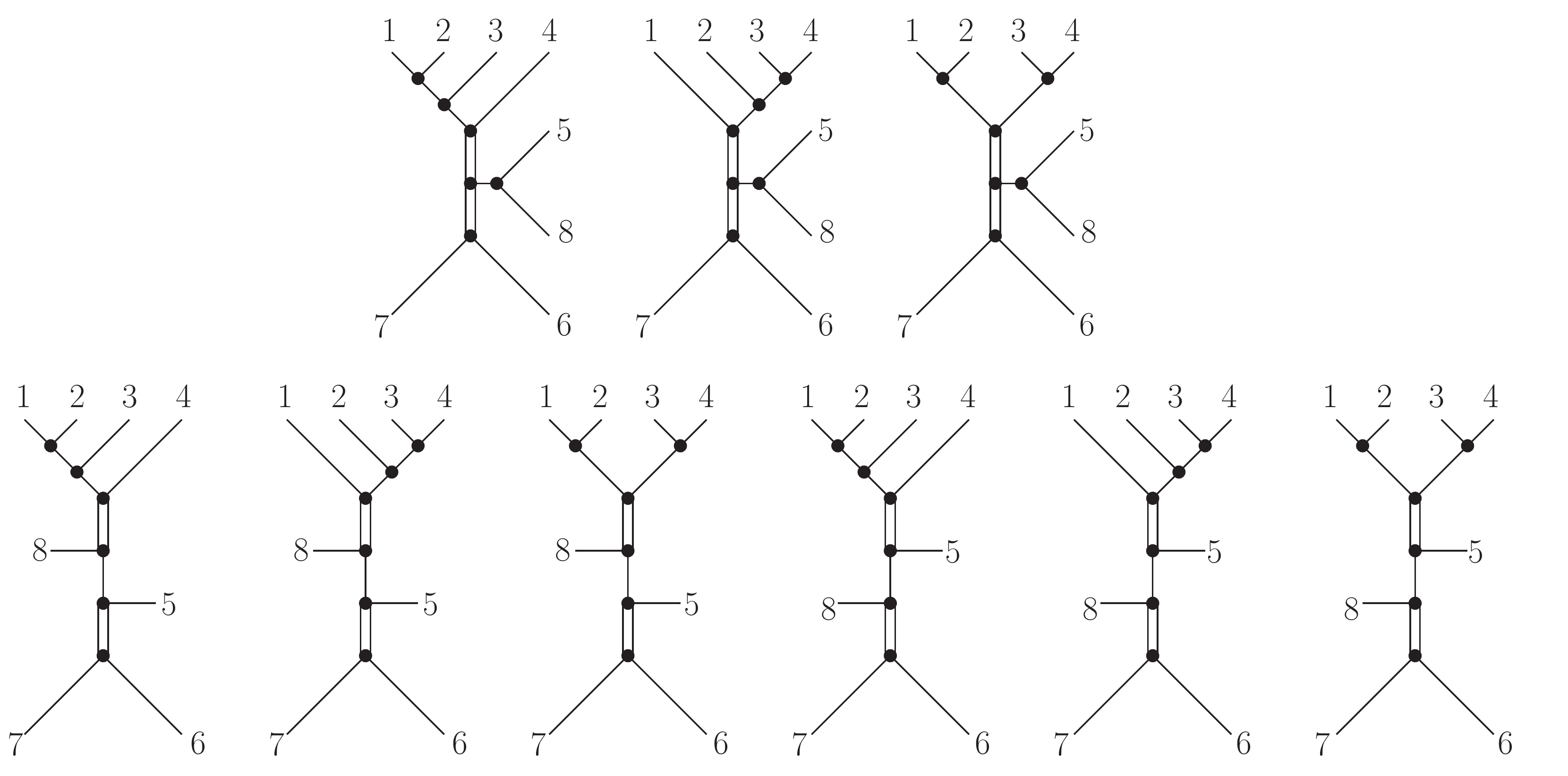}\\
  \caption{The nine Feynman diagrams corresponding to the {\sl 4-regular} graph Figure \ref{Figsupportmix}.c.}\label{FigA86}
\end{figure}

This is an eight-point example with CHY-integrand
\bea {1\over z_{12}^2 z_{34}^2
z_{85}^2z_{67}^3z_{23}z_{31}z_{18}z_{87}z_{65}z_{54}z_{42}}~.~~~\eea
The possible subsets for poles are $\{\underline{1,2,3,4}\}$,
$\{\underline{6,7}\}$, $\{1,2,3\}$, $\{2,3,4\}$, $\{5,6,7\}$, $
\{6,7,8\}$, $\{1,2\}$, $\{3,4\}$, $\{5,8\}$. So we can construct
nine compatible combinations of subsets as
\bea &&\{\{\underline{1,2,3,4}\}, \{\underline{6,7}\}, \{1,2,3\},
\{1,2\},~~\Big[\{5,8\}~~\mbox{or}~~\{5,6,7\}~~\mbox{or}~~\{6,7,8\}\Big]\}~,~~~\nonumber\\
&&\{\{\underline{1,2,3,4}\}, \{\underline{6,7}\}, \{2,3,4\},
\{3,4\},~~\Big[\{5,8\}~~\mbox{or}~~\{5,6,7\}~~\mbox{or}~~\{6,7,8\}\Big]\}~,~~~\nonumber\\
&&\{\{\underline{1,2,3,4}\}, \{\underline{6,7}\}, \{1,2\},
\{3,4\},~~\Big[\{5,8\}~~\mbox{or}~~\{5,6,7\}~~\mbox{or}~~\{6,7,8\}\Big]\}~.~~~\nonumber\eea
From them we can draw nine Feynman diagrams as shown in Figure
\ref{FigA86}.
So according to the rules, the result is given by
{\small \bea &&{1\over
s_{123}s_{12}s_{58}}\ruleIII{\{1,2,3\},\{4\},\{5,8\},\{6\},\{7\}}+{1\over
s_{234}s_{34}s_{58}}\ruleIII{\{1\},\{2,3,4\},\{5,8\},\{6\},\{7\}}\nonumber\\
&&+{1\over
s_{12}s_{34}s_{58}}\ruleIII{\{1,2\},\{3,4\},\{5,8\},\{6\},\{7\}}\nonumber\\
&&+{1\over
s_{123}s_{12}s_{567}}\ruleI{\{1,2,3\},\{4\},\{5,6,7\},\{8\}}\ruleI{\{8,1,2,3,4\},\{5\},\{6\},\{7\}}\nonumber\\
&&+{1\over
s_{234}s_{34}s_{567}}\ruleI{\{1\},\{2,3,4\},\{5,6,7\},\{8\}}\ruleI{\{8,1,2,3,4\},\{5\},\{6\},\{7\}}\nonumber\\
&&+{1\over
s_{12}s_{34}s_{567}}\ruleI{\{1,2\},\{3,4\},\{5,6,7\},\{8\}}\ruleI{\{8,1,2,3,4\},\{5\},\{6\},\{7\}}\nonumber\\
&&+{1\over
s_{123}s_{12}s_{678}}\ruleI{\{1,2,3\},\{4\},\{5\},\{6,7,8\}}\ruleI{\{8\},\{1,2,3,4,5\},\{6\},\{7\}}\nonumber\\
&&+{1\over
s_{234}s_{34}s_{678}}\ruleI{\{1\},\{2,3,4\},\{5\},\{6,7,8\}}\ruleI{\{8\},\{1,2,3,4,5\},\{6\},\{7\}}\nonumber\\
&&+{1\over
s_{12}s_{34}s_{678}}\ruleI{\{1,2\},\{3,4\},\{5\},\{6,7,8\}}\ruleI{\{8\},\{1,2,3,4,5\},\{6\},\{7\}}~.~~~\eea}
%

\subsection*{Example Figure \ref{Figsupportmix}.d:}

\begin{figure}[h]
  \centering
  \includegraphics[width=6in]{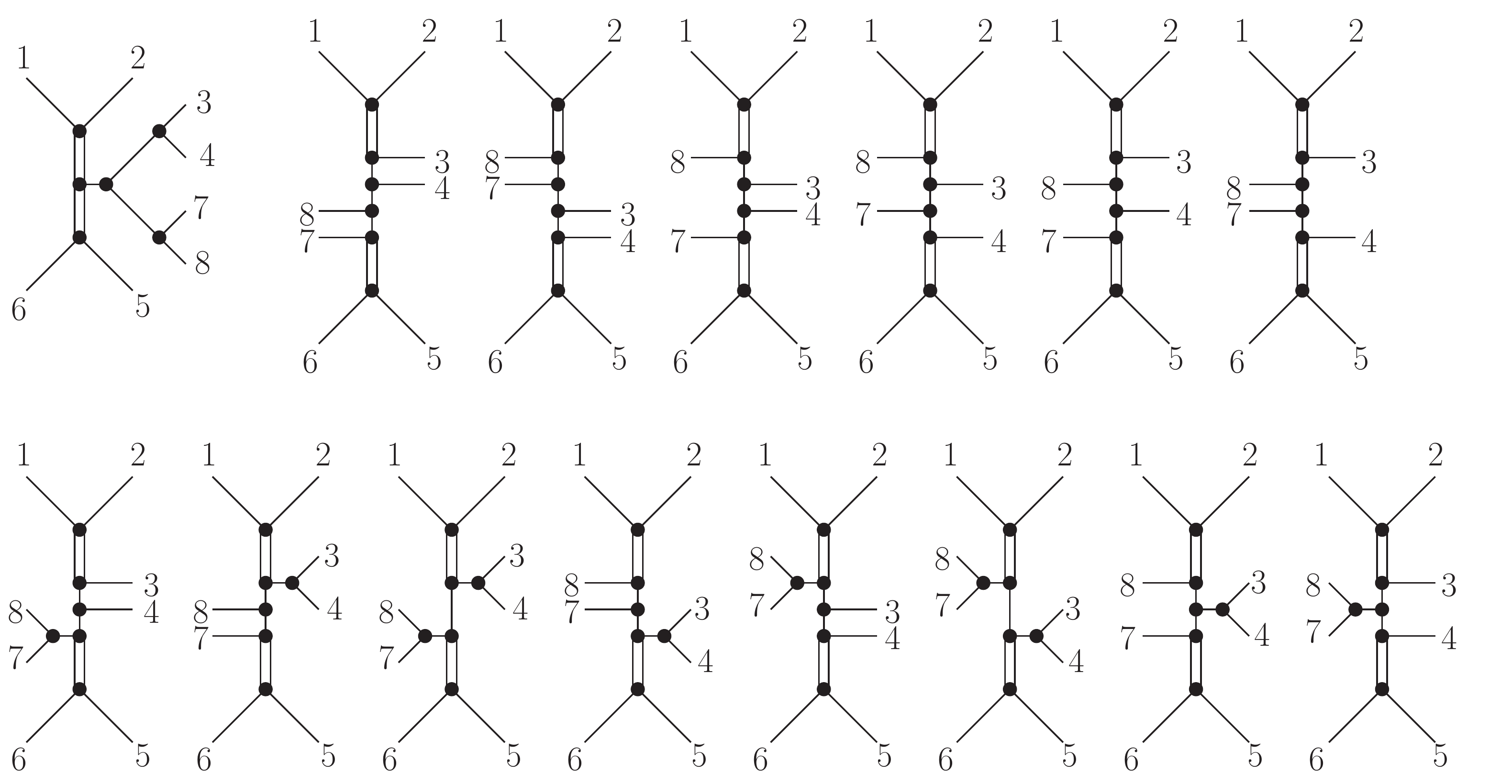}\\
  \caption{The 15 Feynman diagrams corresponding to the {\sl 4-regular} graph Figure \ref{Figsupportmix}.d.}\label{FigA811}
\end{figure}

This is an eight-point example with CHY-integrand
\bea {1\over z_{12}^3z_{56}^3
z_{34}^2z_{78}^2z_{23}z_{45}z_{67}z_{81}z_{47}z_{83}}~.~~~\eea
The possible subsets for poles are $\{\underline{1,2}\}$,
$\{\underline{5,6}\}$, $\{3,4\}$, $\{7,8\}$, $\{4,5,6\}$,
$\{5,6,7\}$, $\{1,2,3\}$, $\{1,2,8\}$, $\{3,4,7,8\}$, $\{1,2,3,4\}$,
$\{1,2,7,8\}$, $\{4,5,6,7\}$. From them we can construct 15
compatible combinations of subsets as
\bea
&&\{\{\underline{1,2}\}~,~\{\underline{5,6}\}~,~\{3,4,7,8\}~,~\{3,4\}~,~\{7,8\}\}~,~~~\nonumber\\
&&\{\{\underline{1,2}\}~,~\{\underline{5,6}\}~,~\{1,2,3,4\}~,~\Big[\{1,2,3\}~~\mbox{or}~~\{3,4\}\Big]~,~\Big[\{7,8\}~\mbox{or}~\{5,6,7\}\Big]~,~\nonumber\\
&&\{\{\underline{1,2}\}~,~\{\underline{5,6}\}~,~\{1,2,7,8\}~,~\Big[\{1,2,8\}~~\mbox{or}~~\{7,8\}\Big]~,~\Big[\{3,4\}~\mbox{or}~\{4,5,6\}\Big]~,~\nonumber\\
&&\{\{\underline{1,2}\}~,~\{\underline{5,6}\}~,~\{4,5,6,7\}~,~\Big[\{1,2,3\}~~\mbox{or}~~\{1,2,8\}\Big]~,~\Big[\{4,5,6\}~\mbox{or}~\{5,6,7\}\Big]~,~\nonumber\\
&&\{\{\underline{1,2}\}~,~\{\underline{5,6}\}~,~\{1,2,8\}~,~\{5,6,7\}~,~\{3,4\}\}~~,~~\{\{\underline{1,2}\}~,~\{\underline{5,6}\}~,~\{1,2,3\}~,~\{4,5,6\}~,~\{7,8\}\}~.~~~\nonumber\eea
The corresponding Feynman diagrams are presented in Figure
\ref{FigA811}.
So accordingly, the result can be written down by summing over all
15 Feynman diagrams, by applying Feynman rules $\rulei$, $\ruleiii$
to corresponding propagators, as
{\footnotesize \bea&&{1\over s_{34} s_{78} s_{3478}} \ruleIII{\{1\},
\{2\}, \{3, 4, 7, 8\}, \{5\}, \{6\}}\nonumber\\
&&+{1\over s_{12 3} s_{1 2 3 4} s_{5 6 7}} \ruleI{\{1\}, \{2\},
\{3\}, \{4, 5,
     6, 7, 8\}} \ruleI{\{7\}, \{8, 1, 2, 3, 4\}, \{5\}, \{6\}}\nonumber\\
&&+{1\over s_{1 2 3} s_{1 2 3 4} s_{7 8}} \ruleI{\{1\}, \{2\},
\{3\}, \{4, 5, 6,
     7, 8\}} \ruleI{\{7, 8\}, \{1, 2, 3, 4\}, \{5\}, \{6\}}\nonumber\\
&&+{1\over s_{3 4} s_{1 2 3 4} s_{5 6 7}} \ruleI{\{1\}, \{2\}, \{3,
4\}, \{5, 6,
     7, 8\}} \ruleI{\{7\}, \{8, 1, 2, 3, 4\}, \{5\}, \{6\}}\nonumber\\
&&+{1\over s_{1 2 3 4} s_{3 4} s_{7 8}} \ruleI{\{1\}, \{2\}, \{3,
4\}, \{5, 6, 7,
     8\}} \ruleI{\{7, 8\}, \{1, 2, 3, 4\}, \{5\}, \{6\}}\nonumber\\
&&+{1\over s_{1 2 7 8} s_{1 2 8} s_{4 5 6}} \ruleI{\{1\}, \{2\},
\{3, 4, 5, 6,
      7\}, \{8\}} \ruleI{\{7, 8, 1, 2, 3\}, \{4\}, \{5\}, \{6\}}\nonumber\\
&&+{1\over s_{1 2 7 8} s_{3 4} s_{1 2 8}} \ruleI{\{1\}, \{2\}, \{3,
4, 5, 6,
     7\}, \{8\}} \ruleI{\{7, 8, 1, 2\}, \{3, 4\}, \{5\}, \{6\}}\nonumber\\
&&+{1\over s_{1 2 7 8} s_{7 8} s_{4 5 6}} \ruleI{\{1\}, \{2\}, \{3,
4, 5,
     6\}, \{7, 8\}} \ruleI{\{7, 8, 1, 2, 3\}, \{4\}, \{5\}, \{6\}}\nonumber\nonumber\\
&& +{1\over s_{12 7 8} s_{7 8} s_{3 4}} \ruleI{\{1\}, \{2\}, \{3, 4,
5, 6\}, \{7,
     8\}} \ruleI{\{7, 8, 1, 2\}, \{3, 4\}, \{5\}, \{6\}}\nonumber\\
&&+{1\over s_{456 7} s_{1 2 8} s_{5 6 7}} \ruleI{\{1\}, \{2\}, \{3,
4, 5, 6,
      7\}, \{8\}} \ruleI{\{7\}, \{8, 1, 2, 3, 4\}, \{5\}, \{6\}}\nonumber\\
&&+{1\over s_{45 6 7} s_{1 2 8} s_{4 5 6}} \ruleI{\{1\}, \{2\}, \{3,
4, 5, 6,
      7\}, \{8\}} \ruleI{\{7, 8, 1, 2, 3\}, \{4\}, \{5\}, \{6\}}\nonumber\\
&&+{1\over s_{4 5 67} s_{12 3} s_{5 6 7}} \ruleI{\{1\}, \{2\},
\{3\}, \{4, 5,
     6, 7, 8\}} \ruleI{\{7\}, \{8, 1, 2, 3, 4\}, \{5\}, \{6\}}\nonumber\\
&&+{1\over s_{4 5 6 7} s_{1 2 3} s_{456}} \ruleI{\{1\}, \{2\},
\{3\}, \{4, 5,
     6, 7, 8\}} \ruleI{\{7, 8, 1, 2, 3\}, \{4\}, \{5\}, \{6\}}\nonumber\\
&&+{1\over s_{1 2 8} s_{5 6 7} s_{3 4}} \ruleI{\{1\}, \{2\}, \{3, 4,
5, 6,
     7\}, \{8\}} \ruleI{\{7\}, \{8, 1, 2, 3, 4\}, \{5\}, \{6\}}\nonumber\\
&&+{1\over s_{1 2 3} s_{4 5 6} s_{7 8}} \ruleI{\{1\}, \{2\}, \{3\},
\{4, 5, 6, 7,
     8\}} \ruleI{\{7, 8, 1, 2, 3\}, \{4\}, \{5\}, \{6\}}~.~~~\eea}
%

\subsection*{Example Figure \ref{Figsupportmix}.e:}

\begin{figure}[h]
  \centering
  \includegraphics[width=5in]{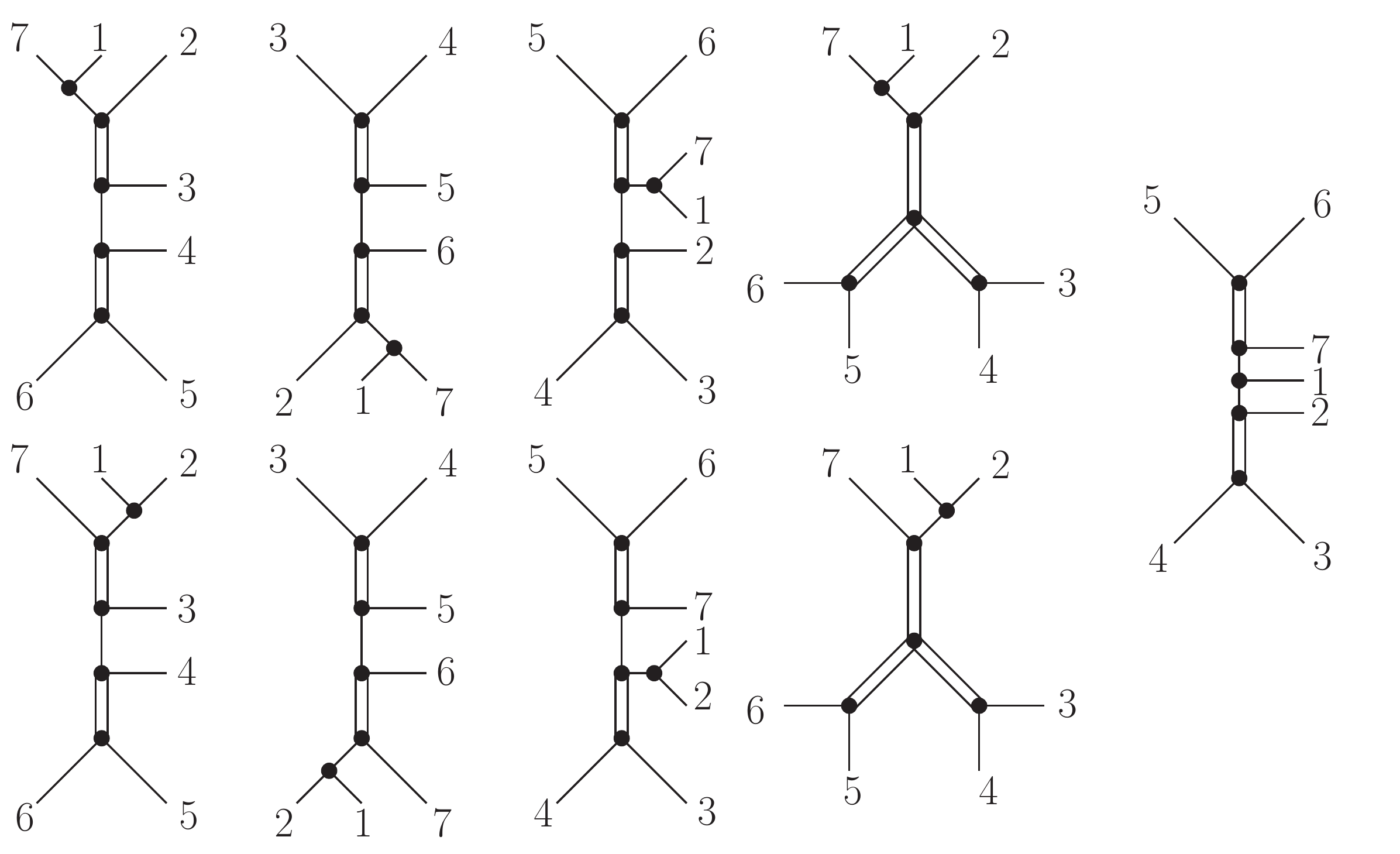}\\
  \caption{The nine Feynman diagrams corresponding to the {\sl 4-regular} graph Figure \ref{Figsupportmix}.e.}\label{FigA78}
\end{figure}

This is a seven-point example with CHY-integrand
\bea -{1\over
z_{71}^2z_{12}^2z_{34}^3z_{56}^3z_{72}z_{23}z_{45}z_{67}}~.~~~\eea
The possible subsets for poles are $\{\underline{1,2,7}\}$,
$\{\underline{3,4}\}$, $\{\underline{5,6}\}$, $\{2,3,4\}$,
$\{3,4,5\}$, $\{4,5,6\}$, $\{5,6,7\}$, $\{1,7\}$, $\{1,2\}$. From
them we can construct nine compatible combinations of subsets as
\bea &&\{\{\underline{1,2,7}\}, \{\underline{3,4}\},
\{\underline{5,6}\},
\Big[\{1,7\}~\mbox{or}~\{1,2\}\Big]\}~~,~~\{\{\underline{3,4}\},
\{\underline{5,6}\}, \{2,3,4\}, \{5,6,7\}\}~,~~~ \nonumber\\
&&\{\{\underline{3,4}\}, \{\underline{5,6}\}, \{2,3,4\},
\{1,7\}\}~~,~~\{\{\underline{3,4}\}, \{\underline{5,6}\}, \{5,6,7\},
\{1,2\}\}\nonumber\\
&&\{\{\underline{1,2,7}\}, \{\underline{3,4}\}, \{3,4,5\},
\Big[\{1,7\}~\mbox{or}~\{1,2\}\Big]\}~~,~~\{\{\underline{1,2,7}\},
\{\underline{5,6}\}, \{4,5,6\},
\Big[\{1,7\}~\mbox{or}~\{1,2\}\Big]\}~.~~~\nonumber\eea
The corresponding nine Feynman diagrams are shown in Figure
\ref{FigA78}.
Applying the rules, we simply get the result
\bea &&{1\over
s_{17}}\ruleIX{\{7,1\},\{2\},\{3\},\{4\},\{5\},\{6\}}+{1\over
s_{12}}\ruleIX{\{7\},\{1,2\},\{3\},\{4\},\{5\},\{6\}}\nonumber\\
&&+{1\over
s_{17}s_{456}}\ruleI{\{7,1\},\{2\},\{3\},\{4,5,6\}}\ruleI{\{7,1,2,3\},\{4\},\{5\},\{6\}}\nonumber\\
&&+{1\over
s_{17}s_{345}}\ruleI{\{3\},\{4\},\{5\},\{5,7,1,2\}}\ruleI{\{3,4,5\},\{6\},\{7,1\},\{2\}}\nonumber\\
&&+{1\over
s_{17}s_{234}}\ruleI{\{5\},\{6\},\{7,1\},\{2,3,4\}}\ruleI{\{5,6,7,1\},\{2\},\{3\},\{4\}}\nonumber\\
&&+{1\over
s_{12}s_{456}}\ruleI{\{7\},\{1,2\},\{3\},\{4,5,6\}}\ruleI{\{7,1,2,3\},\{4\},\{5\},\{6\}}\nonumber\\
&&+{1\over
s_{12}s_{345}}\ruleI{\{3\},\{4\},\{5\},\{6,7,1,2\}}\ruleI{\{3,4,5\},\{6\},\{7\},\{1,2\}}\nonumber\\
&&+{1\over
s_{12}s_{567}}\ruleI{\{5\},\{6\},\{7\},\{1,2,3,4\}}\ruleI{\{5,6,7\},\{1,2\},\{3\},\{4\}}\nonumber\\
&&+{1\over
s_{234}s_{567}}\ruleI{\{5\},\{6\},\{7\},\{1,2,3,4\}}\ruleI{\{5,6,7,1\},\{2\},\{3\},\{4\}}~.~~~\eea
%

\subsection*{Example Figure \ref{Figsupportmix}.f:}

\begin{figure}[h]
  \centering
  \includegraphics[width=7in]{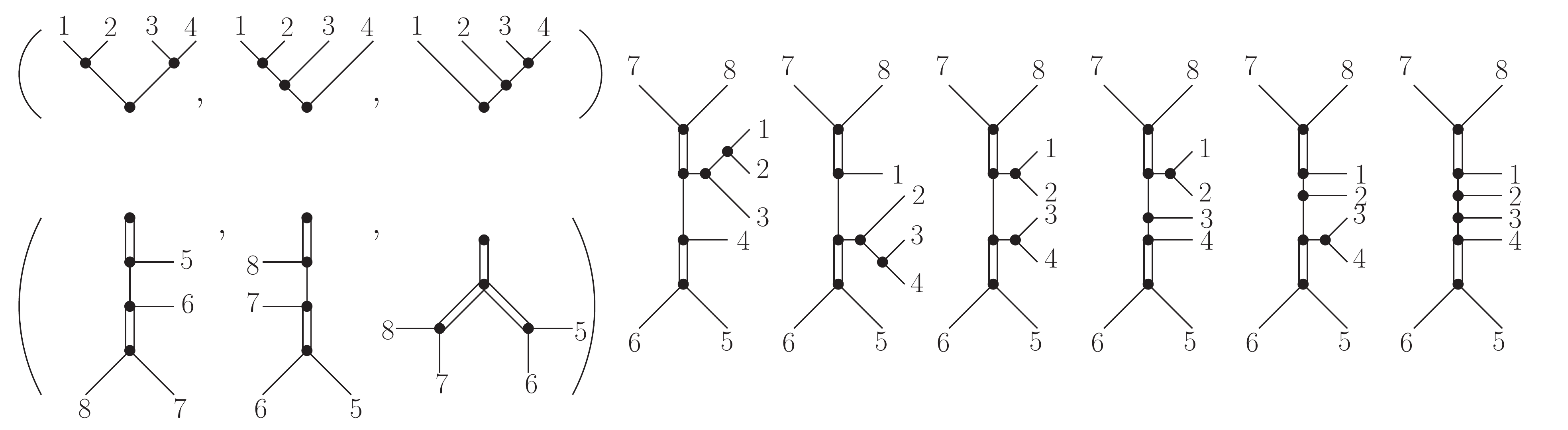}\\
  \caption{The 15 Feynman diagrams corresponding to the {\sl 4-regular} graph Figure \ref{Figsupportmix}.f.}\label{FigA810}
\end{figure}

This is an eight-point example with CHY-integrand
\bea -{1\over
z_{12}^2z_{34}^2z_{56}^3z_{78}^3z_{24}z_{45}z_{67}z_{81}z_{13}z_{32}}~.~~~\eea
The possible subsets for poles are $\{\underline{1,2,3,4}\}$,
$\{\underline{5,6}\}$, $\{\underline{7,8}\}$, $\{3,4,5,6\}$,
$\{4,5,6\}$, $\{5,6,7\}$, $\{6,7,8\}$, $\{7,8,1\}$, $\{1,2,3\}$,
$\{2,3,4\}$, $\{1,2\}$, $\{3,4\}$. There are 15 compatible
combinations of subsets from them, as
\bea &&\{\{\underline{1,2,3,4}\},
\{\underline{5,6}\},~\{\underline{7,8}\}, \Big[\{1,2,3\},
\{1,2\}~\mbox{or}~\{2,3,4\}, \{3,4\}~\mbox{or}~\{1,2\},
\{3,4\}\Big]\}~,~~~\nonumber\\
&&\{\{\underline{1,2,3,4}\}, \{\underline{5,6}\},~\{5,6,7\},
\Big[\{1,2,3\}, \{1,2\}~\mbox{or}~\{2,3,4\},
\{3,4\}~\mbox{or}~\{1,2\}, \{3,4\}\Big]\}~,~~~\nonumber\\
&&\{\{\underline{1,2,3,4}\}, \{\underline{7,8}\},~\{6,7,8\},
\Big[\{1,2,3\}, \{1,2\}~\mbox{or}~\{2,3,4\},
\{3,4\}~\mbox{or}~\{1,2\}, \{3,4\}\Big]\}~,~~~\nonumber\\
&&\{\{\underline{5,6}\}, \{\underline{7,8}\},~\{3,4,5,6\},
\Big[\{3,4\}~\mbox{or}~\{4,5,6\}\Big],
\Big[\{1,2\}~\mbox{or}~\{7,8,1\}\Big]\}~,~~~\nonumber\\
&&\{\{\underline{5,6}\}, \{\underline{7,8}\},~\{4,5,6\}, \{1,2,3\},
\{1,2\} \}~~,~~\{\{\underline{5,6}\},
\{\underline{7,8}\},~\{7,8,1\}, \{2,3,4\},
\{3,4\}\}~.~~~\nonumber\eea
The corresponding 15 Feynman diagrams are illustrated in Figure
\ref{FigA810}.
The result is trivially given by summing over all 15 Feynman
diagrams with $\rulei$, $\ruleix$, and to save space we will not
write it down explicitly here.

\subsection*{Example Figure \ref{Figsupportmix}.g:}

\begin{figure}[h]
  \centering
  \includegraphics[width=7in]{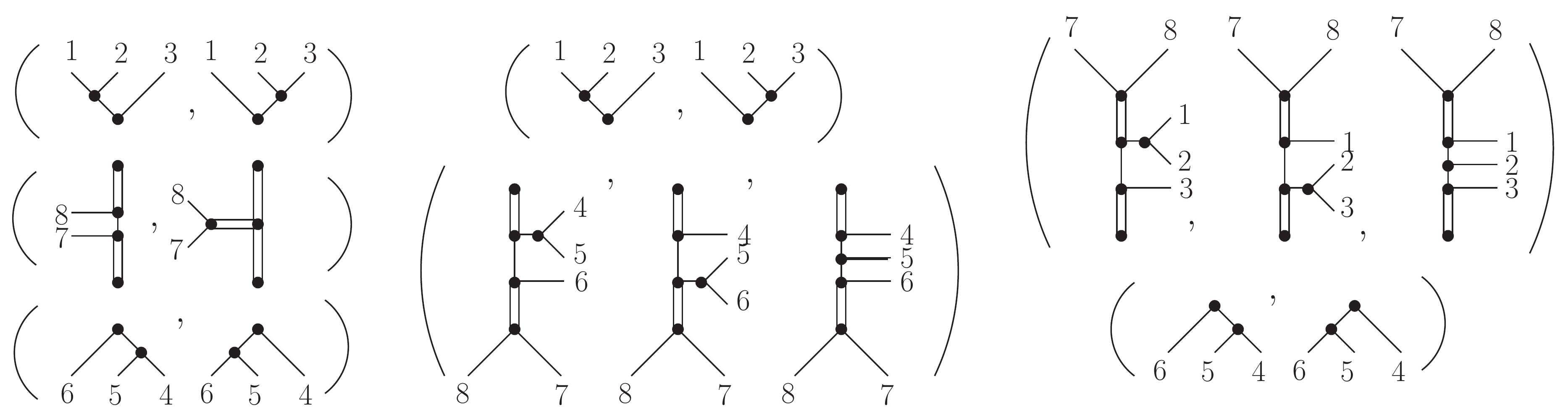}\\
  \caption{The 20 Feynman diagrams corresponding to the {\sl 4-regular} graph Figure \ref{Figsupportmix}.g.}\label{FigA812}
\end{figure}

This is an eight-point example with CHY-integrand
\bea {1\over
z_{12}^2z_{23}^2z_{45}^2z_{56}^2z_{78}^3z_{13}z_{34}z_{46}z_{67}z_{81}}~.~~~\eea
The possible subsets for poles are $\{\underline{1,2,3}\}$,
$\{\underline{4,5,6}\}$, $\{\underline{7,8}\}$, $\{1,2\}$,
$\{2,3\}$, $\{4,5\}$, $\{5,6\}$, $\{1,7,8\}$, $\{6,7,8\}$,
$\{1,2,3,4\}$, $\{1,2,3,8\}$, $\{1,2,7,8\}$. From them we can
construct 20 compatible combinations of subsets as
\bea
&&\{\{\underline{1,2,3}\},\{\underline{4,5,6}\},\{\underline{7,8}\},\Big[\{1,2\}~\mbox{or}~\{2,3\}\Big],\Big[\{4,5\}~\mbox{or}~\{5,6\}\Big]\}~,~~~\nonumber\\
&&\{\{\underline{1,2,3}\},\{\underline{4,5,6}\},\{1,2,3,8\},\Big[\{1,2\}~\mbox{or}~\{2,3\}\Big],\Big[\{4,5\}~\mbox{or}~\{5,6\}\Big]\}~,~~~\nonumber\\
&&\{\{\underline{1,2,3}\},\{\underline{7,8}\},\Big[\{1,2\}~\mbox{or}~\{2,3\}\Big],\Big[\{4,5\},
\{6,7,8\}~\mbox{or}~\{5,6\}, \{1,2,3,4\}~\mbox{or}~\{6,7,8\},
\{1,2,3,4\}\Big]\}~,~~~\nonumber\\
&&\{\{\underline{4,5,6}\},\{\underline{7,8}\},\Big[\{4,5\}~\mbox{or}~\{5,6\}\Big],\Big[\{1,2\},
\{1,2,7,8\}~\mbox{or}~\{2,3\}, \{1,7,8\}~\mbox{or}~\{1,7,8\},
\{1,2,7,8\}\Big]\}~.~~~\nonumber\eea
The corresponding 20 Feynman diagrams for this CHY-integrand is
presented in Figure \ref{FigA812}.
By applying Feynman rules on corresponding propagators, one
immediately get the result, which we will not repeat here.

\subsection*{Example Figure \ref{Figsupportmix}.h:}

\begin{figure}[h]
  \centering
  \includegraphics[width=7in]{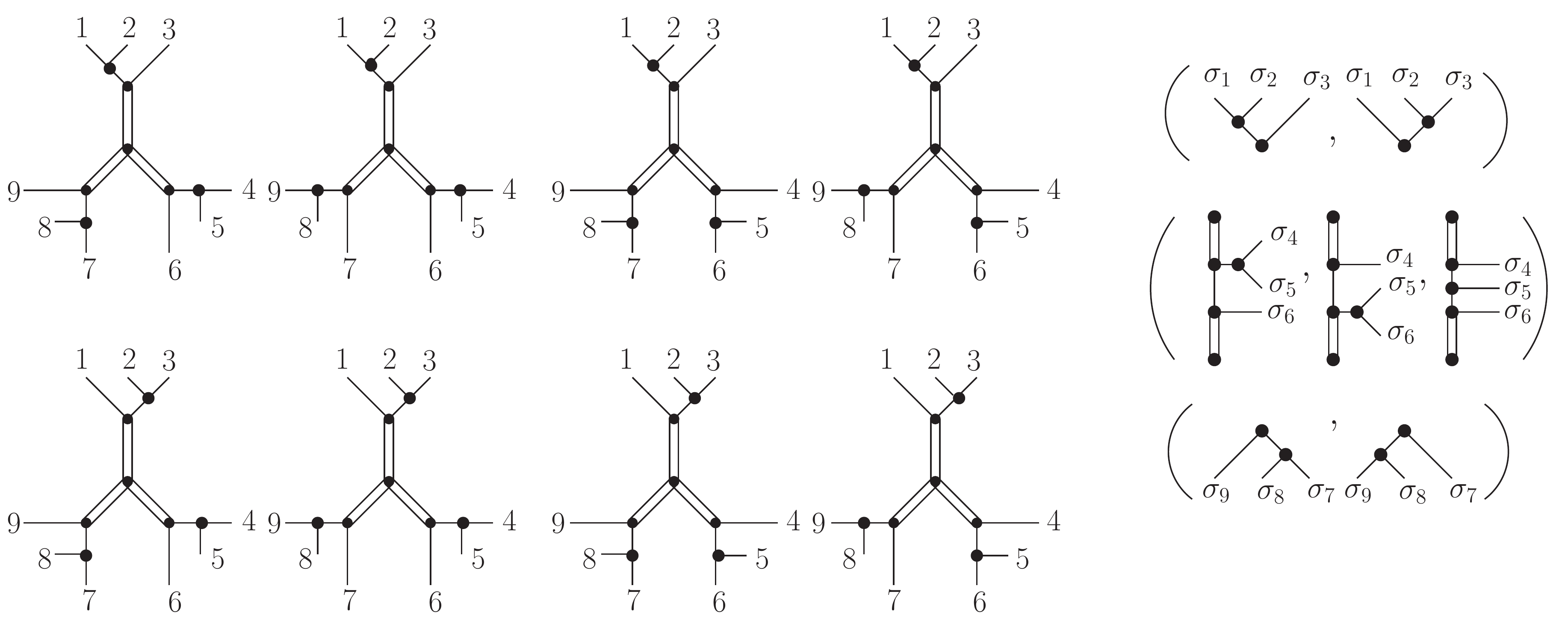}\\
  \caption{The 44 Feynman diagrams corresponding to the {\sl 4-regular} graph Figure \ref{Figsupportmix}.h. $\{\sigma_1,\ldots, \sigma_9\}$ takes
  value of $\{1,2,3,4,5,6,7,8,9\}$, $\{4,5,6,7,8,9,1,2,3\}$ and $\{7,8,9,1,2,3,4,5,6\}$.}\label{FigA93}
\end{figure}

This is a nine-point example with CHY-integrand
\bea -{1\over z_{12}^2 z_{23}^2 z_{45}^2 z_{56}^2 z_{78}^2 z_{89}^2
z_{13} z_{34} z_{46} z_{67} z_{79} z_{91}}~.~~~\eea
The possible subsets for poles are $\{\underline{1,2,3}\}$,
$\{\underline{4,5,6}\}$, $\{\underline{7,8,9}\}$, $\{1,2\}$,
$\{2,3\}$, $\{4,5\}$, $\{5,6\}$, $\{7,8\}$, $\{8,9\}$,
$\{1,2,3,4\}$, $\{1,2,3,9\}$, $\{1,7,8,9\}$, $\{3,4,5,6\}$,
$\{4,5,6,7\}$, $\{6,7,8,9\}$. From them we can construct 44
compatible combinations of subsets as
\bea &&\{\{\underline{1,2,3}\}, \{\underline{4,5,6}\},
\{\underline{7,8,9}\}, \Big[\{1,2\}~\mbox{or}~\{2,3\}\Big],
\Big[\{4,5\}~\mbox{or}~\{5,6\}\Big],
\Big[\{7,8\}~\mbox{or}~\{8,9\}\Big]\}~,~~~\nonumber\\
&&\{\{\underline{1,2,3}\}, \{\underline{4,5,6}\},
\Big[\{1,2\}~\mbox{or}~\{2,3\}\Big],
\Big[\{4,5\}~\mbox{or}~\{5,6\}\Big], \nonumber\\
&&~~~~~~~~~~~~~~~~~~~~~\Big[\{7,8\},
\{1,2,3,9\}~~\mbox{or}~~\{1,2,3,9\},
\{4,5,6,7\}~~\mbox{or}~~\{8,9\},
\{4,5,6,7\}\Big]\}~,~~~\nonumber\\
&&\{\{\underline{1,2,3}\}, \{\underline{7,8,9}\},
\Big[\{1,2\}~\mbox{or}~\{2,3\}\Big],
\Big[\{7,8\}~\mbox{or}~\{8,9\}\Big], \nonumber\\
&&~~~~~~~~~~~~~~~~~~~~~\Big[\{4,5\}, \{6,7,8,9\}~~\mbox{or}~~
\{1,2,3,4\}, \{6,7,8,9\}~~\mbox{or}~~ \{5,6\},
\{1,2,3,4\}\Big]\}~,~~~\nonumber\\
&&\{\{\underline{4,5,6}\}, \{\underline{7,8,9}\},
\Big[\{4,5\}~\mbox{or}~\{5,6\}\Big],
\Big[\{7,8\}~\mbox{or}~\{8,9\}\Big], \nonumber\\
&&~~~~~~~~~~~~~~~~~~~~~\Big[\{1,2\}, \{3,4,5,6\}~~\mbox{or}~~
\{3,4,5,6\}, \{1,7,8,9\}~~\mbox{or}~~ \{2,3\},
\{1,7,8,9\}\Big]\}~.~~~\nonumber\eea
Then we can draw the corresponding 44 Feynman diagrams as in Figure
\ref{FigA93}. When summing over all 44 Feynman diagrams with
$\rulei$, $\ruleix$, it miraculously produces the correct result,
which is confirmed numerically.

\section{Discussion and conclusion}

The {\sl integration rule} presented in \cite{Baadsgaard:2015voa,
Baadsgaard:2015ifa} is quite efficient and elegant. As the word {\sl
rule} suggests, there is actually no practical computation but just
pattern matching. But the {\sl integration algorithm} is limited to
CHY-integrands with simple poles therein. In order to compute
CHY-integrands with higher-order poles, one need to use for example
the Pfaffian identities. Although the Pfaffian identities can relate
a CHY-integrand of higher-order poles with those of simple poles,
the number of terms in the identities suffers from factorial
increasing. Also generally one Pfaffian identity is not enough to
completely decompose a CHY-integrand of higher-order poles into
terms with only simple poles, which makes computation involved. What
is worse, some CHY-integrands, for example the one in Figure
\ref{Figsupport2}.c, as already stated in \cite{Baadsgaard:2015voa},
has no corresponding {\sl 3-regular} graph. So it can not be related
to other CHY-integrand by Pfaffian identities at all.

In this paper, we sharpen the {\sl integration algorithm} by
providing Feynman rules for higher-order poles. With these rules, we
can deal with CHY-integrands with corresponding higher-order poles
exactly in the same sprint as in \cite{Baadsgaard:2015voa,
Baadsgaard:2015ifa}. For example again the CHY-integrand Figure
\ref{Figsupport2}.c, previously not solved, can be instantly
obtained by summing over nine Feynman diagrams of our Feynman rule
$\ruleii$. Ample examples have been checked, not limited to those
provided in \S\ref{secSupporting} of this paper, to support the
validation of these rules. Again as the name {\sl rule} suggests, if
we forget the hard work spent in working out these rules, then the
computation in fact involves no computation but just pattern
matching, and all results come out instantly in Mathematica.

\begin{figure}
  \centering
  \includegraphics[width=6.5in]{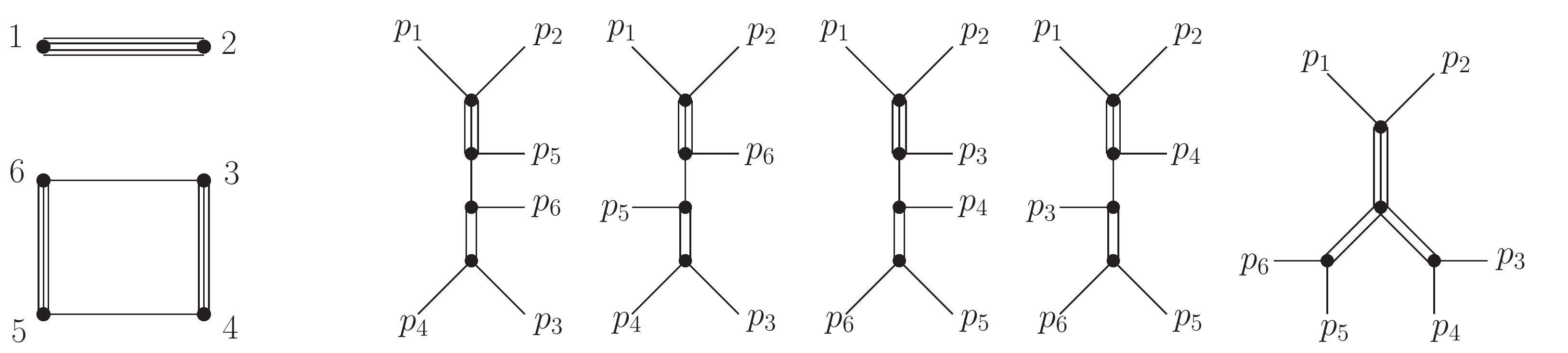}\\
  \caption{An example of CHY-integrand containing new pole structure with Feynman rule unknown.}\label{Figalien}
\end{figure}

In a recent paper \cite{Gomez:2016bmv}, Gomez proposed a so-called
$\Lambda$-algorithm, to compute CHY-integrands with higher-order
poles, in the framework of $\Lambda$ scattering equations. The
$\Lambda$-algorithm is quite elegant and general, but gauge-fixing
is required during computation. One need to choose a proper
gauge-fixing which involves no singular configurations, which is not
always possible. As stated in \cite{Gomez:2016bmv}, for example, the
CHY-integrand Figure \ref{Figsupport2}.c, can only be computed with
the help of general KLT construction. But with {\sl integration
algorithm} strengthened with Feynman rules of higher-order poles,
all the examples therein, e.g., Figures \ref{FigA52}, \ref{FigA53},
\ref{Figsupport2}.b and even eight-point example
\ref{Figsupportmix}.d, can be evaluated without effects. We remark
that, in \cite{Gomez:2016bmv}, $2P_i\cdot P_j$ is chosen to be the
kinematic variables in $\Lambda$-algorithm but not $s_{ij}$, which
is important for it to produce correct answer. In our Feynman rules
of higher-order poles, we are forced to take the same choice. This
would not be a coincidence, and we are eager to find out if there is
any deep connection between these two approaches.

However, the {\sl integration algorithm} also suffers from the
problem that, although the higher-order poles have the standard
structure of Feynman diagrams, they are {\sl quasi-local}, and new
pole structures would appear so that one need to create new rules
for them. For example, when computing the CHY-integrand represented
by {\sl 4-regular} graph as shown in Figure \ref{Figalien}, we end
with five Feynman diagrams. The first four can be computed by using
Feynman rules $\rulei$, $\ruleii$ of double pole and triple pole,
but for the last one, we need a new rule for the pole structure
where a triple propagator and two double propagators are connected
at one vertex. Since we have no generic method to {\sl derive} the
rule, it would be a problem for applying {\sl integration algorithm}
in most generic situations.

Based on current results of Feynman rules for higher-order poles,
there are several directions worth to explore. The first and most
important one is to derive these rules analytically. One could
either achieve this in the CHY framework using the same method as in
\cite{Baadsgaard:2015voa,Baadsgaard:2015ifa}, or use the newly
proposed $\Lambda$-formalism in \cite{Gomez:2016bmv}. The latter one
might be more convenient, since it naturally contains the pinching
picture and iterative construction. The second considers the
question that whether and how Feynman rules of different
higher-order pole structures are related to each other. Although we
have observed different Feynman rules for different higher-order
poles, it would be very interesting to see how far would the {\sl
quasi-local} property preventing us from reducing all other Feynman
rules of higher-order poles(e.g., $\ruleix$), to those with only a
single double or triple pole. Although as shown in
\cite{Baadsgaard:2015voa}, the Pfaffian identities\footnote{There is
a related question whether Pfaffian identities are complete or not,
i.e., could they provide all independent relations among different
pole structures so that all higher-order poles can be related to
simple poles?} can relate one higher-order pole to another, there
might be other better approaches, and one need to figure out the
details. Finally, with the Feynman rules of higher-order poles, one
is ready to apply it in explicit theories and exploit details
therein.

\section*{Acknowledgments}

We would like to thank Song He, Gang Yang and Carlos Cardona for
insightful discussions. This work is supported by Qiu-Shi Funding
and the National Natural Science Foundation of China(NSFC) with
Grant No.11135006, No.11125523 and No.11575156. RH would also like
to acknowledge the supporting from Chinese Postdoctoral
Administrative Committee.

\appendix


\bibliographystyle{JHEP}
\bibliography{mapping}

\providecommand{\href}[2]{#2}\begingroup\raggedright\begin{thebibliography}{10}

\bibitem{Cachazo:2013gna}
F.~Cachazo, S.~He, and E.~Y. Yuan, {\it {Scattering equations and
  Kawai-Lewellen-Tye orthogonality}},  {\em Phys. Rev.} {\bf D90} (2014), no.~6
  065001, [\href{http://arxiv.org/abs/1306.6575}{{\tt arXiv:1306.6575}}].

\bibitem{Cachazo:2013hca}
F.~Cachazo, S.~He, and E.~Y. Yuan, {\it {Scattering of Massless Particles in
  Arbitrary Dimensions}},  {\em Phys. Rev. Lett.} {\bf 113} (2014), no.~17
  171601, [\href{http://arxiv.org/abs/1307.2199}{{\tt arXiv:1307.2199}}].

\bibitem{Cachazo:2013iea}
F.~Cachazo, S.~He, and E.~Y. Yuan, {\it {Scattering of Massless Particles:
  Scalars, Gluons and Gravitons}},  {\em JHEP} {\bf 07} (2014) 033,
  [\href{http://arxiv.org/abs/1309.0885}{{\tt arXiv:1309.0885}}].

\bibitem{Cachazo:2014nsa}
F.~Cachazo, S.~He, and E.~Y. Yuan, {\it {Einstein-Yang-Mills Scattering
  Amplitudes From Scattering Equations}},  {\em JHEP} {\bf 01} (2015) 121,
  [\href{http://arxiv.org/abs/1409.8256}{{\tt arXiv:1409.8256}}].

\bibitem{Cachazo:2014xea}
F.~Cachazo, S.~He, and E.~Y. Yuan, {\it {Scattering Equations and Matrices:
  From Einstein To Yang-Mills, DBI and NLSM}},  {\em JHEP} {\bf 07} (2015) 149,
  [\href{http://arxiv.org/abs/1412.3479}{{\tt arXiv:1412.3479}}].

\bibitem{Dolan:2014ega}
L.~Dolan and P.~Goddard, {\it {The Polynomial Form of the Scattering
  Equations}},  {\em JHEP} {\bf 07} (2014) 029,
  [\href{http://arxiv.org/abs/1402.7374}{{\tt arXiv:1402.7374}}].

\bibitem{Weinzierl:2014vwa}
S.~Weinzierl, {\it {On the solutions of the scattering equations}},  {\em JHEP}
  {\bf 04} (2014) 092, [\href{http://arxiv.org/abs/1402.2516}{{\tt
  arXiv:1402.2516}}].

\bibitem{Du:2016blz}
Y.-j. Du, F.~Teng, and Y.-s. Wu, {\it {CHY formula and MHV amplitudes}},
  \href{http://arxiv.org/abs/1603.08158}{{\tt arXiv:1603.08158}}.

\bibitem{Kalousios:2015fya}
C.~Kalousios, {\it {Scattering equations, generating functions and all massless
  five point tree amplitudes}},  {\em JHEP} {\bf 05} (2015) 054,
  [\href{http://arxiv.org/abs/1502.07711}{{\tt arXiv:1502.07711}}].

\bibitem{Huang:2015yka}
R.~Huang, J.~Rao, B.~Feng, and Y.-H. He, {\it {An Algebraic Approach to the
  Scattering Equations}},  {\em JHEP} {\bf 12} (2015) 056,
  [\href{http://arxiv.org/abs/1509.04483}{{\tt arXiv:1509.04483}}].

\bibitem{Sogaard:2015dba}
M.~Sogaard and Y.~Zhang, {\it {Scattering Equations and Global Duality of
  Residues}},  \href{http://arxiv.org/abs/1509.08897}{{\tt arXiv:1509.08897}}.

\bibitem{Cardona:2015ouc}
C.~Cardona and C.~Kalousios, {\it {Elimination and recursions in the scattering
  equations}},  {\em Phys. Lett.} {\bf B756} (2016) 180--187,
  [\href{http://arxiv.org/abs/1511.05915}{{\tt arXiv:1511.05915}}].

\bibitem{Dolan:2015iln}
L.~Dolan and P.~Goddard, {\it {General Solution of the Scattering Equations}},
  \href{http://arxiv.org/abs/1511.09441}{{\tt arXiv:1511.09441}}.

\bibitem{Cardona:2015eba}
C.~Cardona and C.~Kalousios, {\it {Comments on the evaluation of massless
  scattering}},  {\em JHEP} {\bf 01} (2016) 178,
  [\href{http://arxiv.org/abs/1509.08908}{{\tt arXiv:1509.08908}}].

\bibitem{Lam:2015sqb}
C.~S. Lam and Y.-P. Yao, {\it {The Role of M\"obius Constants and Scattering
  Functions in CHY Scalar Amplitudes}},
  \href{http://arxiv.org/abs/1512.05387}{{\tt arXiv:1512.05387}}.

\bibitem{Lam:2016tlk}
C.~S. Lam and Y.-P. Yao, {\it {Evaluation of the CHY Gauge Amplitude}},
  \href{http://arxiv.org/abs/1602.06419}{{\tt arXiv:1602.06419}}.

\bibitem{Cachazo:2015nwa}
F.~Cachazo and H.~Gomez, {\it {Computation of Contour Integrals on ${\cal
  M}_{0,n}$}},  \href{http://arxiv.org/abs/1505.03571}{{\tt arXiv:1505.03571}}.

\bibitem{Baadsgaard:2015voa}
C.~Baadsgaard, N.~E.~J. Bjerrum-Bohr, J.~L. Bourjaily, and P.~H. Damgaard, {\it
  {Integration Rules for Scattering Equations}},
  \href{http://arxiv.org/abs/1506.06137}{{\tt arXiv:1506.06137}}.

\bibitem{Baadsgaard:2015ifa}
C.~Baadsgaard, N.~E.~J. Bjerrum-Bohr, J.~L. Bourjaily, and P.~H. Damgaard, {\it
  {Scattering Equations and Feynman Diagrams}},
  \href{http://arxiv.org/abs/1507.00997}{{\tt arXiv:1507.00997}}.

\bibitem{Gomez:2016bmv}
H.~Gomez, {\it {$\Lambda$ Scattering Equations}},
  \href{http://arxiv.org/abs/1604.05373}{{\tt arXiv:1604.05373}}.

\bibitem{Baadsgaard:2015hia}
C.~Baadsgaard, N.~E.~J. Bjerrum-Bohr, J.~L. Bourjaily, P.~H. Damgaard, and
  B.~Feng, {\it {Integration Rules for Loop Scattering Equations}},
  \href{http://arxiv.org/abs/1508.03627}{{\tt arXiv:1508.03627}}.

\bibitem{Feng:2016nrf}
B.~Feng, {\it {CHY-construction of Planar Loop Integrands of Cubic Scalar
  Theory}},  \href{http://arxiv.org/abs/1601.05864}{{\tt arXiv:1601.05864}}.

\end{thebibliography}\endgroup

\end{document}